\documentclass[fleqn,usenatbib]{mnras}

\usepackage{newtxtext,newtxmath,siunitx}

\usepackage[T1]{fontenc}
\usepackage{ae,aecompl}

\usepackage{graphicx}	
\usepackage{amsmath}	
\usepackage{amssymb}	

\usepackage{setspace,pdflscape}
\usepackage{titlesec}
\titleformat{\paragraph}{\normalfont\small\bfseries}{}{}{}
\usepackage{booktabs,graphicx,amsmath,dcolumn,xcolor,braket,cleveref}
\usepackage[version=4]{mhchem}

\usepackage{epstopdf}

\newcommand{\Abinitio}{\emph{Ab initio}}
\newcommand{\abinitio}{\emph{ab initio}}
\newcommand{\cm}{cm$^{-1}$}

\newcommand{\X}{X~$^3\Delta$}
\newcommand{\E}{E~$^3\Pi$}
\newcommand{\D}{D~$^3\Sigma^-$}
\newcommand{\A}{A~$^3\Phi$}
\newcommand{\B}{B~$^3\Pi$}
\newcommand{\C}{C~$^3\Delta$}
\newcommand{\F}{F~$^3\Sigma^+$}
\newcommand{\Sa}{a~$^1\Delta$}
\newcommand{\Sd}{d~$^1\Sigma^+$}
\newcommand{\Sb}{b~$^1\Pi$}
\newcommand{\Sg}{g~$^1\Gamma$}
\newcommand{\Sh}{h~$^1\Sigma^+$}
\newcommand{\Sc}{c~$^1\Phi$}
\newcommand{\Sf}{f~$^1\Delta$}
\newcommand{\Si}{i~$^1\Pi$}
\newcommand{\Se}{e~$^1\Sigma^+$}
\usepackage{array}
\newcolumntype{H}{>{\setbox0=\hbox\bgroup}c<{\egroup}@{}}
\newcommand{\mc}{\multicolumn}
\newcommand{\TiO}{\ce{^{48}Ti^{16}O}}
\newcommand{\Duo}{{\sc Duo}}
\newcommand{\Marvel}{{\sc Marvel}}
\newcommand{\PGopher}{{\sc PGopher}}
\newcommand{\LLname}{{\sc Toto}}

\newcommand{\pk}{k}
\newcommand{\m}{m}
\newcommand{\pt}{t}

\newcommand{\pll}{l_2}

\newcommand{\scinot}[1]{\num[scientific-notation=true,round-mode=places,round-precision=4]{#1}}
\newcommand{\noELs}{301,245}
\newcommand{\notrans}{58,983,952}
\newcommand{\noMARVEL}{17,802}
\newcommand{\e}{{\rm e}}

\title[ExoMol XXXIII: TiO]{ExoMol Molecular linelists -- XXXIII. The spectrum of Titanium Oxide}

\author[McKemmish et al.]{Laura K. McKemmish,$^{1,2}$ Thomas Masseron,$^{3,4}$ H. Jens Hoeijmakers,$^{5,6}$, \newauthor
V\'ictor P\'{e}rez-Mesa$^{3,4}$,
Simon L. Grimm$^5$, Sergei N. Yurchenko$^2$ \newauthor
and Jonathan Tennyson.$^2$
\\
$^{1}$ School of Chemistry, University of New South Wales, 2052 Sydney,\\
$^2$ Department of Physics and Astronomy, University College London, Gower Street, WC1E 6BT London, UK\\
$^3$ Instituto de Astrof\'isica de Canarias, E-38205 La Laguna, Tenerife, Spain \\
$^4$ Departamento de Astrof\'isica, Universidad de La Laguna, E-38206 La Laguna, Tenerife, Spain \\
$^5$ University of Bern                                            , Center for Space and Habitability, Gesellschaftsstrasse 6, CH-3012, Bern, Switzerland\\
$^6$ University of Geneva, Geneva Observatory, Chemin des Maillettes 51, CH-1290, Versoix, Switzerland}

\date{Accepted XXX. Received YYY; in original form ZZZ}

\pubyear{2019}

\begin{document}
\label{firstpage}
\pagerange{\pageref{firstpage}--\pageref{lastpage}}
\maketitle

\begin{abstract}
Accurate line lists are crucial for correctly modelling a variety of astrophysical phenomena, including stellar photospheres and the atmospheres of extra-solar planets. This paper presents a new line database \LLname{} for the main isotopologues of titanium oxide (TiO): \ce{^{46}Ti^{16}O}, \ce{^{47}Ti^{16}O}, \ce{^{48}Ti^{16}O}, \ce{^{49}Ti^{16}O} and \ce{^{50}Ti^{16}O}. The \TiO\ line list contains transitions with wave-numbers up to 30,000 cm$^{-1}$ ie long-wards of 0.33 $\mu$m. The \LLname{} line list includes all dipole-allowed transitions between  13 low-lying electronic states (\X, \Sa, \Sd, \E, \A, \B, \C, \Sb, \Sc, \Sf, \Se).
{\it Ab initio} potential energy curves (PECs) are computed at the icMRCI level and combined with  spin-orbit and other coupling curves.
These PECs and couplings are iteratively  refined to match known empirical energy levels.
Accurate line intensities are generated using {\it ab initio} dipole moment curves.
The \LLname{} line lists are appropriate for temperatures below 5000 K and contain 30 million transitions for \TiO;  it is made available in electronic form via the CDS data centre and via www.exomol.com. Tests of the
line lists show greatly improved agreement with observed spectra for objects such as M-dwarfs GJ876 and GL581.
\end{abstract}




\section{Introduction}

Titanium oxide, TiO, is the most important transition metal diatomic in astronomy, dominating the spectra of cool stars in the near-IR and visible region, and was recently found to be present in the atmosphere of at least one hot gas giant exoplanet \citep{17SeBoMa.TiO,17NuKaHa.TiO}.  This makes TiO an important probe of the physics that govern these types astronomical objects, e.g. temperature, radiative properties and chemistry.
However, being able to accurately model opacities of key molecules in stars and planets is critical for thorough testing and refinement of models of their formation and evolution.
Because TiO generates strong opacity across the visible and near-IR, TiO line lists are important ingredients of models of a variety of objects, including cool giant stars \citep{97AlMexx.TiO,03CoEvSa.TiO, 13DaKuPl.TiO,04JuKoxx.TiO},
cool dwarf stars \citep{94PeShBe.TiO,95AlHaxx.TiO,99JoVaKo.TiO,09CuLoBu.TiO,00DaRoxx.TiO}, protostars \citep{00AlHaSc.TiO,12HiKnPa.TiO}, brown dwarfs  \citep{95AlHaxx.TiO,03ClTiHo.TiO,02ClTiCo.TiO,03ClTiHo.TiO,07MaMaFo.TiO,98MaBaZa.TiO,00PaOsRe.TiO,07ReHoHa.TiO}
and hot Jupiter exoplanets \citep{10FoShSh.TiO,08DeVide.TiO,08FoLoMa.TiO,09ShFoLi.TiO,09SpSiBu.TiO}. Brown dwarfs, red dwarfs and hot Jupiter exoplanets will be particularly important objects for astrophysical study in the next few decades as high quality spectroscopic observations of a large sample of these objects are becoming available. TiO is key to answering critical questions about the structure of these objects; for example, \citet{17GhMa.TiO} have shown that the abundance of TiO is crucial in determining the degree of temperature inversion in giant exoplanets.

There have been a number of previous \abinitio{} studies of the potential energy curves and spectroscopy of \TiO, including those of \citet{83BaBaNe.TiO,87DoWeSt.TiO,87SeScxx.TiO,90BaLaKo.TiO,92ScSeCa.TiO,97Laxxxx.TiO,10MiMaxx.TiO}.

The two most important of these studies are by \citet{97Laxxxx.TiO} and from \cite{10MiMaxx.TiO}. 
\cite{97Laxxxx.TiO} performed very sophisticated multi-reference \abinitio{} calculations, especially for the time period, for the potential energy curves and relevant electronic transition moments of TiO.
\cite{10MiMaxx.TiO} performed higher level calculations than \cite{97Laxxxx.TiO}, but considered only the properties of individual states and not the transition moments that are of critical importance to spectroscopy.

Because of its importance, there is a long history of  generating line lists for TiO
\citep{75Coxxxx.TiO,76CoFaxx.TiO,92Plxxxx.TiO,94Joxxxx.TiO,98Plxxxx.TiO,98Scxxxx.TiO}.
Of these we highlight the line lists due to (a)
\cite{98Scxxxx.TiO} as this line list is constructed through a variational treatment of the nuclear motion equation explicitly incorporating spin-orbit coupling between some electronic states,
and (b) \cite{98Plxxxx.TiO}  which  has been continually updated to include molecular constants and experimental transition frequencies \citep{VALD3}.
Among astronomers and through the research conducted for this publication, the updated version of the \cite{98Plxxxx.TiO} line list as detailed in \cite{VALD3} is known to perform the best especially when modeling higher resolution spectra.  
However, recent studies \citep{15HoDeSn.TiO} have emphasised deficiencies in existing line lists in terms of completeness and accuracy which have motivated this current update.

Traditionally, spectra are fit using model or effective Hamiltonian, employing software packages such as \PGopher{} \citep{PGOPHER}. The \cite{98Plxxxx.TiO} line list, and its subsequent incarnations, used this approach.
With the advent of high performance computers, however, it has become possible to solve the nuclear motion Hamiltonian directly using variational methods.  \cite{98Scxxxx.TiO}
was one of the first cases of a sophisticated treatment enabling coupling of different electronic states.

The ExoMol project \citep{jt528} aims to produce high temperature line lists of spectroscopic transitions for key molecular species likely to be significant in the analysis of the atmospheres of extrasolar planets and cool stars. Transition metal (TM) species have proven to be the most difficult  to produce accurate line lists for \citep{jt631} primarily due to their complicated electronic structure, density
of electronic states and the relatively low accuracy of available \abinitio\ electronic structure methods for these multi-reference systems \citep{jt632,jt623}. Nevertheless, ExoMol has produced line lists for some astronomically important transition metal (TM) diatomic species, specifically, VO \citep{jt644} and ScH \citep{jt599}.

TM diatomics are unusual molecules astrophysically because their important spectral features are associated with transitions between electronic states. The accurate treatment of many, coupled electron states remains difficult.  To treat the electronic structure of TM diatomics even reasonably needs high quality {\it ab initio} multi-reference (MR) methods capable of treating multiple electronic states. \citet{97Laxxxx.TiO}
performed state-of-art {\it ab initio} calculations of transition dipole moments for the lowest 11 electronic
states of TiO using a mixture of complete active space self-consistent
        field (CASSCF) and internally contracted
        multireference configuration-interaction (icMRCI) procedures.
Despite the vast increase in computer power since, however, there has been little impetus in the quantum chemistry community to develop improved methods for treating multiple electronic states in molecules such as TiO, and, as we have shown previously \citep{jt632,jt623}, none of the available methods show sufficient spectroscopic accuracy. This is an exciting opportunity for method development in quantum chemistry, but lead time for developing and validating methods is significant and thus we proceeded to present an update to TiO line list at this time.

There is a long history of spectroscopic studies of TiO due to its importance in astronomy and also its experimental characterisation is extensive. \cite{jt672}   collated all available ro-vibrationally resolved and assigned spectroscopic data, then used a \Marvel\ (measured active vibration-rotation energy levels)
approach to extract empirical energy levels.  These data forms the primary data set for the refinement of the potential energy curves and spin-orbit coupling curves in our spectroscopic  of TiO.

\subsection*{Overview}
The full TiO line list developed in this manuscript is called \LLname{}, with data available for all major isotopologues: \ce{^{46}Ti^{16}O}, \ce{^{47}Ti^{16}O}, \ce{^{48}Ti^{16}O}, \ce{^{49}Ti^{16}O} and \ce{^{50}Ti^{16}O}. We develop the components of these line lists systematically:
\begin{itemize}
    \item \textbf{Spectroscopy model: Energies}, presented in \Cref{sec:ESM}.  It incorporates the potential energy curves (PECs), spin-orbit couplings (SOC) and other coupling terms that provide the self-consistent physically-motivated description of the molecule and can be used by our variational diatomic rovibronic nuclear-motion program \Duo{} \citep{jt609} to produce a set of energy levels and rovibronic wavefunctions.
    \item \textbf{Spectroscopic model: Intensities}, presented in \Cref{sec:ISM}. It provides the diagonal and off-diagonal dipole moments.
    \item \textbf{Line list}, presented in \Cref{sec:LL}. It consists of energy levels and transition intensities. Note that the set of energy levels uses empirical known energy levels from our \Marvel{} analysis \citep{jt672} to improve upon some of the \Duo{} energy levels produced using the  spectroscopic model; this thus represents our best current understanding of all the energy levels within TiO, attaining completeness using the \Duo{} energy levels and accuracy using the \Marvel{} energy levels.
   \item \textbf{Partition function}, also developed in \Cref{sec:LL}.
\end{itemize}

Given the importance of TiO astronomically, the quality of existing \abinitio{} methods and the potential for substantial improvements into the future particularly with increased experimental data, we anticipate that this line list will be improved in the future; the component nature of the \LLname{} line list and its systematic presentation in this manuscript and the accompanying supporting materials will substantially assist in these future improvements.



\section{\label{sec:ESM}Spectroscopic Model: Energies}
\subsection{Approach}

The ExoMol line lists are computed using a variational approach based on potential energy curves rather than an effective Hamiltonian approach  based on spectroscopic or equilibrium constants in order to maximise the quality of the extrapolated energy levels.
Our model is self-consistent, incorporating all electronic states in a single description using the \Duo{} program of \citet{jt609}.

\begin{figure}
\includegraphics[width=0.5\textwidth]{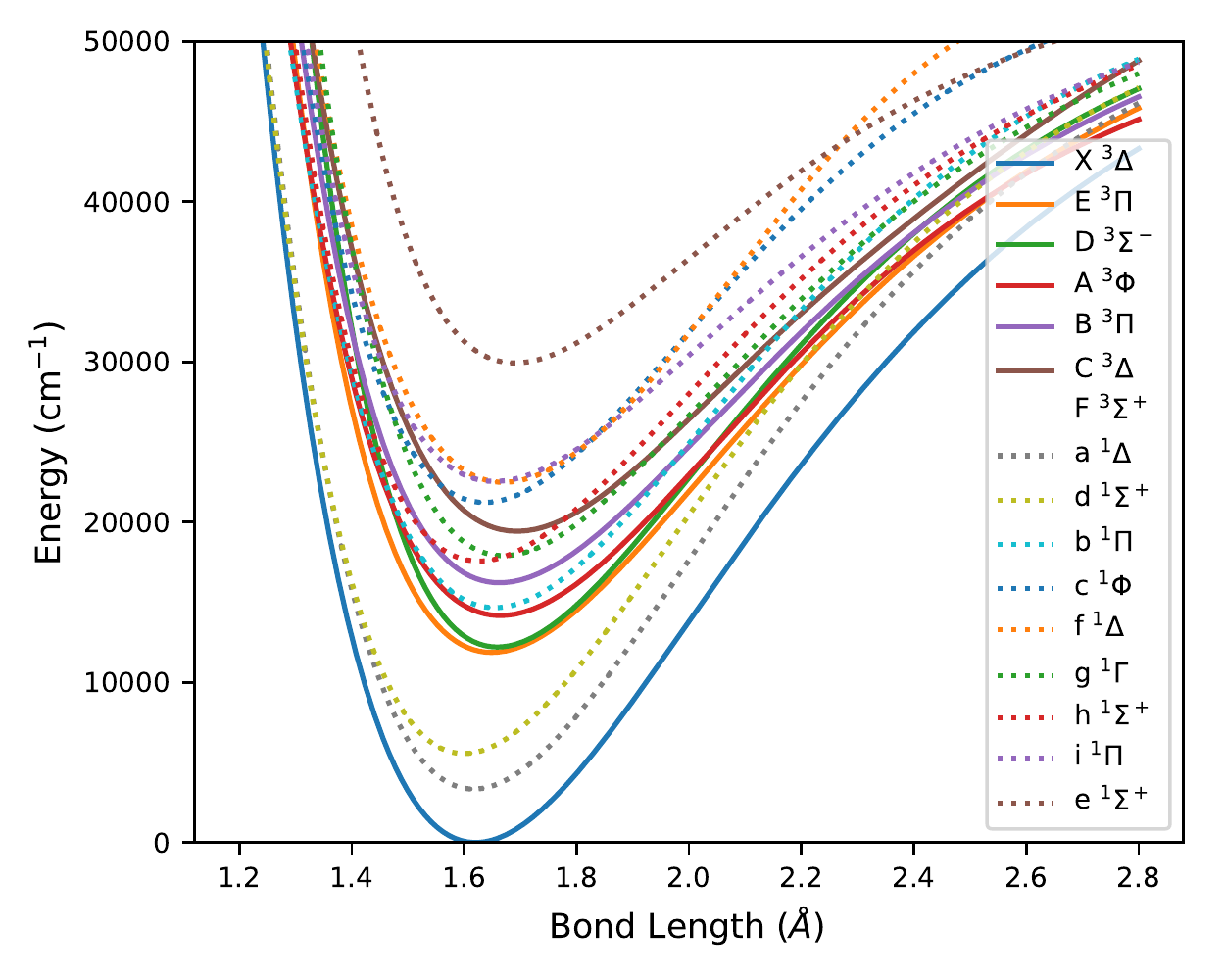}
\caption{\label{fig:PES}Fitted potential energy curves for \TiO.}
\end{figure}

\begin{figure}
\includegraphics[width=0.5\textwidth]{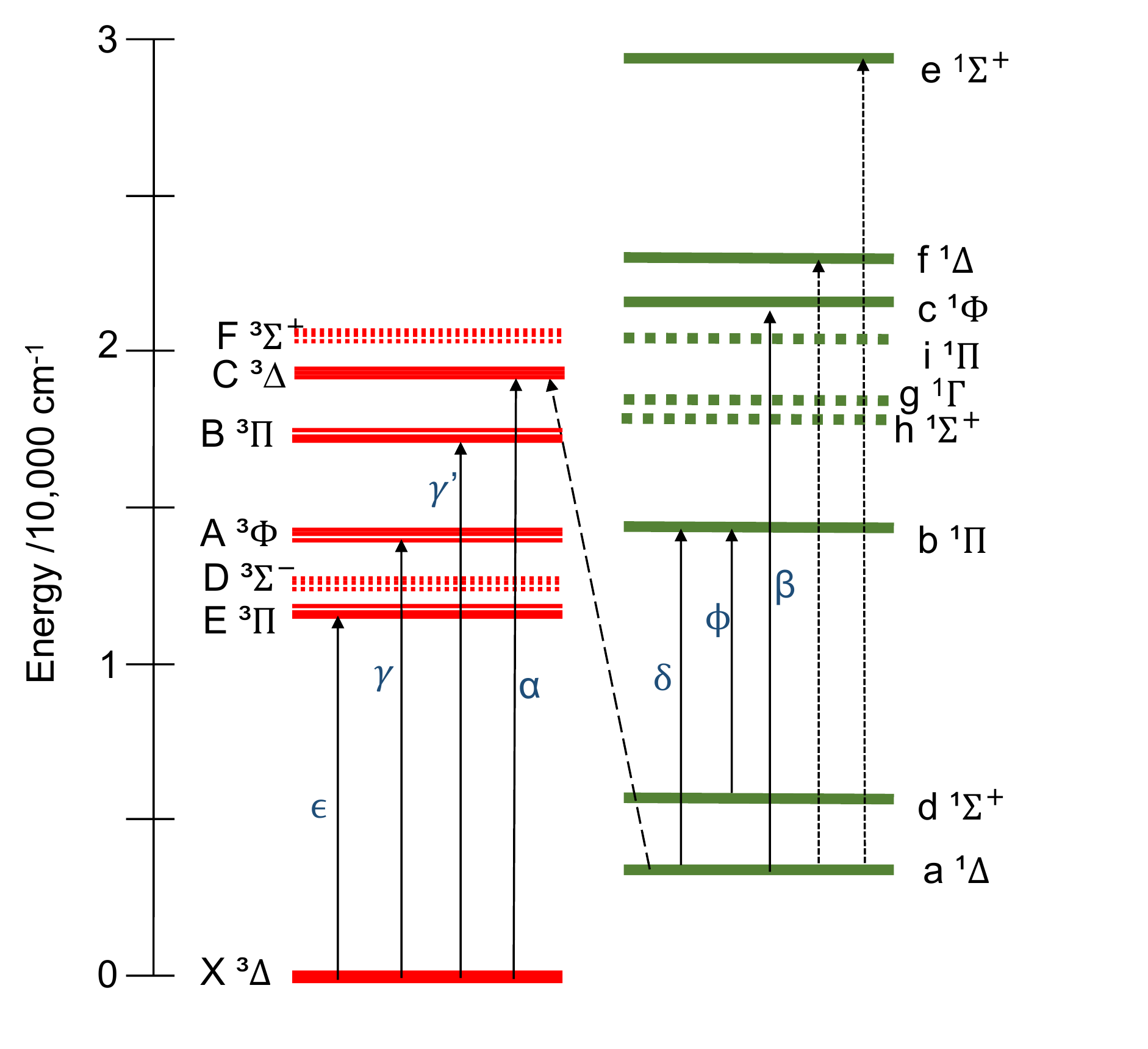}
\caption{\label{fig:TiOelecstates} Electronic states of TiO, all included in our energy spectroscopic model; solid horizontal lines are states that have been observed, whereas the dotted lines are purely theoretical results. The vertical lines denote allowed spectroscopic bands, with solid lines labelled by their historical band designation; the dotted intercombination band has no historical designation.}
\end{figure}

The low-lying electronic states of TiO are reasonably well-known, particularly from calculations of \cite{10MiMaxx.TiO}. These PECs are illustrated in  \Cref{fig:PES}. \Cref{fig:TiOelecstates}
summarizes the major band systems linking these states.

 Many of these levels are well understood, but there are no reliable observations of the \D{}, \F{}, \Sg{}, \Sh{} and \Si{} states. We do have reasonable results for their energies from \cite{10MiMaxx.TiO}, but the errors are likely to be in excess of 100 \cm{}.

The component of the spectroscopic model that determines energy levels consists of the fitted potential energy curves, spin-orbit coupling terms and other coupling terms.
For TiO, \abinitio{} methods perform quite poorly, particularly for the term energies for electronically excited states. Therefore, for spectroscopic accuracy, we need to rely on experimental data. Fortunately, a substantial quantity of experimental data for \TiO\ is available. Therefore, for this molecule, for most curves (particularly potential energy curves) we start with simple flexible models such as extended Morse oscillators rather than \abinitio{} results.

The parameters of these curves were determined by using the {\sc Duo} code to predict the rovibronic energy levels associated with this energy spectroscopic models, then refining the parameters in order to minimize the difference between the \Marvel\ experimental energy levels and the \Duo\ predictions for these energy levels.
This refinement was done by fitting curves related to one electronic state at a time against this electronic state energy levels while keeping other curves fixed to ensure that the parameters remained physically reasonable.
Curve couplings were retained in all fits frozen at their {\it ab initio} values.
Since parameters representing individual curves are linear dependent (see, for example, \citet{73ZaScHa.SH}), this procedure was repeated until convergence of all parameters and guarantee that all parameters are blended in the final spectroscopic model.

\begin{table*}
\caption{\label{tab:PES}Fitted parameters of potential energy curves for TiO from new \Duo{} TiO spectroscopic model, using \Cref{eq:EMO}.  $T_\e$ is given in \cm{}, while the equilibrium bond lengths, $r_\e$, are given in \AA{}; all other
parameters are dimensionless.}
\begin{tabular}{llllllllllHrrrrrrrrrrrrr}
\hline
\mc{1}{c}{State}  & \mc{1}{c}{$T_\e$}  & \mc{1}{c}{$r_\e$} & \mc{1}{c}{$b_0$} & \mc{1}{c}{$b_1$} & \mc{1}{c}{$b_2$} & \mc{1}{c}{$b_3$} & \mc{1}{c}{$b_4$} \\
\hline
\X{} & 0.0 & 1.62028194 & 1.8078496 & 0.16610 & -0.5475 & 0.360 &  \\
\A{} & 14166.887 & 1.6645000 & 1.801512 & 0.1559 & 1.018 & -2.94 &  \\
\B{} & 16211.043 & 1.663127 & 1.861748 &   &  \\
\C{} & 19424.750 & 1.6939134 & 1.860210 & 0.2867 & 1.965 & -19.77 & 36.0 \\
\D{} & 12202 & 1.659 & 2.0 &  & &  &  \\
\E{} & 11865.7334 & 1.649087 & 1.865071 &  &  \\
\F{} & 19771.508 & 1.678 & 2.02747 & 0. & 0. & 0. &  \\
\Sa{} & 3336.346 & 1.6166405 & 1.881696 & 0.2971 & -0.1061 & &  \\
\Sb{} & 14657.030 & 1.654616 & 1.921713 & 0.4932 & 0.037 &  &  \\
\Sc{} & 21225.082 & 1.636462 & 2.09926 & 0.4195 & 2.00 & -1.8 &  \\
\Sd{} & 5552.3390 & 1.5995563 & 1.9321441 & 0.20092 & -0.0052 &  &  \\
\Se{} & 29930.14 & 1.69081 & 2.27702 &  &  \\
\Sf{} & 22479.964 & 1.6699865 & 2.031612 & 0.4108 & 6.064 &  &  \\
\Sg{} & 17900. & 1.67 & 2. & 0. & 0. & 0. &  \\
\Sh{} & 17564.880 & 1.626585028 & 2. & 0. & 0. & 0. &  \\
\Si{} & 22534.4 & 1.65416 & 1.9422 & 0. & 0. & 0. &  \\
\bottomrule
\end{tabular}

\end{table*}

\subsection{Fitted Curves}
The TiO spectroscopic model is presented in the \texttt{48Ti-16O\_\LLname.duo.model} file in the supplementary information to this manuscript as a \Duo{} input file.

\subsubsection{Diagonal terms}

{\bf Potential energy curves: }
The most important component of this element of the spectroscopic model is the potential energy curves. We choose to use extended Morse oscillator (EMO) potentials \citep{EMO}, with functional form,
\begin{equation}
\label{eq:EMO}
V(r) = T_\e + (A_\e - T_\e) \left(1- \exp\left[\beta_\text{EMO} (r-r_\e)\right]\right)^2,
\end{equation}
where
\begin{equation}
    \beta_\textrm{EMO} =  \begin{cases}
      b_0 + b_1 \left(\frac{r^2-r_\e^2}{r^2+r_\e^2} \right) & r\leq r_\e \\
      \sum_{i=0}^4 b_i\left(\frac{r^4-r_\e^4}{r^4+r_\e^4} \right)^i & r > r_\e
   \end{cases},
\end{equation}
$T_\e$ is the term energy, $A_\e$ is the dissociation asymptote fixed at 55410 \cm{} based on experiments \citep{97NaHeCo.TiO}. $(A_\e - T_\e)$ is the dissociation energy of the corresponding potential energy curve, $r_\e$ is the equilibrium bond length  and $b_i$ are vibrational fitting parameters.  Most curves are obtained by fitted against experimentally available energy levels from an updated \Marvel{} analysis (see next section for further details). The exceptions were the \D{}, \F{}, \Sg{}, \Sh{} and \Si{} which have not been observed experimentally; the values chosen here are based on results from \cite{10MiMaxx.TiO}, but we caution that our results for these states are qualitative only.  Note that none of these states are likely to be involved in strong absorption features for TiO; if they were, they would already have been observed experimentally.

\Cref{tab:PES} provides the fitting parameters for the potential energy curves, while \Cref{fig:PES} shows the curves graphically.

 \begin{table}
\caption{\label{tab:sssr} Fitted constants, as defined in \protect\cite{Duo}, affecting energy levels of individual electronic states: diagonal spin-orbit, SO, spin-spin, $\lambda_{\rm SS}$, spin-rotational, $\gamma_\text{SR}$, and lambda doubling $\lambda_\text{opq}$, $\lambda_\text{p2q}$ constants in \cm{}. Note $\lambda_\text{q} = 0$ for all states.}
\begin{tabular}{llHlHlHll}
\hline
State &  \mc{1}{c}{SO} & &  \mc{1}{c}{$\lambda_\text{SS}$} & &  \mc{1}{c}{$\gamma_\text{SR}$} & & \mc{1}{c}{$\lambda_\text{opq}$} & \mc{1}{c}{$\lambda_\text{p2q}$} \\
\hline
\X & 101.2260  & & -0.24720 & & $ $0.000061\\
\E & $ $86.6157  & & $ $0.3156 & & $ $0.0134 & & -0.8605 & -0.0250\\
\A & 173.320  & & -0.5471 & & $ $0.02036\\
\B & $ $20.473  & & -0.585 & & $ $0.02025 & & $ $0.591 & -0.02029\\
\C & $ $98.7122  & & -0.2639 & & -0.04450\\
\hline
\end{tabular}
\end{table}

{\bf Diagonal spin-orbit coupling: }
We did obtain  \abinitio{} diagonal spin-orbit coupling results (e.g. using  icMRCI/aug-cc-pVDZ with a (4,3,3,1) active space based on state-specific or minimal state CASSCF orbitals); our previous study on ScH showed
that these couplings show little sensitivity to basis set used \citep{jt599}. We explored the effect of using fitted forms of these \abinitio{} curves in our spectroscopic model; however, we found that the variation of the spin-orbit coupling across the bond lengths of relevance to our nuclear wave functions was generally small or otherwise unreliable compared to the absolute error of this curve in most cases. Therefore, a constant spin-orbit coupling parameter refined against experimental values was used in preference to the more complex behaviour shown by \abinitio{} spin-orbit curves. These values are given in \Cref{tab:sssr}.

 \begin{table}
\caption{\label{tab:offdiag} Off-diagonal spin-orbit $\langle {\rm State}',\Sigma |{\rm SO}| {\rm State}'' \Sigma'' \rangle $ and electronic angular momentum coupling terms $\langle {\rm State}'|L_x| {\rm State}"  \rangle $, in \cm{} and au respectively. }
\begin{tabular}{lrrrr@{}l}
\hline
Parameter &  State & $\Sigma'$ & $\Sigma''$ & Value \\
\hline
& \mc{2}{l}{\textbf{Spin-orbit}} \\
$\langle ^3\Delta_{z} |{\rm SO}_{x}| ^3\Pi_y \rangle$ & X-E & 0  & 1 & 40.7 & $i$  \\
$\langle ^3\Delta_{z} |{\rm SO}_{x}| ^3\Phi_y \rangle$ & X-a & 1  & 0 & 118.0 & $i$  \\
$\langle ^3\Pi_{x} |{\rm SO}_{x}| ^1\Delta_{xy} \rangle$ & E-a & 1  & 0 & 42.0 & $i$ \\
$\langle ^3\Pi_{x} |{\rm SO}_{z}| ^1\Pi_y \rangle$ & E-b & 0  & 0 & 71.9 & $i$ \\
$\langle ^3\Phi_{x} |{\rm SO}_{x}| ^1\Delta_{xy} \rangle$ & A-a & 1  & 0 & 17.2 & $i$ \\
$\langle ^3\Phi_{x} |{\rm SO}_{z}| ^1\Phi_y \rangle$ & A-c & 0  & 0 & 65.5 & $i$ \\
$\langle ^3\Delta_{z} |{\rm SO}_{z}| ^1\Delta_{xy} \rangle$ & C-a & 0  & 0 & 17.8 & $i$ \\
$\langle ^3\Delta_{z} |{\rm SO}_{x}| ^1\Phi_y \rangle$ & C-c & 1  & 0 & 79.14 & $i$ \\

& \mc{2}{l}{$\bf{L_x}$} \\
$\langle ^3\Delta_{z} |L_{x}| ^3\Pi_y \rangle$  & X-E &&& 0.88 & $i$  \\
$\langle ^3\Delta_{z} |L_{x}| ^3\Phi_y \rangle$ & X-A &&& 0.08 & $i$  \\
$\langle ^1\Delta_{z} |L_{x}| ^1\Pi_y \rangle$ & a-b &&& 0.90 & $i$  \\
$\langle ^1\Delta_{z} |L_{x}| ^1\Phi_y \rangle$ & a-c &&& 0.22 & $i$  \\
$\langle ^1\Sigma^+ |L_{x}| ^1\Pi_y \rangle$ & d-b &&& 0.40 & $i$  \\
\hline
\end{tabular}
\end{table}

{\bf Other diagonal terms: }
To improve the quality of the fit of the \Duo{} levels against the \Marvel{} energies, we included some scalar spin-spin, spin-rotation and $\Lambda$-doubling constants,  tabulated in \Cref{tab:sssr}.

\subsubsection{Off-diagonal coupling terms}

The effect of the off-diagonal spin-orbit couplings and electronic angular momentum coupling terms is dependent on their magnitude and the energy difference between the two interacting states. We chose to take a minimalist approach and only incorporate couplings for which we were reasonably confident of our \abinitio{} results and which had a non-negligible impact on the final line list. Further, we used explicit lambda doubling terms rather than incorporating $\Pi-\Sigma$ spin orbit couplings as these proved more effective at modelling the observed lambda doubling patterns; this only influenced the E-d coupling which was neglected despite having a magnitude of around 25$i$. These are shown in \Cref{tab:offdiag}. Their magnitudes were determined using state-specific or minimal state CASSCF/aug-cc-pVDZ with a (4,3,3,1) active space. We used constant values rather than bond-length dependent values to reflect the reliability of these results and their relatively small influence on the final line list.

\begin{table}
\setlength{\tabcolsep}{0.5em}
\caption{\label{tab:rmserror1} Quality of our calculated energy levels and those of \citet{98Plxxxx.TiO} and \citet{98Scxxxx.TiO}
compared to the empirical \Marvel{} energy levels \citep{jt672} for rotational
states with $J \leq J^{\rm max}$.}
\begin{tabular}{lcrccccccc}
State  & $v$  & $J^{\rm max}$ & \mc{2}{c}{ExoMol} &  \mc{2}{c}{\citet{98Plxxxx.TiO}} & \mc{2}{c}{\footnotesize{\citet{98Scxxxx.TiO}}}  \\
&& & rmsd & max & rmsd & max & rmsd & max\\
\hline
\vspace{-0.5em} \\
															
\X	&	0	& 95 &	 0.041 	&	0.135	&	0.624	&	3.442	&	0.572	&	1.126	\\
	&	1	& 95 &	 0.013 	&	0.04	&	0.599	&	3.309	&	0.611	&	1.282	\\
	&	2	& 95 &	 0.025 	&	0.056	&	0.317	&	2.344	&	0.607	&	1.579	\\
	&	3	& 95 &	 0.022 	&	0.054	&	0.497	&	2.741	&	0.744	&	2.416	\\
	&	4	& 95 &	 0.045 	&	0.238	&	1.101	&	5.612	&	1.173	&	5.457	\\
	&	5	& 95 &	 0.065 	&	0.141	&	0.741	&	4.764	&	0.820	&	4.017	\\
\vspace{-0.5em} \\															
\E	&	0	& 35 &	 0.022 	&	0.075	&	0.720	&	1.698	&	1.657	&	2.998	\\
	&	1	& 25 &	 0.275 	&	0.552	\\								
\vspace{-0.5em} \\															
\A	&	0	& 95 &	 0.454 	&	0.583	&	0.628	&	3.632	&	0.746	&	2.113	\\
	&	1	& 95 &	 0.208 	&	0.380	&	0.602	&	3.702	&	0.790	&	2.257	\\
	&	2	& 95 &	 0.052 	&	0.164	&	0.594	&	3.997	&	0.806	&	2.239	\\
	&	3	& 95 &	 0.303 	&	0.510	&	0.300	&	2.182	&	0.717	&	1.955	\\
	&	4	& 95 &	 0.499 	&	0.646	&	0.412	&	4.175	&	0.747	&	3.584	\\
	&	5	& 95 &	 0.743 	&	1.042	&	0.492	&	3.078	&	0.733	&	2.228	\\
\vspace{-0.5em} \\															
\B	&	0	& 95 &	 0.143 	&	0.338	&	5.901	&	14.999	&	18.602	&	49.894	\\
	&	1	& 95 &	 0.236 	&	0.946	&	0.988	&	3.854	&	27.948	&	49.896	\\
\vspace{-0.5em} \\															
\C	&	0	& 95 &	 0.093 	&	0.159	&	3.693	&	10.452	&	0.998	&	3.521	\\
	&	1	& 95 &	 0.229 	&	0.400	&	1.944	&	4.868	&	1.027	&	3.865	\\
	&	2	& 95 &	 0.142 	&	0.220	&	2.074	&	6.221	&	1.119	&	3.37	\\
	&	3	& 95 &	 0.143 	&	0.311	&	2.148	&	11.849	&	2.475	&	8.545	\\
	&	4	& 95 &	 0.782 	&	4.503	&	1.401	&	3.741	&	2.845	&	9.8	\\
	&	5	& 91 &	 2.181 	&	7.871	&	3.03	&	9.874	&	3.318	&	11.107	\\
	&	6	& 86 &	 4.043 	&	11.124	&	2.424	&	8.545	&	7.523	&	19.053	\\
	&	7	& 49 &	 0.931 	&	3.754	&	1.381	&	5.869	&	10.036	&	16.303	\\
\vspace{-0.5em} \\															
\Sa{}&	0   & 95 &	 0.006 	&	0.015	&	0.156	&	0.276	&	0.19	&	0.385	\\
	&	1	& 92 &	 0.017 	&	0.060	&	8.66	&	19.956	&	2.219	&	11.846	\\
	&	2	& 60 &	 0.236 	&	1.390	&	14.223	&	14.328	&	0.195	&	0.517	\\
	&	3	& 59 &	 0.038 	&	0.088	&		&		&		&		\\
\vspace{-0.5em}									&		&		&		\\
\Sd	&	0	& 92 &	 0.021 	&	0.105	&	0.414	&	1.064	&	0.121	&	0.288	\\
	&	1	& 85 &	 0.018 	&	0.034	&	8.279	&	8.454	&	0.116	&	0.291	\\
	&	2	& 75 &	 0.015 	&	0.048	&	15.81	&	15.932	&	0.103	&	0.316	\\
	&	3	& 70 &	 0.027 	&	0.092	&	22.841	&	22.933	&	0.08	&	0.201	\\
	&	4	& 50 &	 0.026 	&	0.049	&	29.444	&	29.497	&	0.059	&	0.215	\\
	&	5	& 55 &	 0.101 	&	0.194	&	35.565	&	35.61	&	0.051	&	0.087	\\
\vspace{-0.5em}									&		&		&		\\
\Sb	&	0	& 95 &	 0.089 	&	0.338	&	1.425	&	4.08	&	2.984	&	3.55	\\
	&	1	& 86 &	 0.084 	&	0.343	&	6.519	&	6.902	&	3.227	&	3.839	\\
	&	2	& 71 &	 0.083 	&	0.347	&	12.848	&	13.251	&	3.189	&	3.74	\\
	&	3	& 73 &	 0.092 	&	0.484	&	17.632	&	18.093	&	2.943	&	3.28	\\
	&	4	& 66 &	 0.064 	&	0.277	&	21.108	&	21.527	&	2.603	&	2.999	\\
\vspace{-0.5em}									&		&		&		\\
\Sc	&	0	& 95 &	 0.164 	&	0.939	&	1.245	&	5.402	&	0.205	&	0.383	\\
	&	1	& 93 &	 0.035 	&	0.146	&	9.394	&	19.004	&	2.201	&	10.538	\\
	&	2	& 60 &	 0.157 	&	0.877	&	15.244	&	15.41	&	0.783	&	1.947	\\
	&	3	& 59 &	 0.077 	&	0.329	&		&		&		&		\\
\vspace{-0.5em}									&		&		&		\\
\Sf	&	0	& 71 &	 0.030 	&	0.143	&	4.594	&	4.645	&	0.155	&	0.458	\\
	&	1	& 62 & 	 0.028 	&	0.095	&	4.540	&	4.612	&	0.345	&	0.771	\\
	&	2	& 23 &	 0.308 	&	0.593	&	4.211	&	4.565	&	0.301	&	0.432	\\
\vspace{-0.5em}									&		&		&		\\
\Se	&	0	& 49 &	 0.718 	&	1.288	&		&		&		&		\\
	&	1	& 59 &	 1.095 	&	2.309	&		&		&		&		\\
	\bottomrule														
\end{tabular}
\end{table}

\subsection{Discussion}

\paragraph{Accuracy of Duo Energy Levels from the Energy Spectroscopic Model}

To explore the success of this optimisation process, in \Cref{tab:rmserror1}, we tabulate the root mean squared errors for individual energy levels for each spin-rovibronic band against empirical \Marvel{} energy levels. The quality of our fit to \Marvel{} energies is very good for low lying electronic states, particularly the \X{} state; which can be attributed to low mixing between states. The highest lying singlet and triplet states (\Se{} and \C{}) in our energy spectroscopic model have significant errors, particularly the \C{} state at high vibrational levels. This is almost certainly due to neglected interactions between electronic states, probably including those states not incorporated into our energy spectroscopic model. Many attempts to improve this fitting using additional couplings (e.g. between the \Si--\C{} state{}) were unsuccessful and did not lead to an improvement in fit quality. 
Overall, the singlet state energy levels are better reproduced than the triplet state energy levels as their energy spacings are more regular in the absence of spin-coupling interactions.

\Cref{tab:rmserror1} compares the analogous results from the Plez and Schwenke line lists. This table clearly shows that our new TiO energy spectroscopic model reproduces the \Marvel{} empirical energy levels better than those of Plez and Schwenke, which is expected because we performed explicit fitting to these results. Note that more recently, the Plez line list has updated to explicitly include empirical energy levels \citep{VALD3}; this is not considered in this section as our focus here is on the underlying spectroscopic model itself.

\begin{table*}
\caption{\label{tab:pop} Population of different electronic states relative to total population for \TiO\ at different temperatures (K). Calculated based on partition function for the full line list compared to the partition function from the individual state(s).}
\begin{tabular}{lrrrrrrrrrrr}
\toprule
State & \mc{6}{c}{\emph{Population rel. to total population} }\\
 & 1000 & 1500 & 2000 & 3000 & 5000 & 8000 \\
 \midrule

\X & \num{9.97E-01} & \num{9.86E-01} & \num{9.68E-01} & \num{9.21E-01} & \num{7.91E-01} & \num{5.76E-01} \\
\E & \num{4.40E-08} & \num{1.28E-05} & \num{2.17E-04} & \num{3.57E-03} & \num{3.01E-02} & \num{7.88E-02} \\
\D & \num{1.26E-08} & \num{4.37E-06} & \num{8.03E-05} & \num{1.43E-03} & \num{1.29E-02} & \num{3.51E-02} \\
\A & \num{1.79E-09} & \num{1.55E-06} & \num{4.53E-05} & \num{1.29E-03} & \num{1.68E-02} & \num{5.59E-02} \\
\B & \num{9.14E-11} & \num{2.14E-07} & \num{1.03E-05} & \num{4.80E-04} & \num{9.34E-03} & \num{3.90E-02} \\
\C & \num{9.86E-13} & \num{1.07E-08} & \num{1.11E-06} & \num{1.11E-04} & \num{4.01E-03} & \num{2.35E-02} \\
\F & \num{2.66E-13} & \num{3.45E-09} & \num{3.89E-07} & \num{4.27E-05} & \num{1.65E-03} & \num{1.02E-02} \\
\Sa & \num{2.65E-03} & \num{1.31E-02} & \num{2.87E-02} & \num{6.09E-02} & \num{9.93E-02} & \num{1.04E-01} \\
\Sb & \num{2.64E-10} & \num{2.95E-07} & \num{9.74E-06} & \num{3.12E-04} & \num{4.48E-03} & \num{1.58E-02} \\
\Sc & \num{2.03E-14} & \num{5.31E-10} & \num{8.50E-08} & \num{1.32E-05} & \num{6.78E-04} & \num{4.93E-03} \\
\Sd & \num{5.35E-05} & \num{7.63E-04} & \num{2.85E-03} & \num{1.03E-02} & \num{2.57E-02} & \num{3.42E-02} \\
\Se & \num{4.31E-20} & \num{7.31E-14} & \num{9.44E-11} & \num{1.20E-07} & \num{3.30E-05} & \num{6.08E-04}1 \\
\Sf & \num{3.70E-15} & \num{1.75E-10} & \num{3.75E-08} & \num{7.82E-06} & \num{4.98E-04} & \num{4.03E-03} \\
\Sg & \num{2.58E-12} & \num{1.36E-08} & \num{9.79E-07} & \num{6.85E-05} & \num{1.84E-03} & \num{9.27E-03} \\
\Sh & \num{1.95E-12} & \num{8.81E-09} & \num{5.85E-07} & \num{3.78E-05} & \num{9.58E-04} & \num{4.68E-03} \\
\Si & \num{3.56E-15} & \num{1.73E-10} & \num{3.78E-08} & \num{8.08E-06} & \num{5.33E-04} & \num{4.41E-03} \\
\midrule
$T_\e < 17,000$ \cm{} &
\num{1.00E+00} & \num{1.00E+00} & \num{1.00E+00} & \num{1.00E+00} & \num{9.91E-01} & \num{9.47E-01} \\
$T_\e > 17,000$ \cm{} & \num{5.79E-12} & \num{3.68E-08} & \num{3.10E-06} & \num{2.69E-04} & \num{9.03E-03} & \num{5.26E-02} \\
\midrule
D+F+g+h+i & \num{1.26E-08} & \num{4.40E-06} & \num{8.23E-05} & \num{1.59E-03} & \num{1.79E-02} & \num{6.37E-02} \\
 \bottomrule
\end{tabular}
\end{table*}

\paragraph{Population of different electronic states}
\Cref{tab:pop} shows the population of different electronic states as a function of temperature, as calculated by considering each state's contribution to the total partition function. 99.7\% of the total population is in the ground electronic state at 1000 K, decreasing to 92.3\% at 3000 K and 80.7\% at 5000 K. The \Sa{} and \Sd{} states have populations of 6.1\% and 1.0\% at 3000 K respectively, while the other states have contributions less than 1\%. In particular, up to 3000 K, consideration of all electronic states with $T_\e < 17,000$ \cm{} recovers all but 0.01\% of the total population.


\section{\label{sec:ISM}Spectroscopic Model: Intensities}
The intensity component of the spectroscopic model refers to the  set of diagonal and off-diagonal dipole moment curves that are used to produce a particular set of transition intensities.
Experimental measurements of absolute intensities for TiO are non-existent or extremely unreliable; however, there are a number of experimental measurements of lifetimes of certain states and various attempts at obtaining relative transition intensities either from astrophysical sources or in the laboratory, although these are fraught with uncertainties, particularly over larger spectral regions.

{\it Ab initio} calculations can provide reasonable results for absolute intensities based on diagonal and off-diagonal dipole moment curves. These should not be relied upon for anywhere near sub-1\% accuracy as for the main group molecules like \ce{CO2} \citep{jt613} and \ce{H2O} \citep{jt509}, but can be trusted to within a factor of two or better, depending on the particular states involved.


Rather than using the \abinitio{} results as grid point inputs for the nuclear motion calculation, it has been shown \citep{16MeMeSt} that it is preferable to represent the dipoles using a functional forms. This also ensures the results show physical behaviour at long bond lengths and don't display unphysical kinks \citep{jt573}.

In our intensity spectroscopic model, we include all diagonal and off-diagonal dipole moments involving observed electronic states with $T_\e < 17,000$ \cm{}, i.e. \X{}, \E{}, \A{}, \B{}, \Sa{}, \Sd{}, \Sb{}. As shown in \Cref{tab:pop}, the population of the more highly excited states at 3000 K (the highest temperature for which TiO is dominant in stars) is only around 1 in 10,000. This approximation therefore has negligible effect, compared to the other errors within the line list.

\begin{figure}
\includegraphics[width=0.45\textwidth]{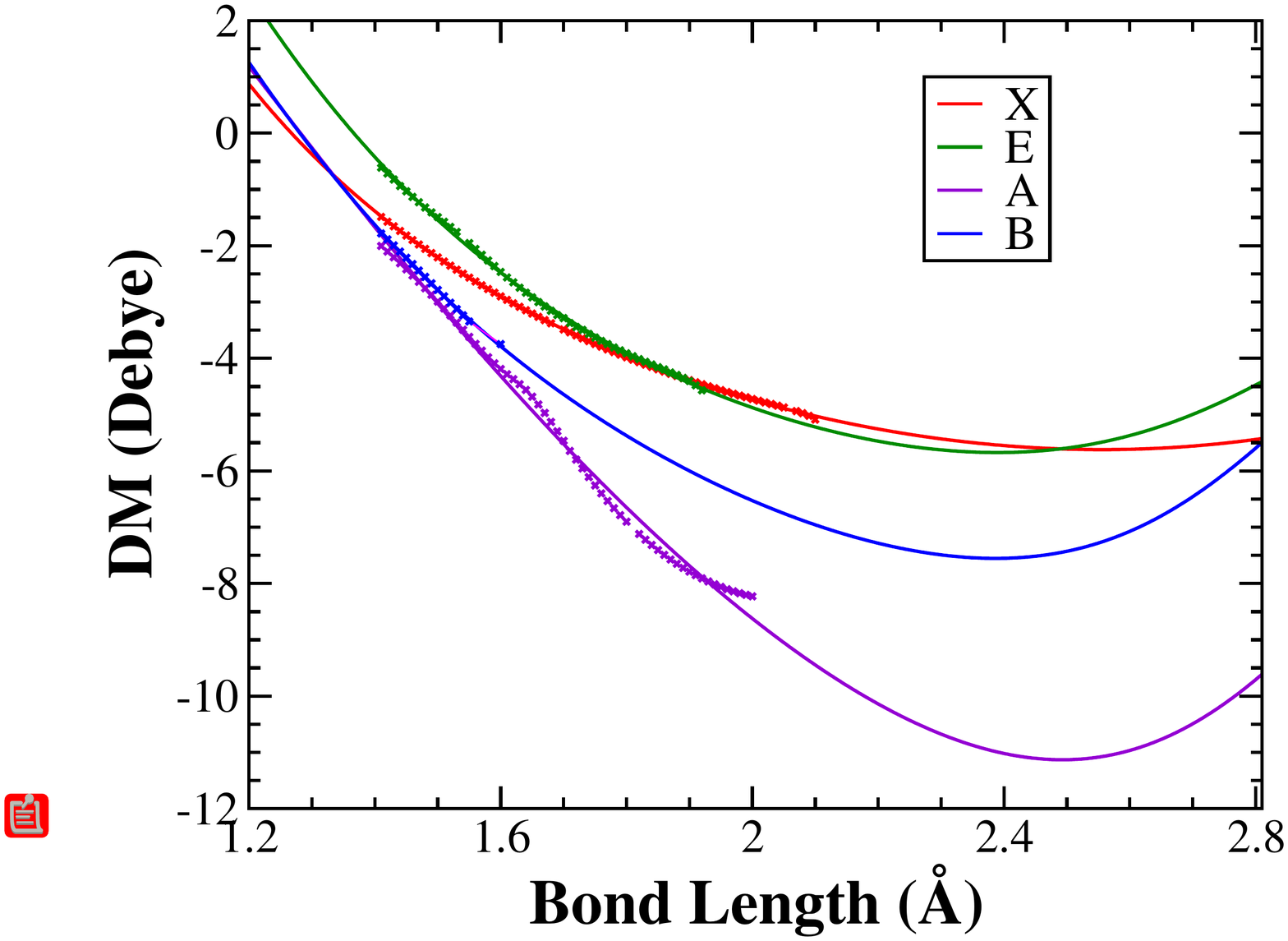}

\includegraphics[width=0.45\textwidth]{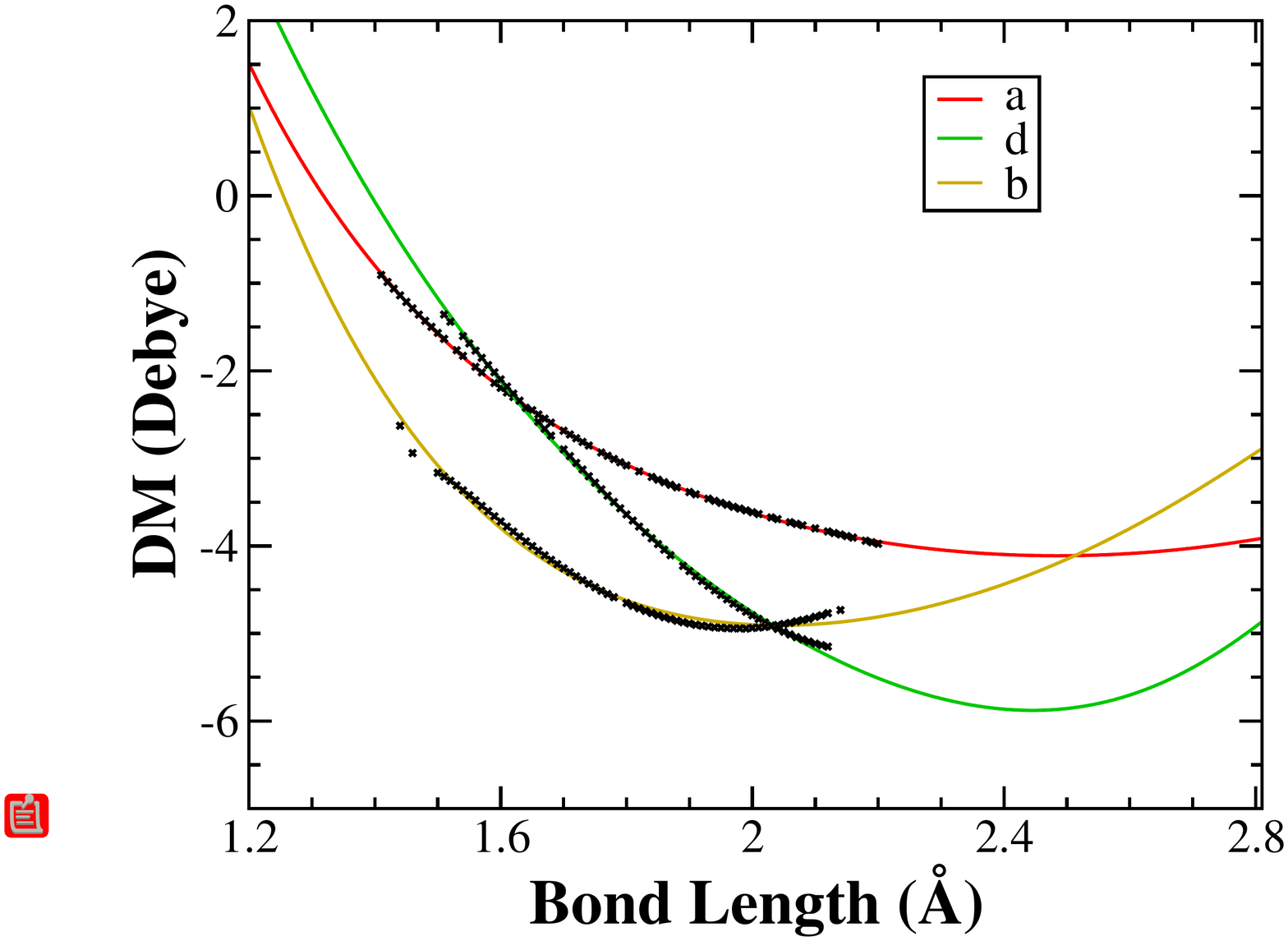}
\caption{\label{fig:diagDM} \Abinitio{} diagonal dipole moment in crosses, and the fitted curves as lines.}
\end{figure}

\begin{table*}
\caption{\label{tab:diagDM} Specifications for diagonal dipole moment in Duo, where dipole moment is given in Debye. The  equilibrium dipole moment is taken at: 1.60 \AA{} for \Sd{}, 1.62 \AA{} for \X{} and \Sa{} states and 1.65 \AA{} for all other states.  Experimental permanent electric dipole moments are obtained from \citet{03StVixx.TiO} who used optical Stark spectra. Uncertainties in the experimental values are given in brackets. }
 \begin{tabular}{llrrrrrrrrrrrrrrrr}
 \hline
&  \mc{1}{c}{Exp} &  \mc{1}{c}{{Ab Initio}} & \mc{5}{c}{{Extrapolation Parameters}} & \mc{4}{c}{{Fit Characteristics}}\\
 \cmidrule(r){2-2} \cmidrule(r){3-3}  \cmidrule(r){4-8} \cmidrule(r){9-12}
& \mc{1}{c}{$\mu_{r_\text{eq}}$} &  \mc{1}{c}{$\mu_{r_\text{eq}}$} &    \mc{1}{c}{$j$} &  \mc{1}{c}{$m$} &  \mc{1}{c}{$t$} &  \mc{1}{c}{$l_1$} &  \mc{1}{c}{$l_2$}  &  \mc{1}{c}{$\mu_{r_\text{eq}}$}  &  \mc{1}{c}{${\mu^\prime}_{r_\text{eq}}$}  &  \mc{1}{c}{$|\mu|_\text{max}$} &  \mc{1}{c}{$r$ for $|\mu|_\text{max}$} \\
 \hline
 \X & -3.34(1) &  -3.214 & 1 &  -1.533 &  -2.680 &  -6.589 &  17.146 & -3.223 & -6.210 & -5.378 & 2.346\\
 \E & -3.2(4) & -3.146 & 1 & -1.562 &  -0.268 &  -22.210 &  43.049 & -3.213 & -7.597 & -4.848 & 2.166 \\
 \D &  & -7.372 & 1 & -1.397 &  -3.390 &  -19.057 & 30.732 & -7.526 & -7.355 & -9.387 & 2.213 \\
 \A & -4.89(5) & -5.109 & 1 & -1.497 &  -9.145 &  11.893 &  4.613 & -5.135 & -12.034 & -11.312 & 2.493 \\
 \B & -4.9(2) & -4.292 & 1 & -1.493 & -1.341 & -29.652 & 59.303$^\dagger$ &-5.135 & -11.101 & -7.829 & 2.209 \\
 \hline
 \Sa & & -2.300 & 1 &  -0.561 & -2.010 &  -9.719 & 22.956 & -2.300 & -5.215& -4.112 & 2.485 \\
 \Sd & & -2.098 & 1 &  -1.537 & -3.820 & -1.978 &  18.271 & -2.121 & -8.800 & -2.312 & 2.445  \\
\Sb{} &  & -4.002 & 1 & -1.464 &  -0.055 &  -21.501 &  33.970 & -4.069 & -4.960 & -4.907 & 2.043 \\
 \hline
 \end{tabular}

 $^\dagger$ Constrained $l_2 = -2 l_1$
\end{table*}

\subsection{Diagonal Dipole Moment Curves}
The \abinitio\ methodology is MRCI/aug-cc-pVQZ based on state-selective or minimal-state (4s,3d)$_{/\textrm{Ti}}$ (2p)$_{/\textrm{O}}$ active space with core + (3s,3p)$_{/\textrm{Ti}}$ and (2s)$_{/\textrm{O}}$ closed and doubly occupied orbitals (see Supplementary Information for example input files). The dipoles are calculated using {\sc Molpro} \citep{MOLPRO} and the finite field difference expression (see \cite{jt623} for details); a Davidson correction was not included. Scalar relativistic corrections were also not considered: there is considerable
evidence \citep{jt424,jt607}, at least for molecules not containing a transition metal, that scalar relativistic
and core corrections nearly cancel for dipole moments; we therefore followed the recommendation \citep{jt573}
that calculations should not include one without the other.

The \abinitio\ data points for the diagonal dipole moments are indicated by the dark dots in Fig. \ref{fig:diagDM}, with key quantities given in Table \ref{tab:diagDM}. It is clear that all fitted curves are smooth. Quantitative results are provided in the Supplementary Information.

A good benchmark comparison for our data is the results from \cite{10MiMaxx.TiO}, which provide good data for equilibrium diagonal dipole moments from a variety of high level ab-initio methods. The finite field dipole moments obtained in this study are generally consistent with the \cite{10MiMaxx.TiO} finite field results, within calculation errors of about 0.4 Debye (this estimate is based on the spread of values from \cite{10MiMaxx.TiO} for a single finite field dipole moment with varying theoretical methods).

For smooth interpolation and extrapolation, we fit \abinitio\ data to functional forms. The functional form of the diagonal dipole moment curves is derived from a diabatic model for an avoided crossing between a neutral and ionic electronic state, as discussed by \cite{jt644}.
The resulting functional form is
\begin{equation}
\mu_{GS}(r) = \frac{\left(\sqrt{4 j^2+\lambda ^2}+\lambda \right)^2 \mu _{\text{Ionic}}(r)}{\left(\sqrt{4 j^2+\lambda ^2}+\lambda \right)^2+4 j^2}.
\end{equation}
The $j$ parameter controls the interaction between the ionic and covalent diabatic potentials and therefore how sharp the transition is. It is fixed at 1 for all TiO diagonal dipole moments.
 The $\lambda$ parameter controls the distance between the two diabatic states and their crossing points and is of the form $\lambda=\m(r-d)(r-\pk)$, where $\pk = 2.75$~\AA\ and $d= 1$~\AA\ are the crossing points of the ionic and covalent diabatic
curves (assuming that the ionic curve is a quadratic function of energy and the covalent curve is a linear or constant function). $m$ together with $j$ control the depth of the ionic well. $\mu_\text{ionic}$ is a functional form of the dipole moment of the ionic diabatic state, empirically expanded in the form $\mu_\text{ionic}(r) = \pt r + l_1 r^{-1} + \pll r^{-3}$, where $\pt, \pll, l_1$ are fitting parameters.

The fitted curves are given in Fig. \ref{fig:diagDM}. The parameters of these fits are given in Tables \ref{tab:diagDM}. All fits are satisfactory.



\begin{figure*}
\includegraphics[width=0.48\textwidth]{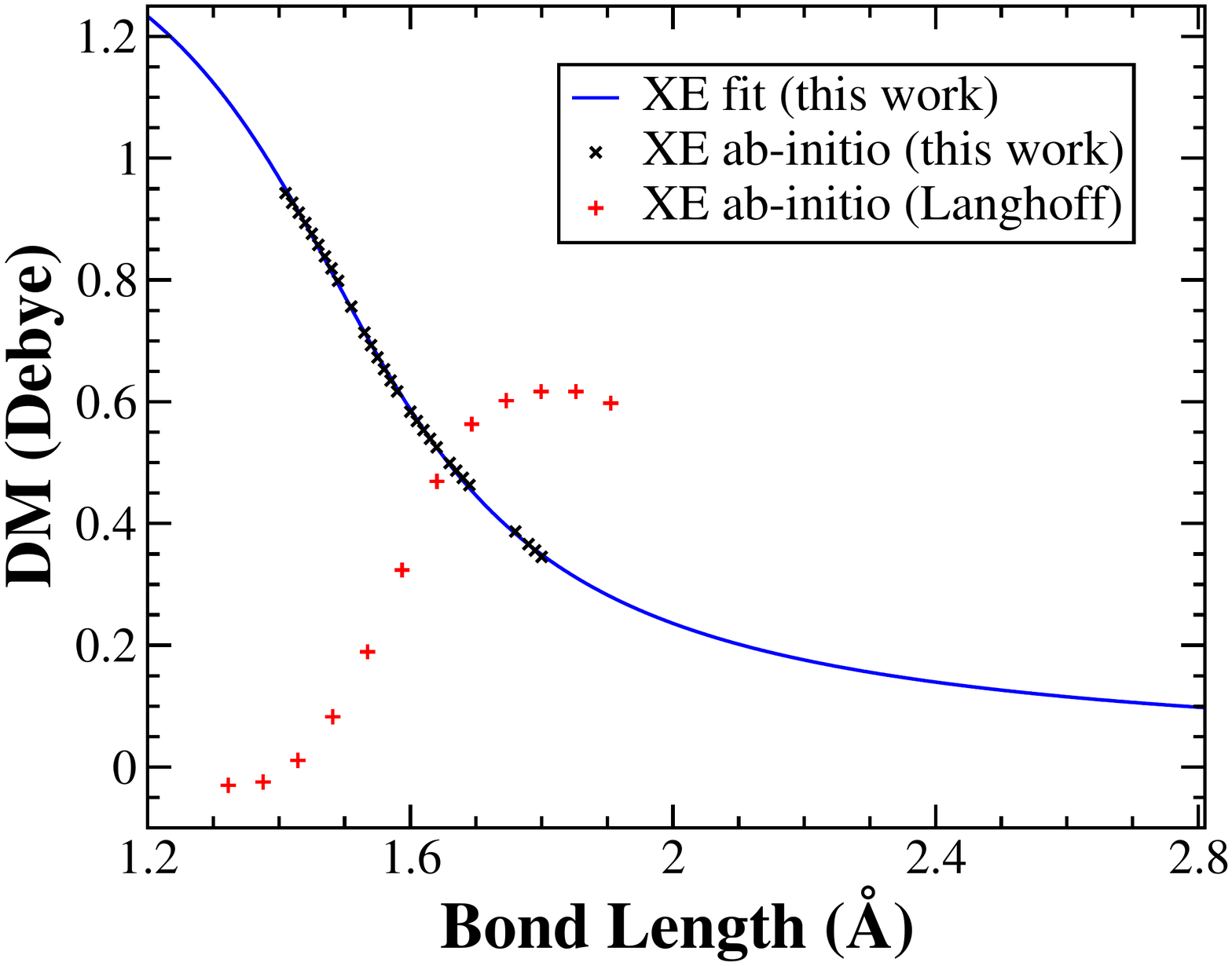}
\includegraphics[width=0.48\textwidth]{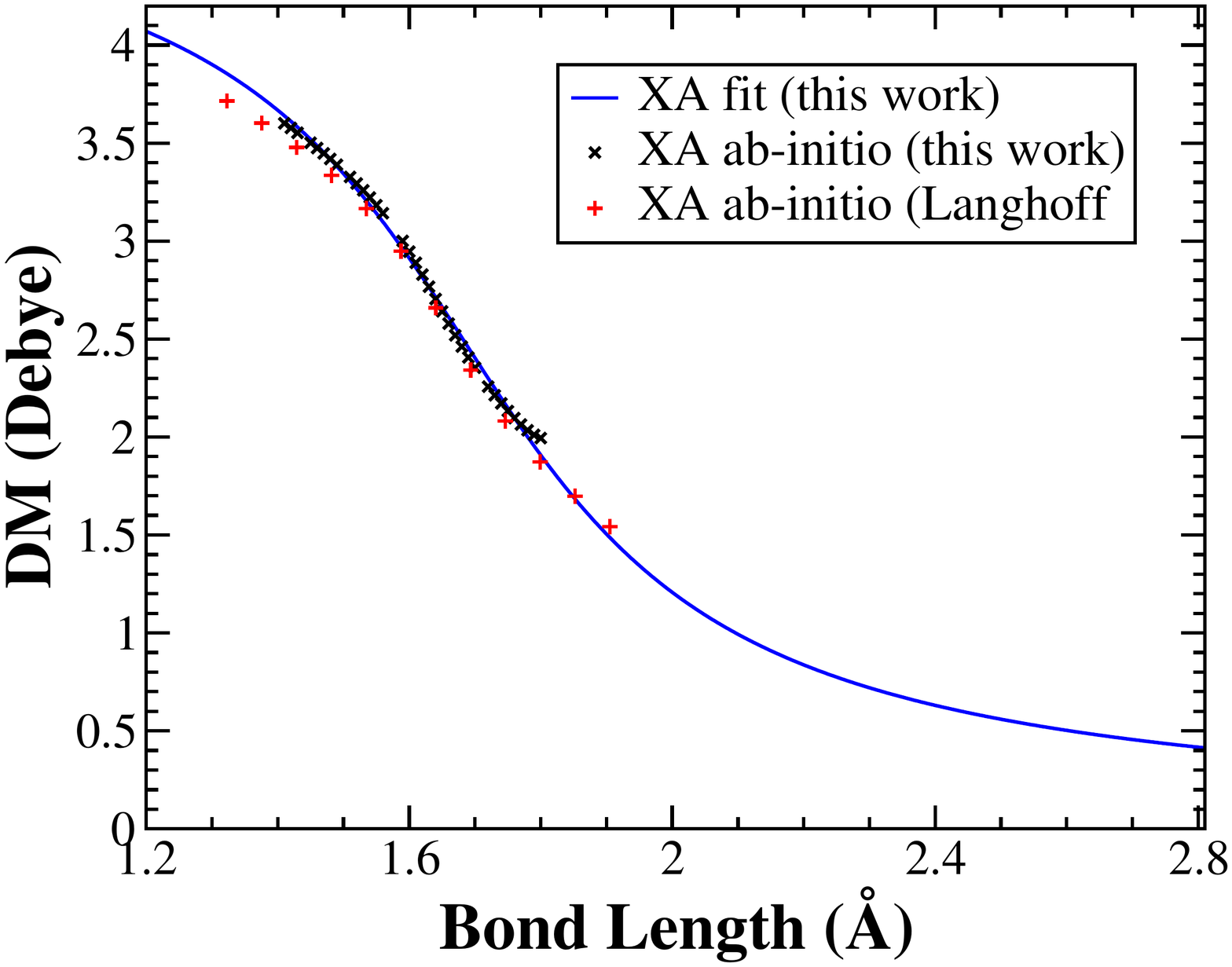}
\includegraphics[width=0.48\textwidth]{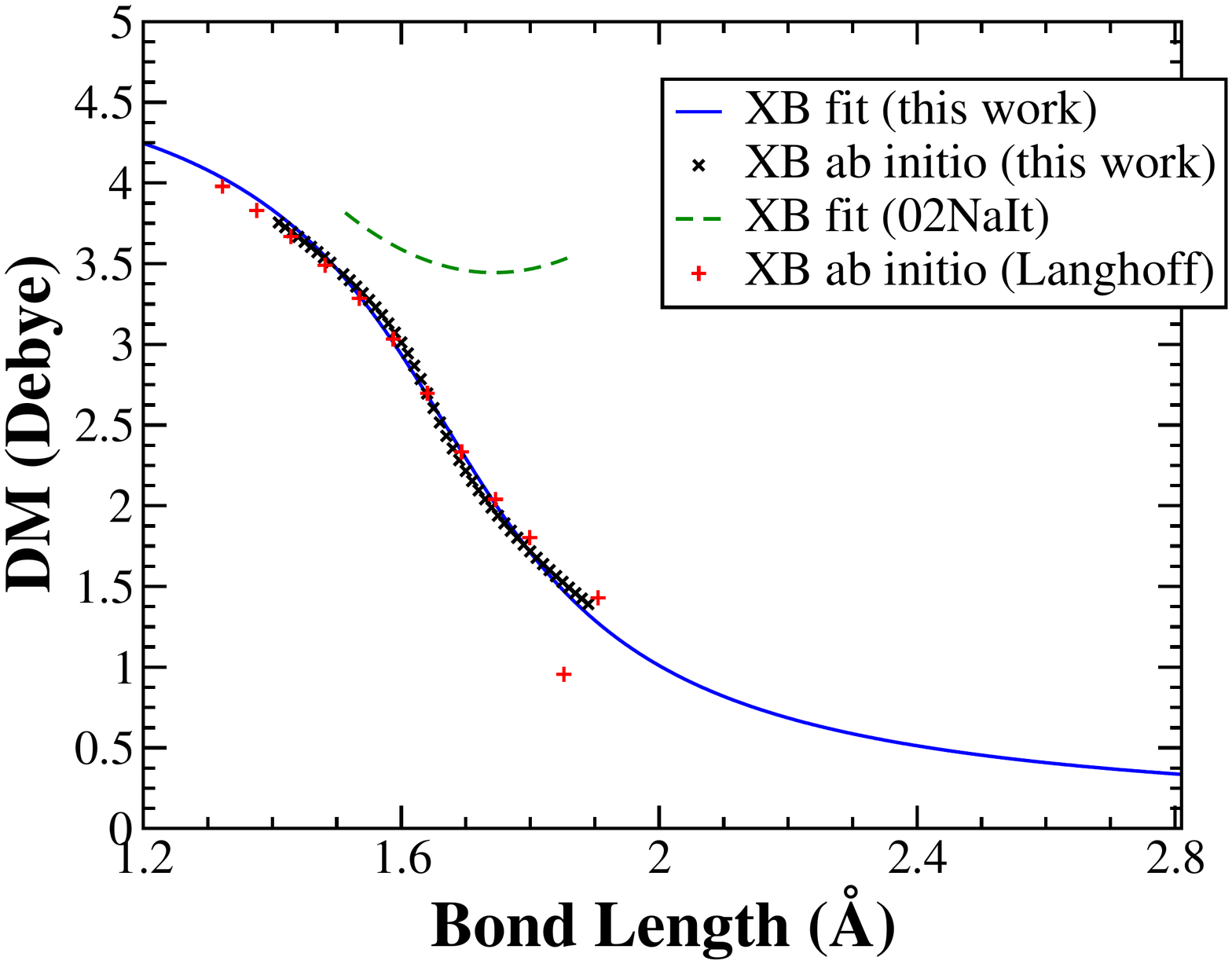}
\includegraphics[width=0.48\textwidth]{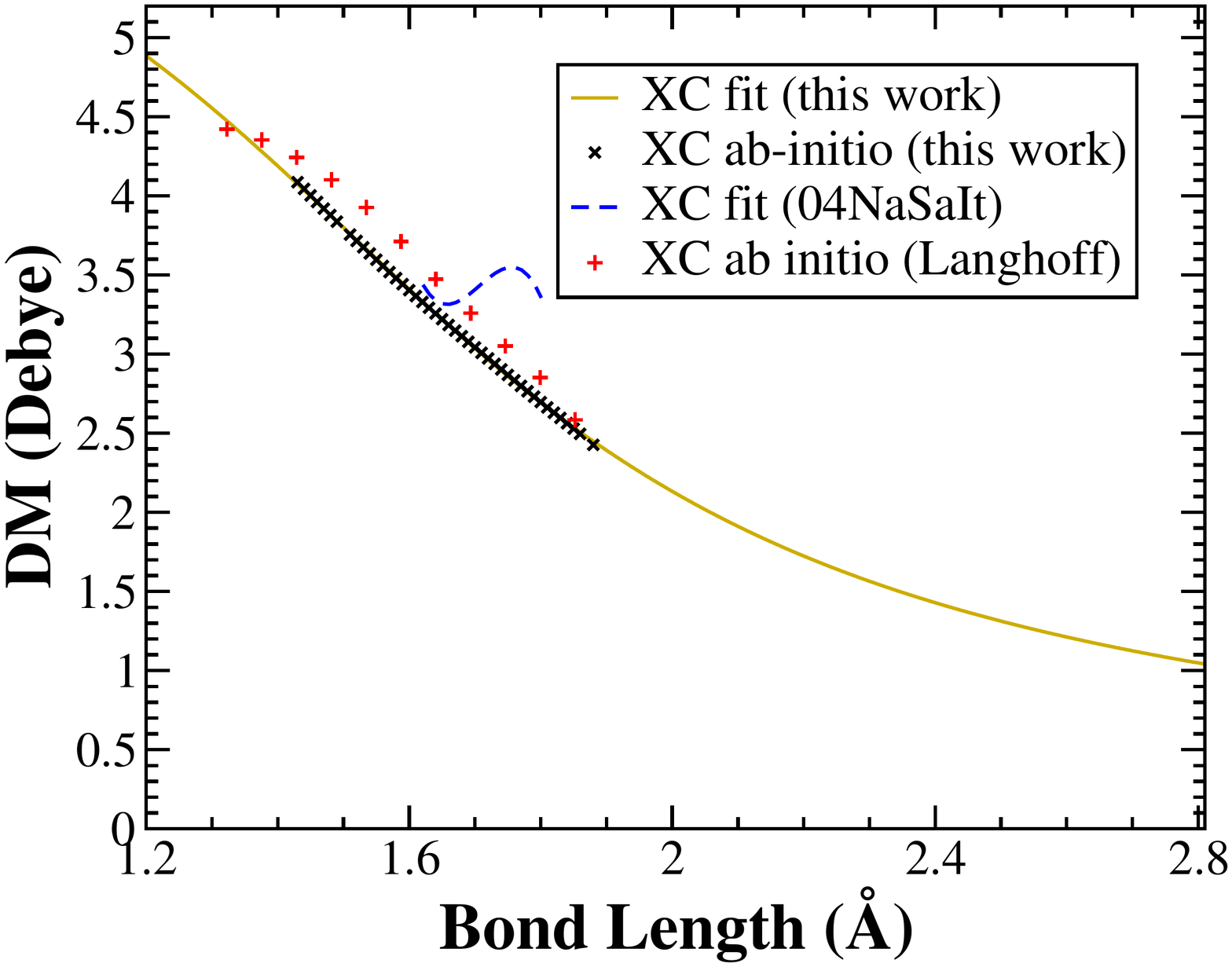}
\caption{\label{fig:offdiagDMTripX} \Abinitio{} off-diagonal dipole moments involving \X{} electronic state for triplets in crosses, and the fitted curves in lighter continuous curves.}
\end{figure*}

\begin{table*}
\caption{\label{tab:offdiagDMTrip}\Abinitio\ off-diagonal dipole moments in Debye at $r$=1.60 \AA, $\mu_\text{eq}$, and parameters for fitting curve in units of Debye and \AA. }
\begin{tabular}{llrrrrrH}
\hline
&  &  \mc{1}{c}{{\Abinitio}} & \mc{4}{c}{{Fitting Parameters}}  \\
  \cmidrule(r){3-3} \cmidrule(r){4-7}
& & \mc{1}{c}{$|\mu|_\text{eq}$} &  \mc{1}{c}{$|\mu|_\text{eq}^\text{fitted}$} &  \mc{1}{c}{$c$} & \mc{1}{c}{$a$} & \mc{1}{c}{$r_m$}  \\
\hline
$\langle ^3\Delta_{z} |\mu_{x}| ^3\Pi_x \rangle$ & X-E & 0.584 & 0.587 &  1.608 &  3.907 & 1.4848 & $-$  \\ 
$\langle ^3\Delta_{z} |\mu_{x}| ^3\Phi_x \rangle$ &X-A & 2.955 &2.913 & 4.495 &  3.312 & 1.687 & $-$\\ 
$\langle ^3\Delta_{z} |\mu_{x}| ^3\Pi_x \rangle$ &X-B & 3.011 &2.937 &  5.013 & 4.101 & 1.667 & $-$ \\
$\langle ^3\Delta_{z} |\mu_{z}| ^3\Delta_{z} \rangle$ &X-C & 3.404 & 3.409 & 7.482 &  1.652 &  1.515  & $?$ \\
 $\langle ^3\Pi_{y} |\mu_{x}| ^3\Sigma^- \rangle$ &  E-D & 0.295 & 0.299 & 1.010 & 2.549 &  1.258  \\
$\langle ^3\Pi_{x} |\mu_{z}| ^3\Pi_x \rangle$ &E-B & 0.622 & 0.361 & 0.721 & 4$^*$ & 1.6$^*$  & $+$\\
  $\langle ^3\Sigma^- |\mu_{x}| ^3\Pi_y \rangle$ &  D-B  & 1.903 & 1.866 & 3.114 & 3.756 & 1.686 \\
$\langle ^3\Pi_{x} |\mu_{x}| ^3\Delta_{z} \rangle$ &E-C & 0.129 & 0.126 & 0.593 & 8.229 & 1.530 & $+$\\
$\langle ^3\Phi_{x} |\mu_{x}| ^3\Delta_{z} \rangle$ &A-C &  0.268 & 0.296 & 0.652 & 4$^*$ & 1.6$^*$   & $?$\\
$\langle ^3\Pi_{x} |\mu_{x}| ^3\Delta_{z} \rangle$ &B-C & 0.198 & 0.165 & 0.330 & 4$^*$ & 1.6$^*$  & $?$\\
\hline
\end{tabular}

$^*$ Constrained
\end{table*}

\subsection{Off-diagonal Dipole (Transition) Moment Curves}

\subsubsection{Method}
\paragraph{The reference ab initio results}
The best existing calculations are from \cite{97Laxxxx.TiO} and were based on state-averaged (SA) CASSCF orbitals with (4s,3d)$_{/\textrm{Ti}}$ (2p)$_{/\textrm{O}}$ active space with core + (3s,3p)$_{/\textrm{Ti}}$ and (2s)$_{/\textrm{O}}$ using an augmented triple zeta basis set, with dynamic electron correlation effects subsequently included using icMRCI for all electronic states except the \C{} state.

The triplet off-diagonal dipole moments involving the \X{} state are the most important parameters of our intensity spectroscopic model, so  we choose to perform new \abinitio{} calculations for these parameters; results for other triplet off-diagonal dipole moments are naturally obtained through this method. These were done using {\sc Molpro}. The active space was identical to \cite{97Laxxxx.TiO}, as was the icMRCI methodology (for completeness, we note that the value of $n$ in icMRCI($n$) notation introduced in \cite{jt632} is  equal to one plus the number of states with the same $C_{\infty v}$ spin-symmetry lower in energy than the target state). We used the more modern and slightly larger basis set aug-cc-pVQZ, but the effect of this will be modest. The most substantial methodology change from \cite{97Laxxxx.TiO} is that we used state-selective or minimal state (SS/MS) CASSCF calculations (not SA CASSCF orbitals). As described in \cite{jt632}, the use of SS/MS CASSCF orbitals is theoretically preferable as it enables these orbitals to optimally describe each electronic state (whereas for SA CASSCF orbitals, there is a compromise between describing many electronic states) and has been shown to have an qualitatively important effect in practice, though SS/MS methodology can be more time consuming and lead to convergence difficulties especially at longer bond lengths. We make one major improvement in methodology since \cite{jt632} that reduces calculation difficulty using SS/MS CASSCF orbitals: for off-diagonal dipole moment calculations, we could describe each electronic state using its MRCI wavefunction built from \emph{different non-orthogonal} CASSCF orbitals; this was made possible using the biorth keyword introduced in the 2015 version of {\sc Molpro}. This meant that we no longer needed to calculate MS CASSCF orbitals for each off-diagonal dipole moment individually.

For singlet off-diagonal dipole moments (which are less important to the overall spectroscopy of TiO because less than 7\% of a population of \TiO\ are in singlet states even at 3000 K, see \Cref{tab:pop}), we chose to use the results from \cite{97Laxxxx.TiO} as the basis of our fit. Note that this means we do not have any dipole moments incorporating the \Se{} state, and thus no associated transitions; we retain the \Se{} state in our model because we have good experimentally-derived \Marvel{} energies for this state to aid our fitting, and want to enable future work to introduce dipole moments associated with this state.

\paragraph{Functional Form}

Here, we choose to fit all off-diagonal dipole moments to the functional form
\begin{equation}
\mu_\text{fit}(r) = \frac{c}{2\pi} \left(\pi +2\tan^{-1}(-a(r-r_m))\right)
\end{equation}
where $c$, $a > 0$ and $r_m$ are free parameters that are modified to optimise the quality of the fit. Usually $m$ is around $r_\e$. This functional form smoothly goes to 0 as the bond length increases while not increasing substantially at low bond lengths (as an exponential fit would).

\subsubsection{Results and Discussion}

\paragraph{Triplet off-diagonal dipole moments involving the \X{} state}
Our results for the triplet off-diagonal dipole moments involving the \X{} state are shown in \Cref{fig:offdiagDMTripX} and compared against results from \cite{97Laxxxx.TiO}, \cite{02NaItxx.TiO} (02NaIt) and \cite{04NaSaIt.TiO} (04NaSaIt) where available. Our results and Langhoff's agree semi-quantitatively in all cases except the \X-\E{} dipole moment. Our investigations showed that the \E{} state from Langhoff may be qualitatively incorrectly modelled. Also the behaviour of our dipole moment (i.e. decreasing as bond length increases) is more physically realistic in the absence of curve crossings, which are not expected for these low lying states. Given these two factors combined with the improvements in our \abinitio\ methodology compared to Langhoff, we choose to use our \abinitio{} results for our new line lists.

The parameters of the fitted curves are in \Cref{tab:offdiagDMTrip}, while the solid lines in \Cref{fig:offdiagDMTripX} show the fit. Qualitatively, the X-A, X-B and X-C off-diagonal dipole moments are much larger than the X-E result, as expected given that the latter is a much weaker transition than the previous three strongly allowed bands that dominate the spectra of the photospheres of M stars.

\begin{figure*}
\includegraphics[width=0.48\textwidth]{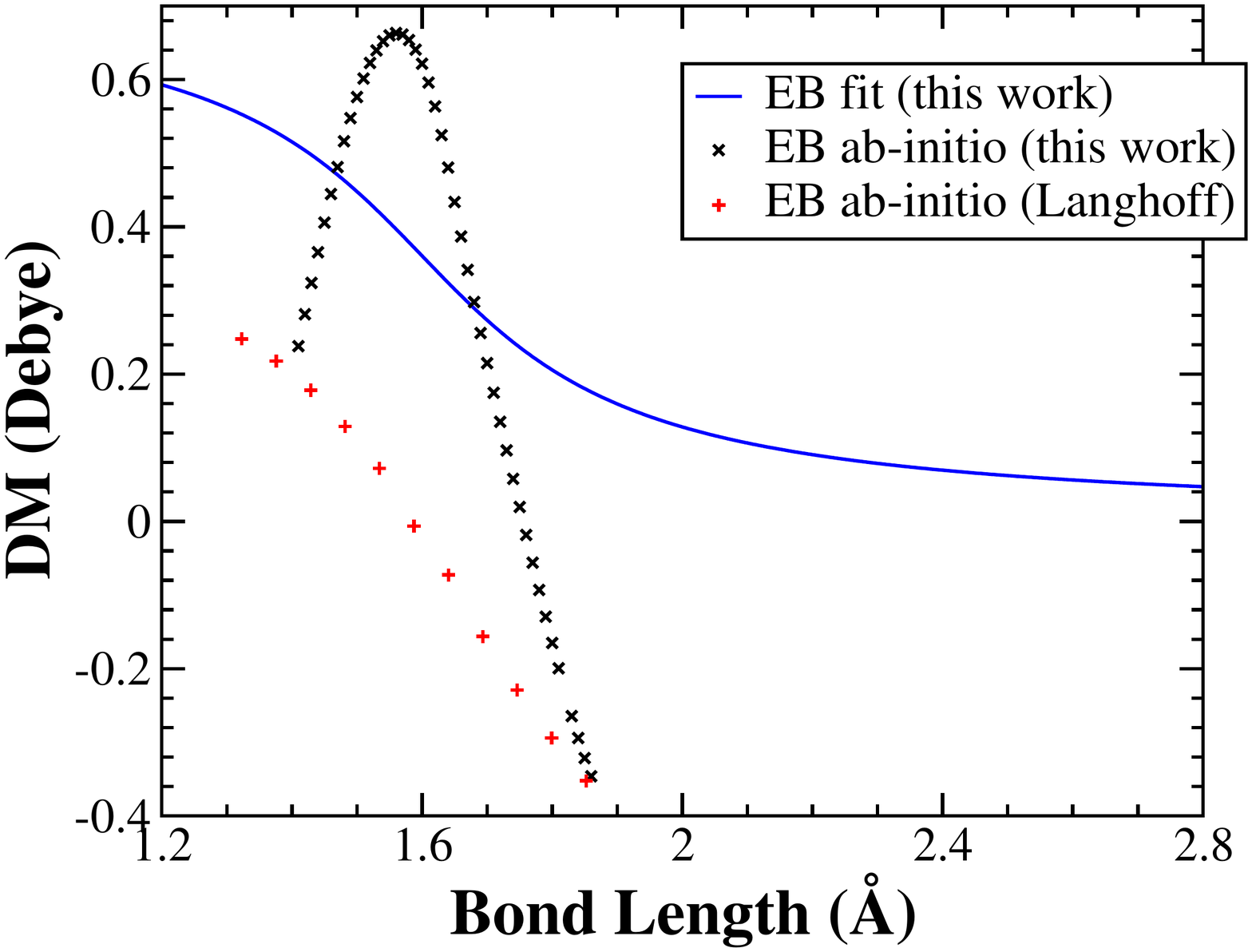}
\includegraphics[width=0.48\textwidth]{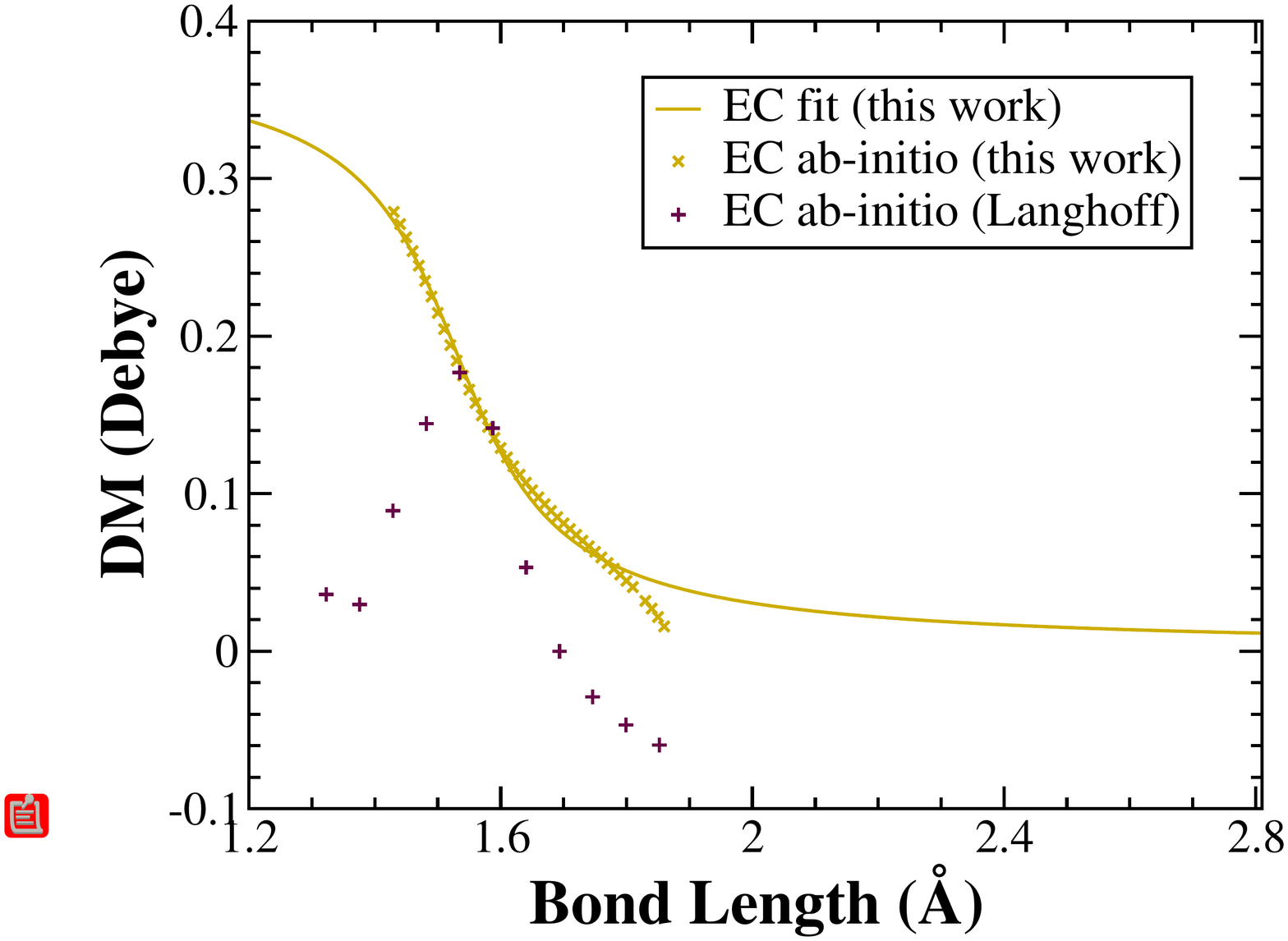}
\includegraphics[width=0.48\textwidth]{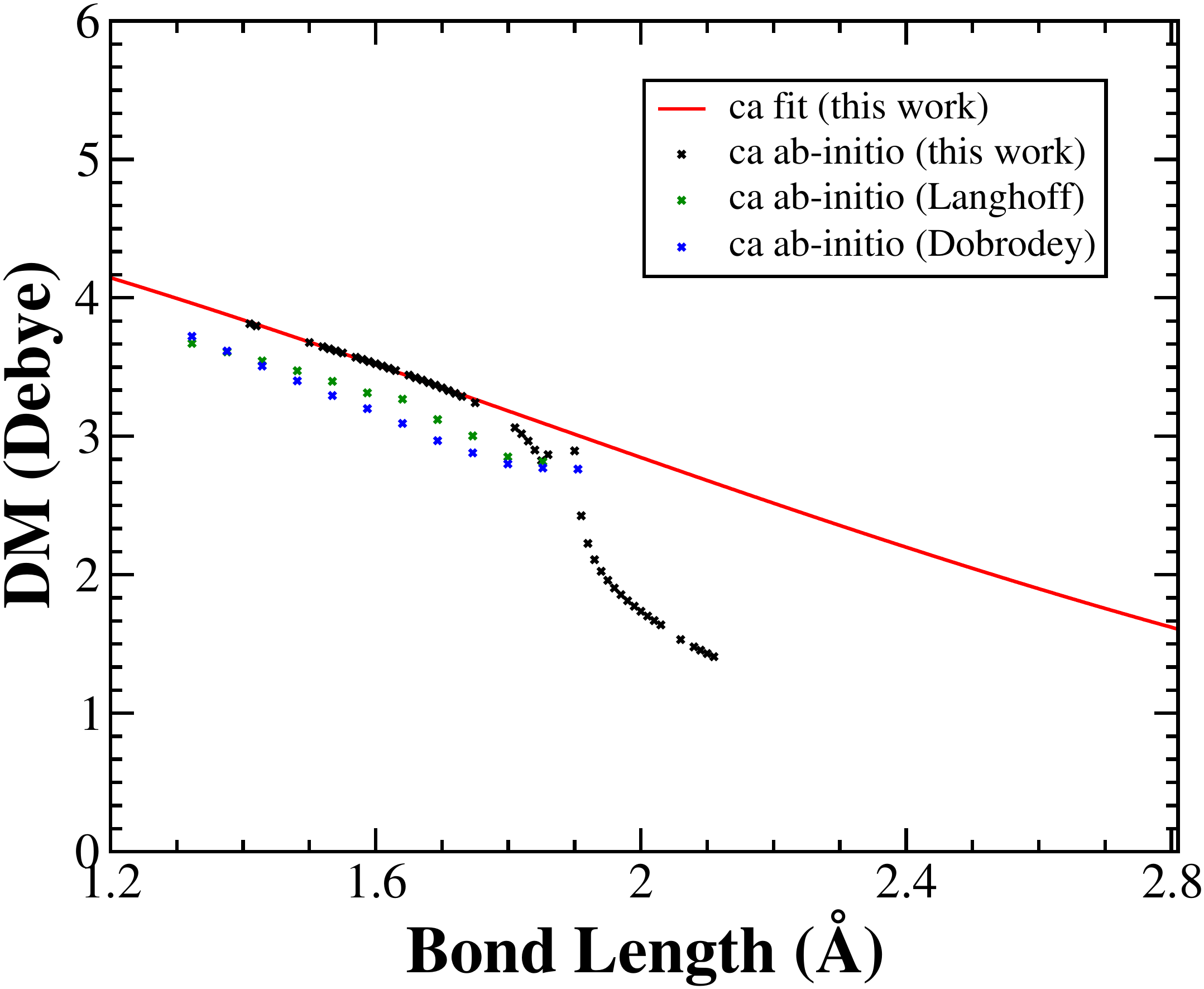}
\includegraphics[width=0.48\textwidth]{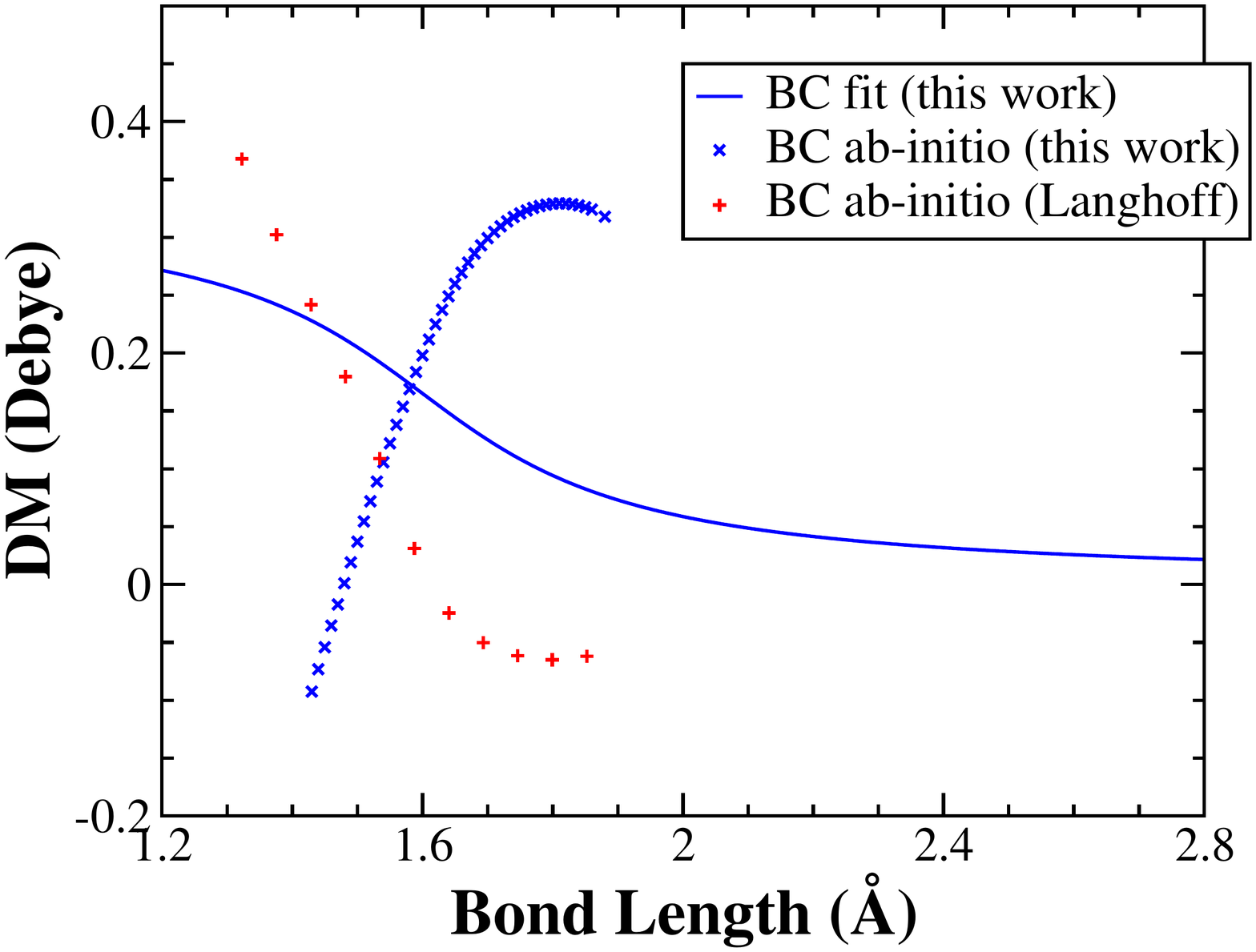}
\caption{\label{fig:offdiagDMTrip} Other \abinitio{} off-diagonal dipole moment curves for triplets in crosses, and the fitted curves in continuous curves.}
\end{figure*}

\paragraph{Other triplet off-diagonal dipole moments}

We also performed \abinitio{} calculations for other triplet off-diagonal dipole moments, with results compared against Langhoff's results in \Cref{fig:offdiagDMTrip}. Qualitatively, only our result for the E--C curve and the Langhoff B--C curve smoothly converge towards zero at long bond length because these states go to the same atomic limits. All other curves from both us and Langhoff show clear signs of the effect of curve crossings or computational convergence issues (e.g. changing CASSCF orbital type).

An arctan fit to our E-C results was reasonable. However, for the E--B, A--C and B--C curves, the {\it ab initio} results do not have the correct long-range behaviour, so the values of two of the parameters in the fit ($r_\e$ and $a$) were constrained with only $c$ allowed to vary meaning the results were of the right order of magnitude.  Overall, this is an unsatisfactory result, but the effect on the final line list should be negligible as none of these lines are particularly strong and the thermal population of the lower states is minimal. Quantitatively, fewer than 0.5\% of \TiO\ molecules at 3000 K are in the \E{}, \A{} or \B{} state (see \Cref{tab:pop}).

\paragraph{Singlet off-diagonal dipole moments}

\begin{table*}
\caption{\label{tab:offdiagDMSing}Singlet \abinitio\ off-diagonal dipole moments $ \langle {\rm State}' |\mu_x| {\rm State}'' \rangle $ in Debye from \citet{97Laxxxx.TiO} at $r$=1.59 \AA, $\mu_\text{eq}$, and new parameters for fitting curve  (this work). }
\begin{tabular}{llrrrrrH}

\hline
 & &  \mc{1}{c}{{\Abinitio}} & \mc{4}{c}{{Fitting Parameters}}  \\
  \cmidrule(r){3-3} \cmidrule(r){4-7}
& &  \mc{1}{c}{$|\mu|_\text{eq}$} &  \mc{1}{c}{$|\mu|_\text{eq}^\text{fitted}$} &  \mc{1}{c}{$c$} & \mc{1}{c}{$a$} & \mc{1}{c}{$r_m$} & $f$  \\
\hline
$\langle ^1\Delta_{z} |\mu_{x}| ^1\Pi_x \rangle$ & a-b & 3.108 & 3.101 & 3.996 &  3.334 & 1.953   \\
$\langle ^1\Delta_{z} |\mu_{x}| ^1\Phi_x \rangle$ &a-c & 3.314 & 3.295 & 4.723 & 1.705 &  2.019 \\
$\langle ^1\Delta_{z} |\mu_{z}| ^1\Delta_z \rangle$ & a-f & 3.829 & 3.808 & 5.608 & 1.911 & 1.930 & \\
$\langle ^1\Sigma^+ |\mu_{x}| ^1\Pi_x \rangle$ &d-b & 2.099 & 2.114 & 4.227 & 1.146 & 1.6$^*$&   \\
$\langle ^1\Pi_{y} |\mu_{x}| ^3\Delta_{xy} \rangle$ &b-f & 0.851 & 0.814 & 1.518 & 6.621 & 1.617 \\
\hline
\end{tabular}

$^*$ Constrained
\end{table*}

\begin{figure}
\includegraphics[width=0.48\textwidth]{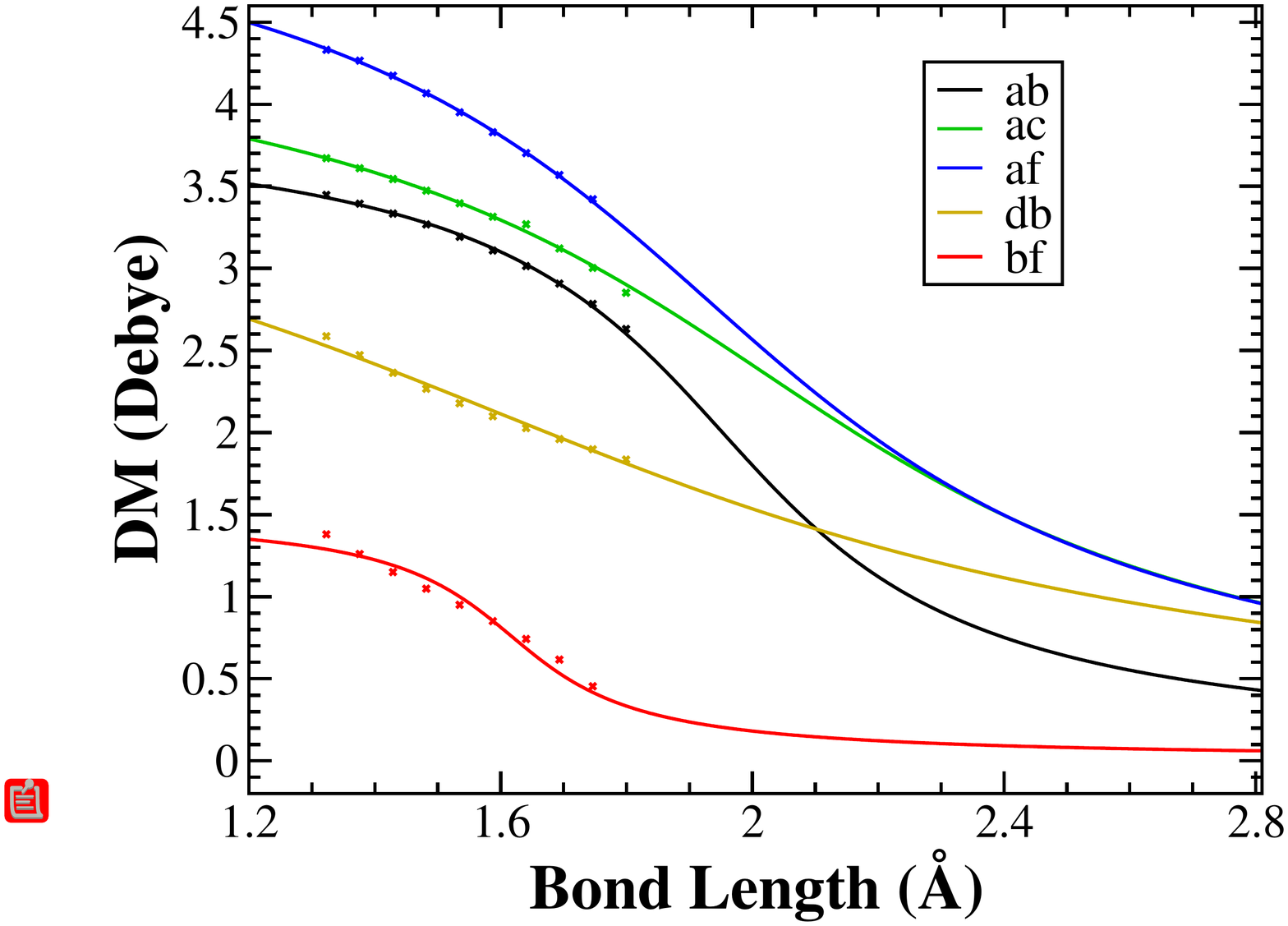}
\caption{\label{fig:offdiagDMSing} \Abinitio{} off-diagonal dipole moment for singlets in crosses from \citet{97Laxxxx.TiO}, and the fitted curves (this work) in continuous curves.}
\end{figure}

For the singlet off-diagonal dipole moments, we chose to fit the results from \cite{97Laxxxx.TiO} to our arctan functional form. Parameters of the fit and original results are shown in \Cref{tab:offdiagDMSing}, while the fits and \abinitio\ data points are shown in \Cref{fig:offdiagDMSing}. The curves all monotonically decrease, thus adverse effects of the fit to the desired functional form are minimal.

\section{\label{sec:LL}The Complete Line List}

\subsection{Method}
The nuclear-motion Schr\"odinger equation for diatomics systems with many coupled electronic states was solved directly using the program \Duo\ \citep{jt609} with the  \LLname{} spectroscopic model to produce a set of Duo energy levels, i.e. the \texttt{*.states} file, and transition intensities, i.e. the \texttt{*.trans} file following standard ExoMol format \citep{jt548}.  One key input parameter for this calculation include the value of $\max(E_{\rm lower})$, which is the maximum energy of the lower energy level in the transitions. We chose $\max{E_{\rm lower}} = 20,000$
\cm{} because this meant that the lower energy levels contained 95\% of the total population at 5000 K according to the estimated partition function. The upper energy threshold was set at 50,000 \cm{}. The frequency range considered is up to 30,000 \cm{}; however, at higher frequencies in this range, our line list will be incomplete (e.g. the \Se-\Sa{} transition is not present in our model). \Duo\ uses a grid-based sinc DVR method to solve the coupled Schr\"odinger equation; for our final line list calculations we used 301 grid points uniformly distributed between 1.2 \AA{}  and 4.0 \AA, which were fully converged (within $10^{-6}$ \cm)  with respect to the number of points.
The vibrational basis set was contracted to 30 for each electronic state. We calculated rotational levels up to $J$ = 200.

We supplemented the completeness offered by the Duo spectroscopic model with the accuracy of the \Marvel{} energy levels by replacing individual \TiO\ \Duo{} energy levels with \Marvel{} empirical energy levels.
Furthermore, the overall \Duo\ error is expected to be larger than the isotopologue error. Therefore, the initial line list was improved by substituting quasi-experimental isotopic energy levels obtained by adding the \Marvel\ \TiO\ energies to the \Duo\ predicted isotopic energy difference,
see \citet{jt665}.

This step maximises the accuracy of the \LLname\ linelist for absolute frequency descriptions of line positions. This is particularly important for very high resolution spectroscopy of exoplanets currently being pursued by many groups. Note that the \Marvel{} energy levels have been updated since the original publication in \citet{jt672} with three key sets of experimental data, namely: \cite{18BiBexx.TiO} with singlet data, \cite{18HoBexx.TiO} with \C-\X{} data and  infrared spectra from \citet{18WiGiFu.TiO} and \citet{19BrWaFu.TiO}.  We have updated the original \Marvel{} TiO files to v2.0. These can be found in the Supplementary Information of this manuscript.

The lifetimes for each quantum state are calculated using {\sc ExoCross} \citep{jt708}, and reported as part of the \texttt{*.states} file. Note that the \Se{}, \F{}, \Sg{}, \Sh{} and \Si{} quantum states have infinite lifetime in this line list, as we have not included any transitions involving this electronic state due to lack of reliable \abinitio{} results.

\subsection{\label{sec:pf}Partition Function}

\begin{landscape}
\begin{table}
\small
\def\arraystretch{1.5}
\caption{\label{tab:pf} Partition Functions for \TiO. The \citet{98Scxxxx.TiO} partition function was found at \texttt{http://kurucz.harvard.edu/molecules/tio/tiopart.dat}.}
\begin{tabular}{rrrHrrrrrrrrrrr}
\toprule
T(K) & \TiO& \TiO & & \TiO& \TiO  & \TiO& \TiO  & \ce{^{46}Ti^{16}O} & \ce{^{47}Ti^{16}O} & \ce{^{49}Ti^{16}O} & \ce{^{50}Ti^{16}O} \\
& \cite{66Tatum} & \cite{98Scxxxx.TiO} &&  Duo ELs & Marvelised ELs   & Duo All States & Rec. PF    & Rec. PF& Rec. PF & Rec. PF & Rec. PF  \\
\midrule
0    & - &- & & 6 & 6  &6 & 6 & 6 & 21      &    27   & 6  &  \\
1 & -  &   -    &  & \scinot{6.48613600E+00} & \scinot{6.48609738E+00} & \scinot{6.48613600E+00} & \scinot{6.48613600E+00} & \scinot{6.47022688E+00} & \scinot{3.88697610E+01} & \scinot{5.19505061E+01} & \scinot{6.50127992E+00} \\
5    & -       & - &  & \scinot{1.60603589E+01} & \scinot{1.60776044E+01} & \scinot{1.60603589E+01} & \scinot{1.60603589E+01} & \scinot{1.59232214E+01} & \scinot{9.59578298E+01} & \scinot{1.29007464E+02} & \scinot{1.61893422E+01}  \\
10   & -      &  \scinot{29.107} &  &\scinot{2.91092180E+01} & \scinot{2.91640372E+01} & \scinot{2.91092180E+01} & \scinot{2.91092180E+01} & \scinot{2.88311298E+01} & \scinot{1.73835482E+02} & \scinot{2.33937278E+02} & \scinot{2.93706939E+01} \\
50   & -     &   \scinot{142.564} &  & \scinot{1.42572858E+02} & \scinot{1.42724492E+02} &\scinot{1.42572858E+02} & \scinot{1.42572858E+02} & \scinot{1.41088346E+02} & \scinot{8.51060783E+02} & \scinot{1.14625991E+03} & \scinot{1.43968563E+02}  \\
80   & -    &  \scinot{254.774} &  & \scinot{2.54787772E+02} & \scinot{2.54966536E+02} &\scinot{2.54787772E+02} & \scinot{2.54787772E+02} & \scinot{2.52109012E+02} & \scinot{1.52082960E+03} & \scinot{2.04854622E+03} & \scinot{2.57306274E+02}   \\
100  & -   &   \scinot{344.021} &  & \scinot{3.44039920E+02} & \scinot{3.44230911E+02} & \scinot{3.44039920E+02} & \scinot{3.44039920E+02} & \scinot{3.40409769E+02} & \scinot{2.05353778E+03} & \scinot{2.76620167E+03} & \scinot{3.47452891E+02}\\
200  & -  &   \scinot{908.862}  &  &\scinot{9.08922330E+02} & \scinot{9.09140164E+02} & \scinot{9.08922330E+02} & \scinot{9.08922330E+02} & \scinot{8.99230104E+02} & \scinot{5.42496030E+03} & \scinot{7.30844653E+03} & \scinot{9.18035733E+02}\\
300  & -   &    \scinot{1589.41}  &  &\scinot{1.58952822E+03} & \scinot{1.58974798E+03} &\scinot{1.58952822E+03} & \scinot{1.58952822E+03} & \scinot{1.57224071E+03} & \scinot{9.48619753E+03} & \scinot{1.27823660E+04} & \scinot{1.60579113E+03}\\
500  & -  &   \scinot{3208.50}    &  &\scinot{3.20876868E+03} & \scinot{3.20897132E+03} & \scinot{3.20876868E+03} & \scinot{3.20876868E+03} & \scinot{3.17154989E+03} & \scinot{1.91428407E+04} & \scinot{2.58126678E+04} & \scinot{3.24382125E+03} \\
800  & - &  \scinot{6370.03}      &  & \scinot{6.37060671E+03} & \scinot{6.37077265E+03} & \scinot{6.37060671E+03} & \scinot{6.37060671E+03} & \scinot{6.29033119E+03} & \scinot{3.79868150E+04} & \scinot{5.12724874E+04} & \scinot{6.44628816E+03} \\
1000 & \scinot{1.027e4}        & \scinot{9024.67}  &  &\scinot{9.02559121E+03} & \scinot{9.02573394E+03} &\scinot{9.02559132E+03} & \scinot{9.02559132E+03} & \scinot{8.90744444E+03} & \scinot{5.38049588E+04} & \scinot{7.26576364E+04} & \scinot{9.13702208E+03}   \\
1500 & \scinot{1.912e4}       & \scinot{17781.8}   &  & \scinot{1.77840501E+04} & \scinot{1.77842547E+04} & \scinot{1.77841283E+04} & \scinot{1.77841283E+04} & \scinot{1.75370780E+04} & \scinot{1.05975740E+05} & \scinot{1.43220504E+05} & \scinot{1.80172740E+04}\\
2000 & \scinot{3.081e4}      & \scinot{29880.5}    &  & \scinot{2.98832545E+04} & \scinot{2.98848006E+04} &\scinot{2.98857138E+04} & \scinot{2.98857138E+04} & \scinot{2.94564986E+04} & \scinot{1.78047587E+05} & \scinot{2.40732229E+05} & \scinot{3.02909037E+04}  \\
3000 & \scinot{6.308e4}     &  \scinot{65735.9}    &  &\scinot{6.56551326E+04} & \scinot{6.56752956E+04} & \scinot{6.57596895E+04} & \scinot{6.57596895E+04} & \scinot{6.47815147E+04} & \scinot{3.91671123E+05} & \scinot{5.29830781E+05} & \scinot{6.66833929E+04}\\
5000 & \scinot{1.707e5}    &   \scinot{202508}    &  & \scinot{1.99656221E+05} & \scinot{1.99805750E+05} & \scinot{2.03289397E+05} & \scinot{2.03289397E+05} & \scinot{2.00219106E+05} & \scinot{1.21067594E+06} & \scinot{1.63809092E+06} & \scinot{2.06187086E+05} \\
8000& \scinot{4.990e5} &   -   &      & \scinot{6.47082111E+05} & \scinot{6.47493129E+05} &\scinot{6.91078817E+05} & \scinot{6.91078817E+05} & \scinot{6.81357539E+05} & \scinot{4.11780632E+06} & \scinot{5.56583749E+06} & \scinot{7.00227532E+05} \\
\bottomrule
\end{tabular}
\end{table}
\end{landscape}


The partition function for each set of energy levels was calculated using {\sc ExoCross} \citep{jt708} at 1 K intervals up to 8000 K. Results at representative temperature intervals are shown in \Cref{tab:pf}. We compare the partition function from \citet{66Tatum}, \cite{98Scxxxx.TiO} against one computed for \TiO\ with Duo energy levels (\Duo\ EL) excluding the \D{}, \F{}, \Sg{}, \Sh{} and \Si{} states, one with \Marvel{} energy levels also excluding these states, one with Duo energy levels with approximate values for these additional electronic states and five recommended partition functions for each major isotopologue of TiO which utilise these additional electronic states and \Marvel ised energy levels. 

The \cite{98Scxxxx.TiO} and \LLname{} partition functions are very similar for temperatures, and certainly within the mutual errors of both models. There is, however, a substantial improvement from \cite{66Tatum} results.

It is clear that replacing the Duo energy levels with \Marvel{} energy levels  has a negligible effect on the partition function (less than 0.1\%). Therefore, as expected this replacement of energy levels with experimentally derived \Marvel{} energy levels is much more important for ensuring accuracy than completeness of data.

Adding the \D{}, \F{}, \Sg{}, \Sh{} and \Si{} electronic states has an influence of 0.2\% at 3000 K, 1.8\% at 5000 K and 6.4\% at 8000 K. This gives confidence that at astrophysically relevant temperatures for TiO (i.e. usually 3000 K or below), the effects of neglecting these electronic states is negligible in the context of the current accuracy of TiO opacity measurements.

ExoMol follows the HITRAN convention of explicitly including the full nuclear spin multiplicity in the partition function \citep{jt692}.
The nuclear degeneracy factors mean that the partition function for \ce{^{47}Ti^{16}O} and \ce{^{47}Ti^{16}O} are approximately 6 and 8 times larger than the \ce{^{48}Ti^{16}O} partition function at a given temperature. Apart from this, the partition function slightly increases with isotopologue mass. This can be attributed to the slight decrease in vibrational and rotational energy level spacings. This effect is modest, around 2-3\%.




\subsection{Characteristics of \LLname{}  line list. }

\subsubsection{States and Trans Files}

\begin{table*}
\caption{\label{t:Energy-file}  Extract from the state file for \ce{^{48}Ti^{16}O}. Full tables are available from \url{www.exomol.com}.}
\footnotesize \tabcolsep=5pt
\begin{tabular}{rrrrrcrcrrcccr}
\toprule
   \mc{1}{c}{1}     &        \mc{1}{c}{2}        &   \mc{1}{c}{3}  &   \mc{1}{c}{4}  &   \mc{1}{c}{5}  &   \mc{1}{c}{6}  &   \mc{1}{c}{7}  &   \mc{1}{c}{8}  &   \mc{1}{c}{9}  &  \mc{1}{c}{10}  &  \mc{1}{c}{11} &  \mc{1}{c}{12} &  \mc{1}{c}{13} &  \mc{1}{c}{14}\\ 
    \midrule
     \mc{1}{c}{$n$} & \multicolumn{1}{c}{$\tilde{E}$} &  \mc{1}{c}{$g$}    & \mc{1}{c}{$J$}  & \mc{1}{c}{$\tau$} & \mc{1}{c}{$g$} &  \multicolumn{1}{c}{$+/-$} &  \multicolumn{1}{c}{$e/f$} & \mc{1}{c}{State} & \mc{1}{c}{$v$}    & \mc{1}{c}{$\Lambda$}& \mc{1}{c}{$\Sigma$} & \mc{1}{c}{$\Omega$}&  \multicolumn{1}{c}{$\tilde{E}_\Duo$}\\
     \midrule
 385 & 23432.014308 & 3 & 1 & 1.3511E-07 & 0.5000 & + & f & b1Pi & 10 & 1 & 0 & 1 & 23432.014308 \\
  386 & 23654.113628 & 3 & 1 & 6.3865E-08 & 0.0845 & + & f & B3Pi & 9 & 1 & -1 & 0 & 23654.113628 \\
  387 & 23675.747870 & 3 & 1 & 6.4765E-08 & 0.4154 & + & f & B3Pi & 9 & 1 & 0 & 1 & 23675.747870 \\
  388 & 23731.604159 & 3 & 1 & 3.0397E-06 & 0.0211 & + & f & E3Pi & 14 & 1 & -1 & 0 & 23731.604159 \\
  389 & 23768.803862 & 3 & 1 & 1.9771E-01 & -0.0011 & + & f & X3Delta & 27 & 2 & -1 & 1 & 23768.803862 \\
  390 & 23816.500357 & 3 & 1 & 3.0184E-06 & 0.4788 & + & f & E3Pi & 14 & 1 & 0 & 1 & 23816.500357 \\
  391 & 24085.108637 & 3 & 1 & 1.0677E-02 & 1.0011 & + & f & D3Sigma- & 13 & 0 & 1 & 1 & 24085.108637 \\
  392 & 24179.682010 & 3 & 1 & 5.3403E-08 & -0.0011 & + & f & C3Delta & 6 & 2 & -1 & 1 & 24179.682010 \\
  393 & 24187.550885 & 3 & 1 & -1.0000E+00 & 2.0023 & + & f & F3Sigma+ & 5 & 0 & 1 & 1 & 24187.550885 \\
  394 & 24188.111429 & 3 & 1 & -1.0000E+00 & 0.5000 & + & f & i1Pi & 2 & 1 & 0 & 1 & 24188.111429 \\
  395 & 24190.413889 & 3 & 1 & -1.0000E+00 & -1.0011 & + & f & F3Sigma+ & 5 & 0 & 0 & 0 & 24190.413889 \\
  396 & 24254.153701 & 3 & 1 & 1.3701E-07 & 0.5000 & + & f & b1Pi & 11 & 1 & 0 & 1 & 24254.153701 \\
  397 & 24430.660917 & 3 & 1 & 6.4350E-08 & 0.0838 & + & f & B3Pi & 10 & 1 & -1 & 0 & 24430.660917 \\
 \bottomrule
\end{tabular}

\begin{tabular}{cll}
\\
  Column       &    Notation      &      \\
\midrule
   1 &   $n$   &       Energy level reference number (row)    \\
   2 & $\tilde{E}$        &       Term value (in \cm) \\
   3 &  $g_{\rm tot}$     &       Total degeneracy   \\
   4 &  $J$    &       Rotational quantum number    \\
   5 & $\tau$ & Lifetime (s) \\
   6 & $g$ & Land\'e factors \\
   7 & $+/-$ & Total parity  \\
   8 & $e/f$ & Rotationless parity \\
   9 & State & Electronic state \\
   10 & $v$ & State vibrational quantum number \\
  11 &  $\Lambda$ &   Projection of the electronic angular momentum \\
 12 & $\Sigma$ &   Projection of the electronic spin \\
13 & $\Omega$ & Projection of the total angular momentum ($\Omega=\Lambda+\Sigma$) \\
14 & $\tilde{E}_{\rm \Duo}$ & Energy from \Duo{} using \LLname{} spectroscopic model \\
\bottomrule
\end{tabular}
\end{table*}

  \begin{table}
\caption{\label{t:Transit-file} Extract from the transition file for \ce{^{48}Ti^{16}O}. Full tables
are available from \url{www.exomol.com}.}
\centering
\begin{tabular}{rrr}
\toprule
$f$  &  $i$  & $A({\rm f}\leftarrow {\rm i})$ / s$^{-1}$ \\
\midrule
48953  & 48319  & 4.4282E-03 \\
48324  & 48949  & 4.4283E-03 \\
83765  & 84381  & 4.7035E-20 \\
17291  & 19040  & 4.7189E-07 \\
30933  & 31554  & 6.7524E-11 \\
698    & 1169   & 6.3331E-12 \\
26823  & 26176  & 2.1256E-07 \\
105306 & 104729 & 1.6203E-02 \\
104733 & 105301 & 1.6200E-02 \\
\bottomrule
\end{tabular}

\noindent
 $f$: Upper (final) state counting number;

$i$: Lower (initial) state counting number;

$A({\rm f}\leftarrow {\rm i})$:  Einstein $A$ coefficient in s$^{-1}$.

\end{table}

The final \texttt{48Ti-16O\_\LLname.states} file contains \noELs{} energy levels. For \noMARVEL{} of these energy levels, the original \Duo{} energy values were replaced by the \Marvel{} data.
Our \texttt{48Ti-16O\_\LLname.trans} file contains \notrans{} transitions. 26,540,994 and 33,005,825 of these transitions have intensities above $10^{-32}$ cm/molecule  at 1500 K and 3000 K respectively, while there are 374,819 and 898,090 transitions with intensities above 10$^{-20}$ cm/molecule.
Tables~\ref{t:Energy-file} and \ref{t:Transit-file} give extracts of the states and transitions files.
Note that the states file also contains Land\'e $g$ factors computed using our \Duo\ wavefunctions
\citep{jt656}; these can be used to model the spectrum of TiO in a magnetic field.

\begin{figure*}
\includegraphics[width=0.48\textwidth]{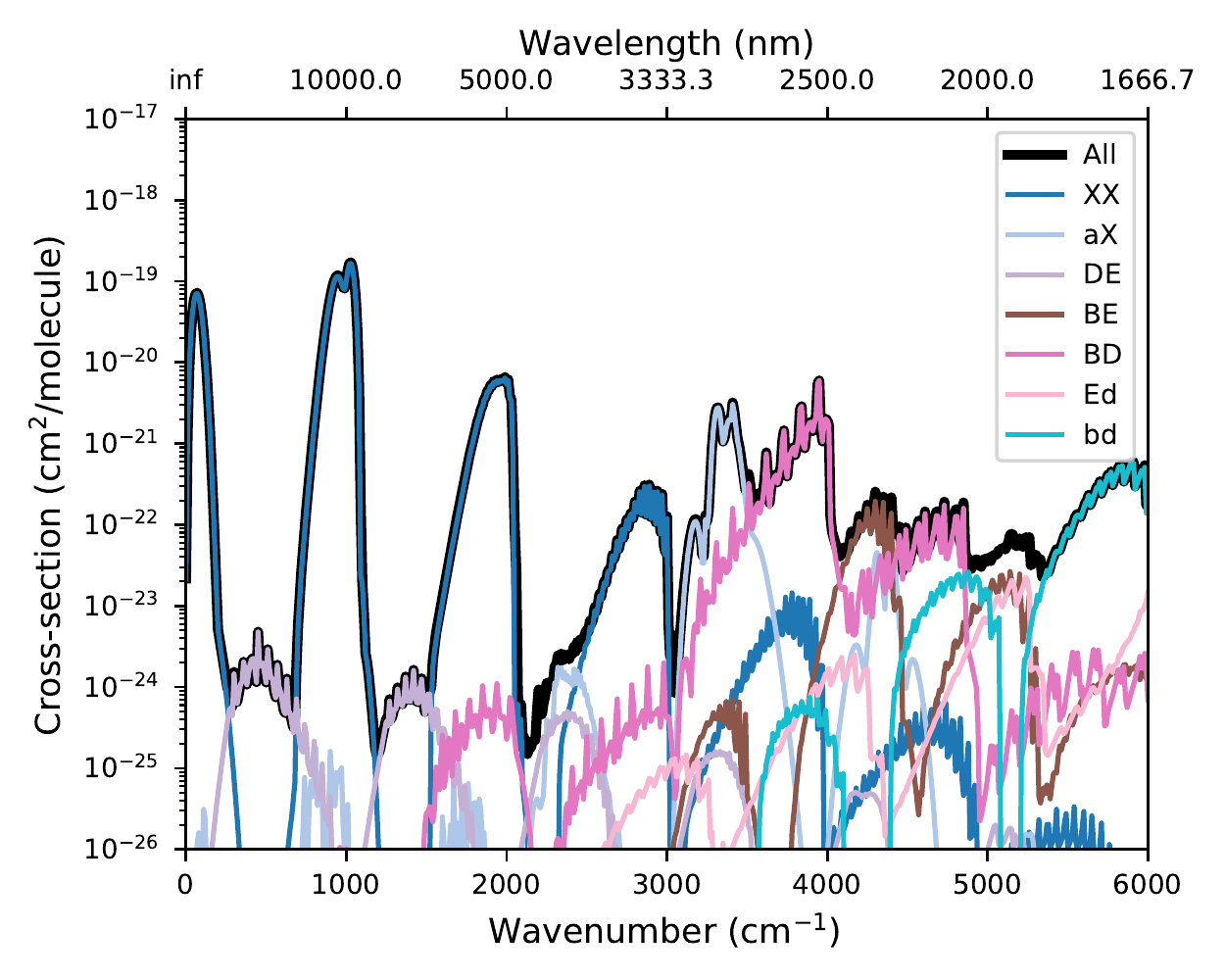}
\includegraphics[width=0.48\textwidth]{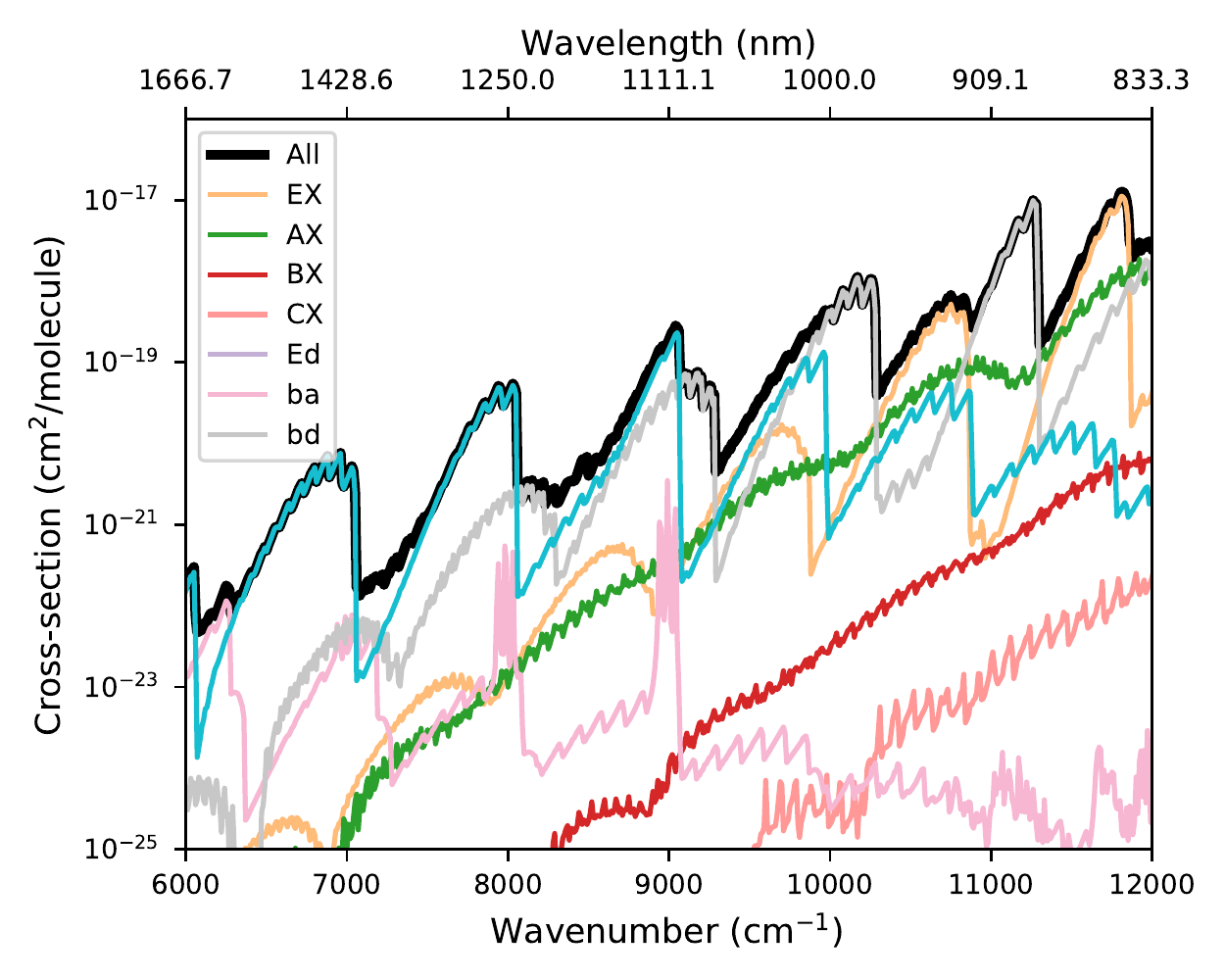}

\includegraphics[width=0.48\textwidth]{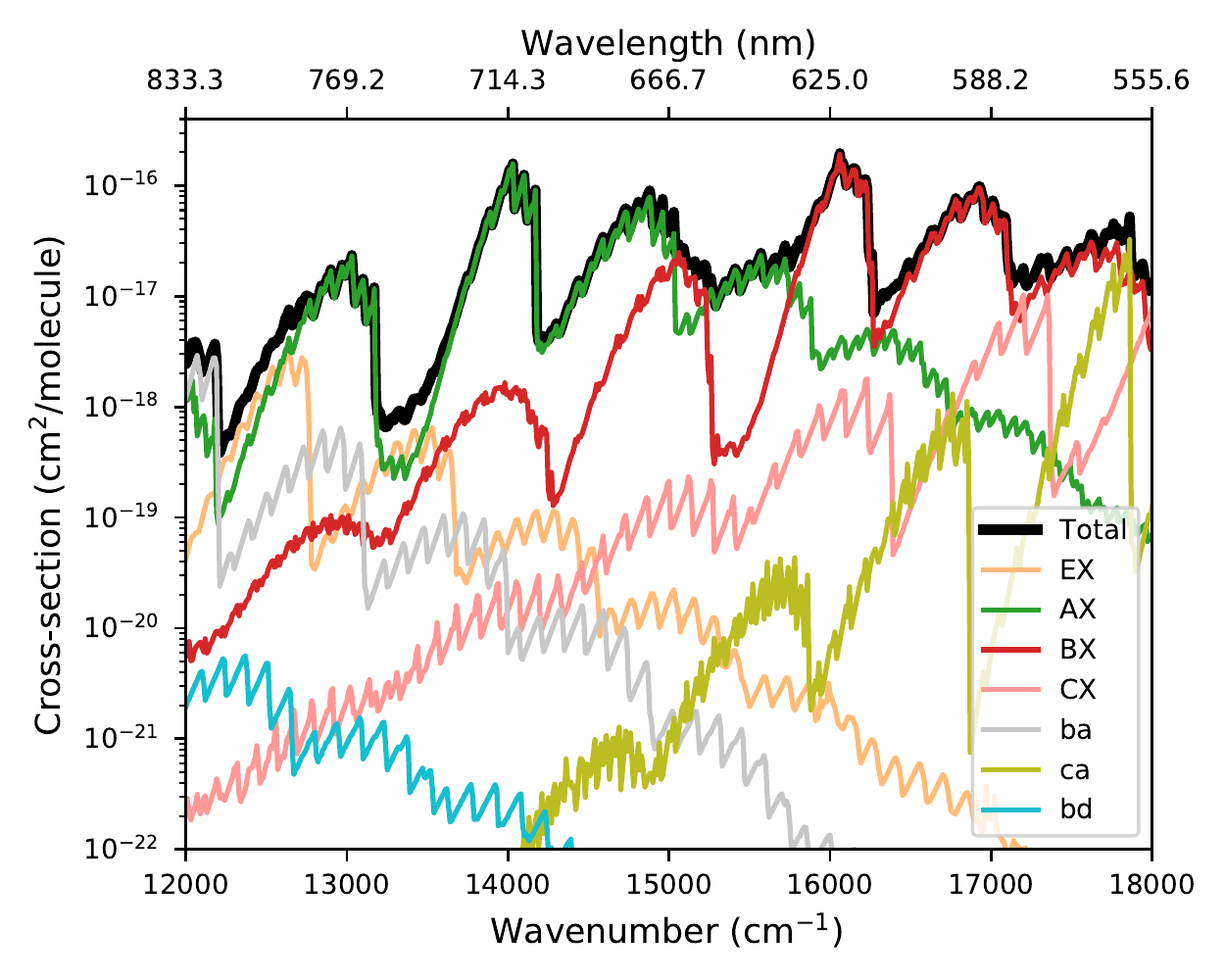}
\includegraphics[width=0.48\textwidth]{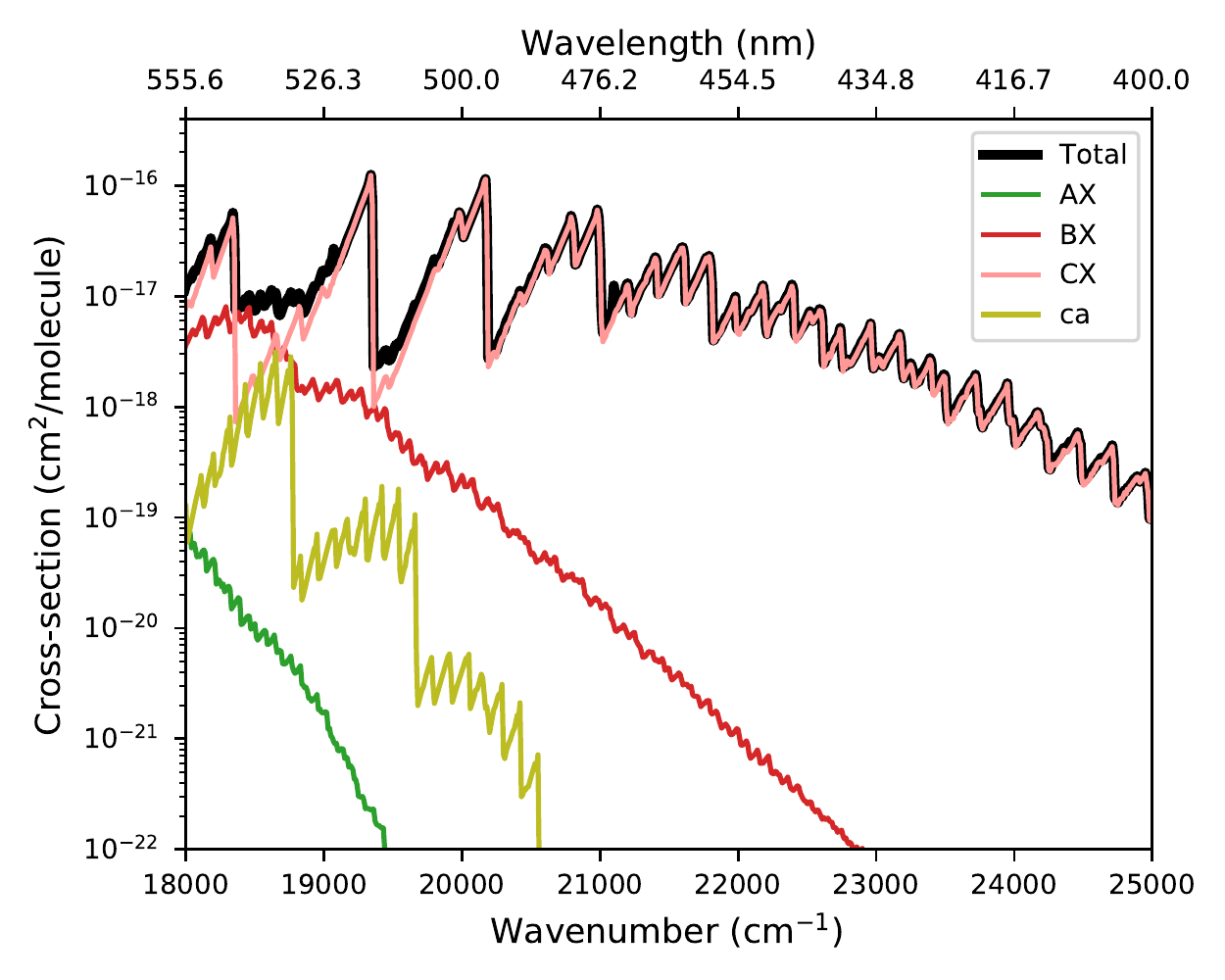}

\caption{\label{fig:cmfull} Decomposition of the total cross-section of TiO using \LLname{} below 25,000 \cm{} into the main bands for an assumed temperature of 2000 K. Individual lines have been broadened with a
Gaussian profile with a half-width half-maximum (HWHM) of 2 \cm{}, T=2000 K. Note that different $y$-axis scales have been used for each sub-figure.}
\end{figure*}

Figure \ref{fig:cmfull} identifies the main bands in the absorption spectrum of TiO at 2000 K. The strongest visible bands have cross-sections up to 10$^{-16}$ cm$^2$/molecule, while the infrared and microwave bands are much weaker, peaking at just above 10$^{-20}$ cm$^2$/molecule. The \A\ -- \X{}, \B\ -- \X{} and \C\ -- \X{} bands have similar peak cross-sections, while the \E\ -- \X{} band is much weaker and does not dominate the total absorption in any spectral region. The singlet bands dominate the spectra in the region from 4,500 \cm{} to 11,000 \cm{}, particularly the \Sb\ -- \Sa{} and \Sb\ -- \Sd{} transitions. Interestingly, the  \Sa\ -- \X{} inter-combination band is observable in the range 2000 - 4500 \cm{}. In the visible region, there are many bands that contribute to the overall spectra, though there is usually a dominant triplet progression.


\begin{figure*}
\includegraphics[width=0.48\textwidth]{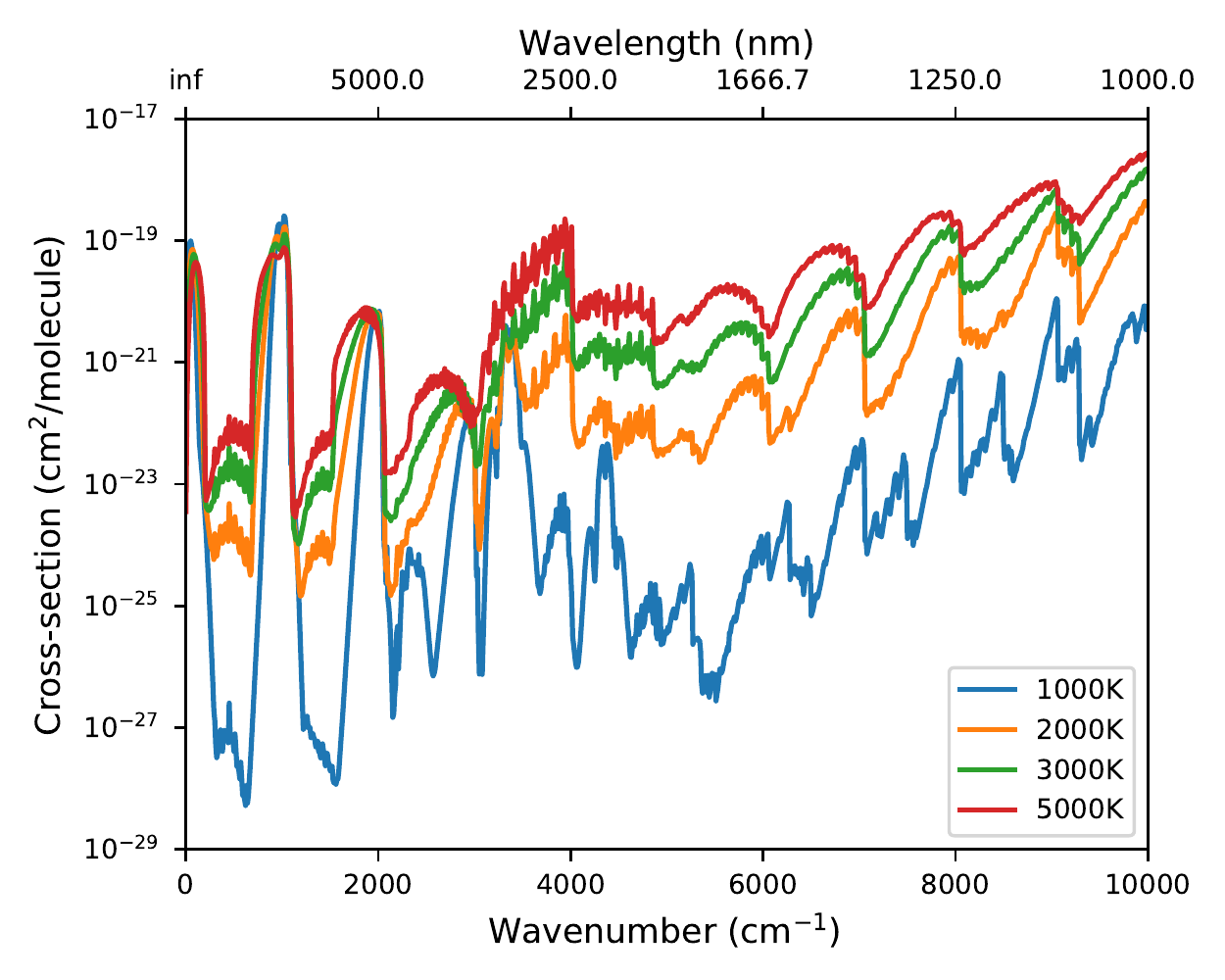}
\includegraphics[width=0.48\textwidth]{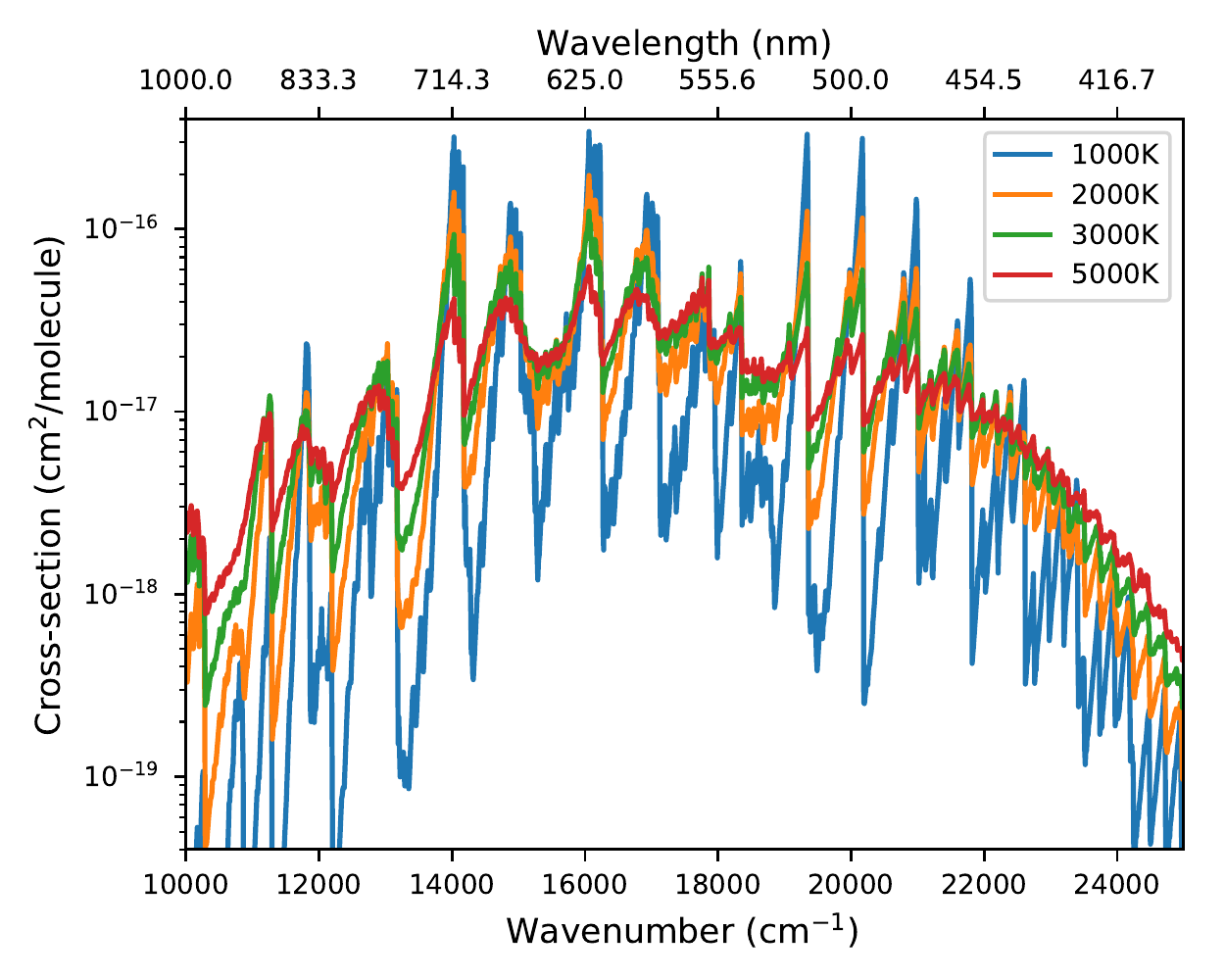}
\caption{Overview of the full spectrum of TiO using \LLname{} as a function of temperature for $T$ = 1000, 2000, 3000 and 5000~K, absorption cross-sections with HWHM = 2 cm$^{-1}$.}
\label{fig:TempEffect}
\end{figure*}

\Cref{fig:TempEffect} shows the TiO absorption spectrum as a function of temperature. As expected, the spectra becomes less defined and  broader at higher temperatures. The near-infrared bands between 5000 and 10,000 \cm{} (mostly \Sb-\Sa{} and \Sb-\Sd{} transitions) are particularly pronounced, as the lower \Sa{} and \Sd{} singlet states become increasingly populated at higher temperatures.


\subsection{Comparisons}

\begin{table}
\centering
\caption{\label{tab:lifetime} Lifetimes of \ce{^{48}Ti^{16}O} excited rovibronic states from \citet{95HeNaCo.TiO} compared to lifetimes for \sc{Toto}.}
\begin{tabular}{lcrrcr}
\toprule
State & $v$ &  \mc{2}{c}{Experiment} & \mc{1}{c}{\LLname} \\
& & $\tau$ (ns) & $\Delta\tau$ (ns) &  $\tau$ (ns)  \\
\midrule
\A$_2$ & 0 & 103.3 & 3.6  & 110.9\\
\A$_3$ & 0 & 101.9 & 3.6  & 109.1\\
\A$_4$ & 0 & 98.7 & 4.4  & 107.5\\
\A$_2$ & 1 & 112.9 & 1.8  & 111.3\\
\A$_3$ & 1 & 109.2 & 2.8 & 109.5\\
\A$_4$ & 1 & 105.6 & 3.6  & 107.9\\
\\
\B$_0$ & 0 & 65.6 & 1.2  & 66.7\\
\B$_1$ & 0 & 64.6 & 1.6  & 62.2\\
\B$_2$ & 0 & 66.1 & 1.2  & 63.1\\
\B$_0$ & 1 & 66.5 & 1.0 & 61.4\\
\B$_1$ & 1 & 67.6 & 1.2  & 62.3 \\
\B$_2$ & 1 & 69.0 & 1.2 & 63.8 \\
\B$_0$ & 2 & 63.8 & 1.4 & 61.7 \\
\B$_1$ & 2 & 67.8 & 1.4 & 62.5\\
\B$_2$ & 2 & 68.1 & 2.0 & 63.5\\
\\
\C & 0 & 43.3 & 1.0  & 53.3\\
\C & 1 & 43.0 & 1.2  & 53.2 \\
\\
\Sc & 0 & 38.3 & 1.6  & 27.9 \\
\Sc & 1 & 37.3 & 5.2  & 28.3\\
\Sf & 0 & 43.2 & 2.0  & 37.5 \\

\bottomrule
\end{tabular}
\end{table}

\subsubsection{Lifetimes}

\Cref{tab:lifetime} compares the TiO lifetimes obtained using the \LLname{} line list against the experimental measurements of \cite{95HeNaCo.TiO}; they are quite close, particularly for the \A{} and \B{} states which will have the biggest effect on the observed astronomical spectra of TiO.

\begin{figure}
\includegraphics[width=0.48\textwidth]{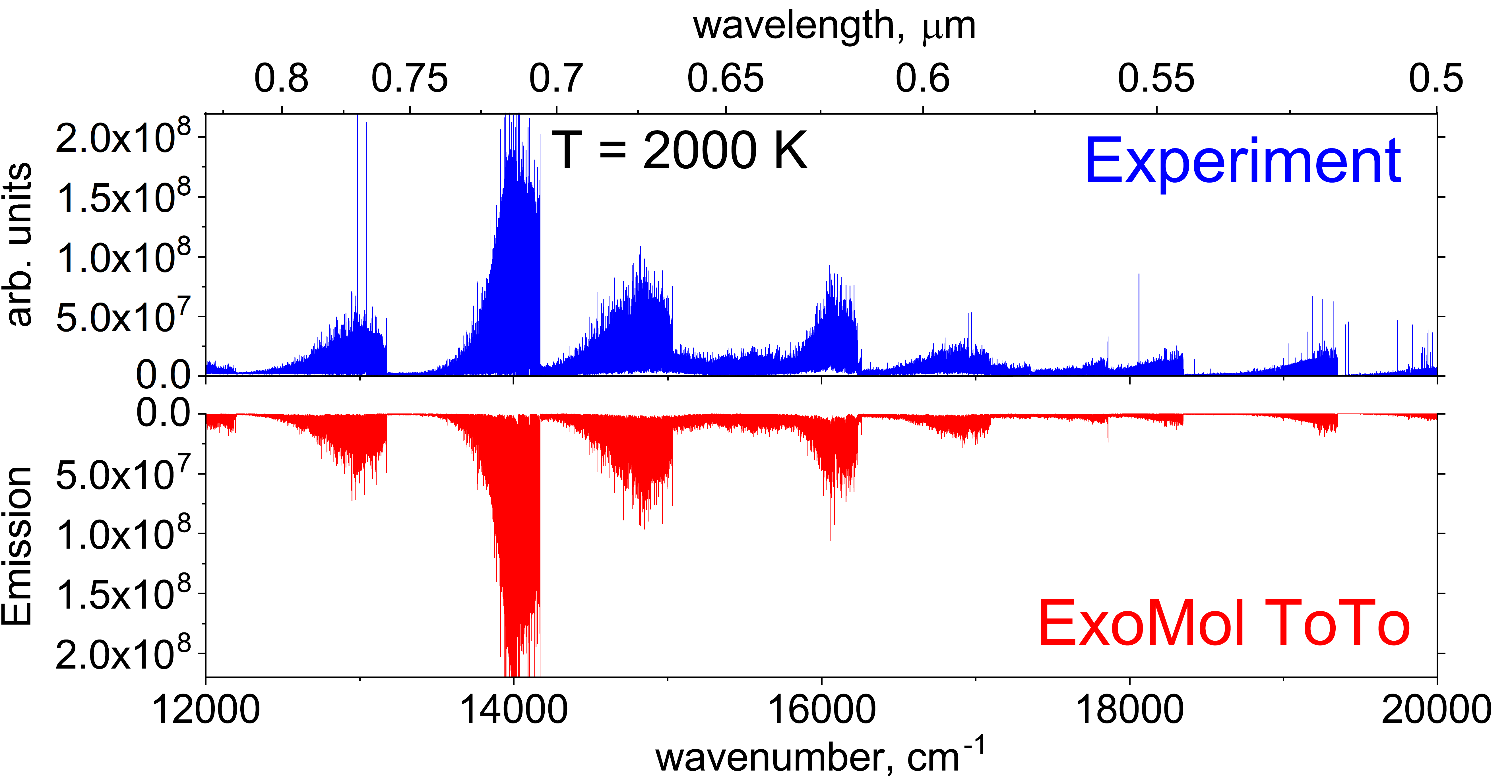}
\caption{\label{fig:Bernathoverview} Overview comparison between the \LLname{} line list and experimental spectra from the Kitt Peaks archive. The experimental Kitt Peak spectra were measured with the
1 m Fourier transform interferometer associated with the
McMath-Pierce Solar Telescope of the National Solar
emission. 
\LLname{} emission spectrum (photon/s) were computed using the Gaussian line profile  (HWHM=0.05 cm$^{-1}$)  at $T = 2000$ K and terrestrial isotopic abundances of the five stable isotopologues of TiO considered here. The theoretical spectrum was re-scaled to match the experimental spectrum.
}
\end{figure}

\begin{figure*}
\includegraphics[width=0.48\textwidth]{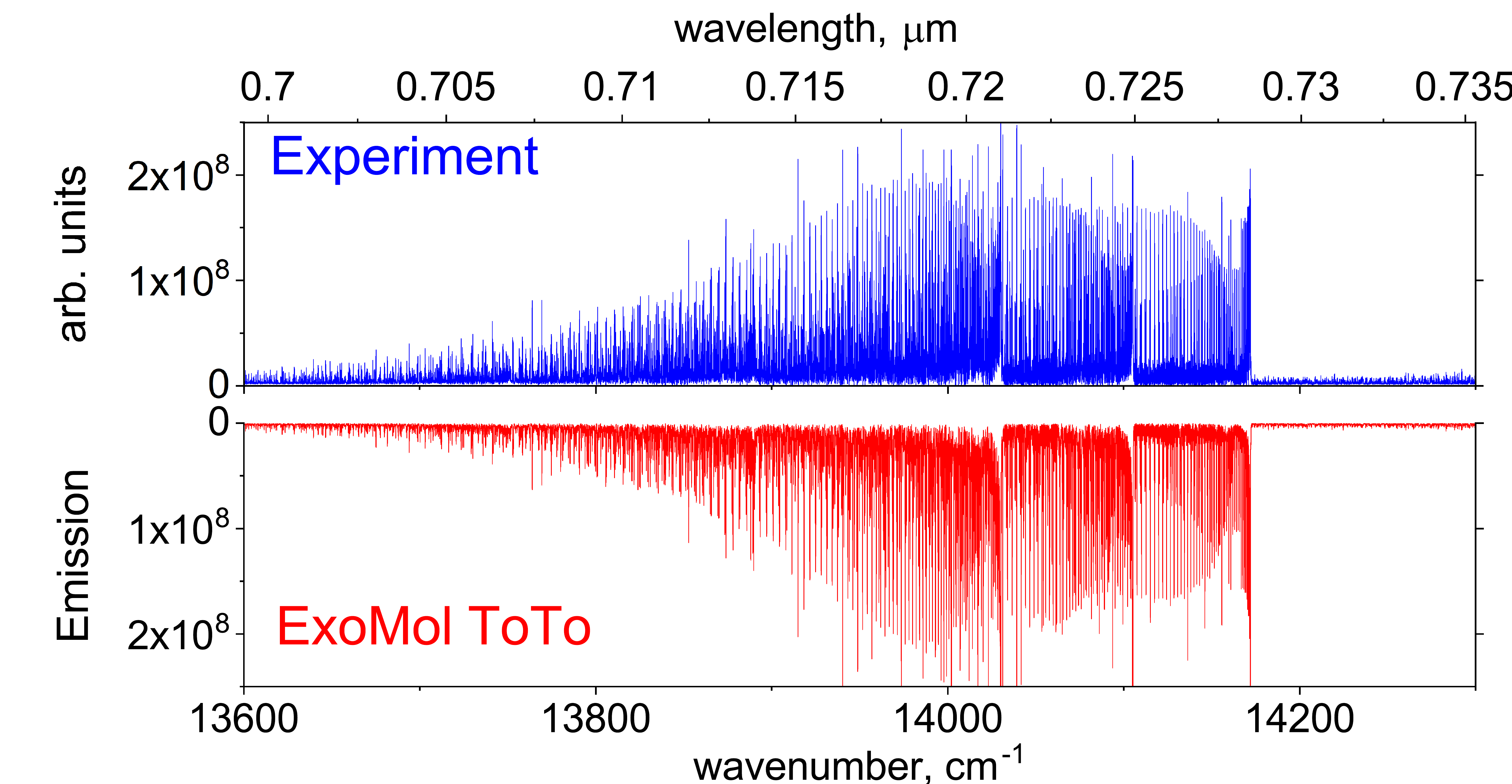}
\includegraphics[width=0.48\textwidth]{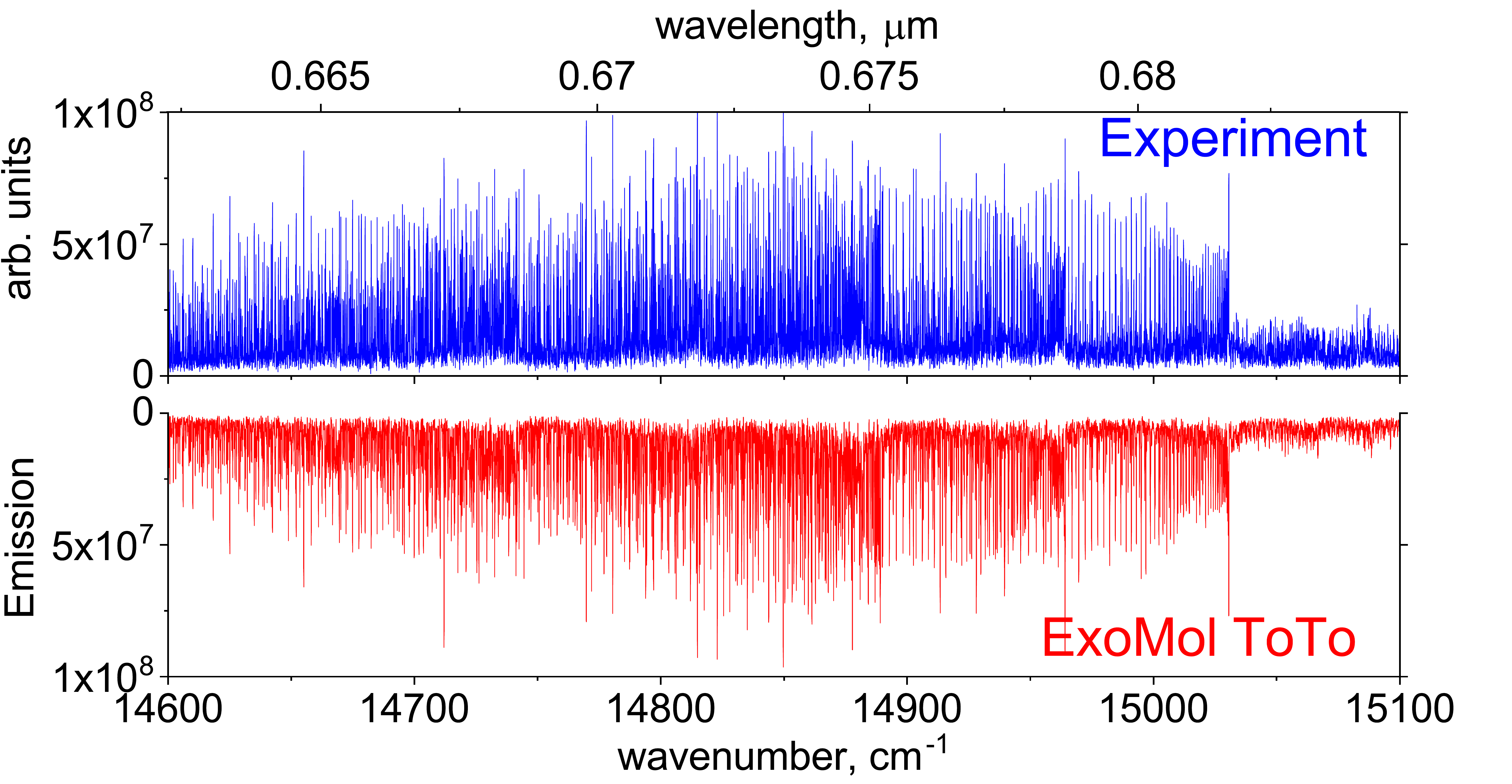}
\hspace{2em}

\includegraphics[width=0.48\textwidth]{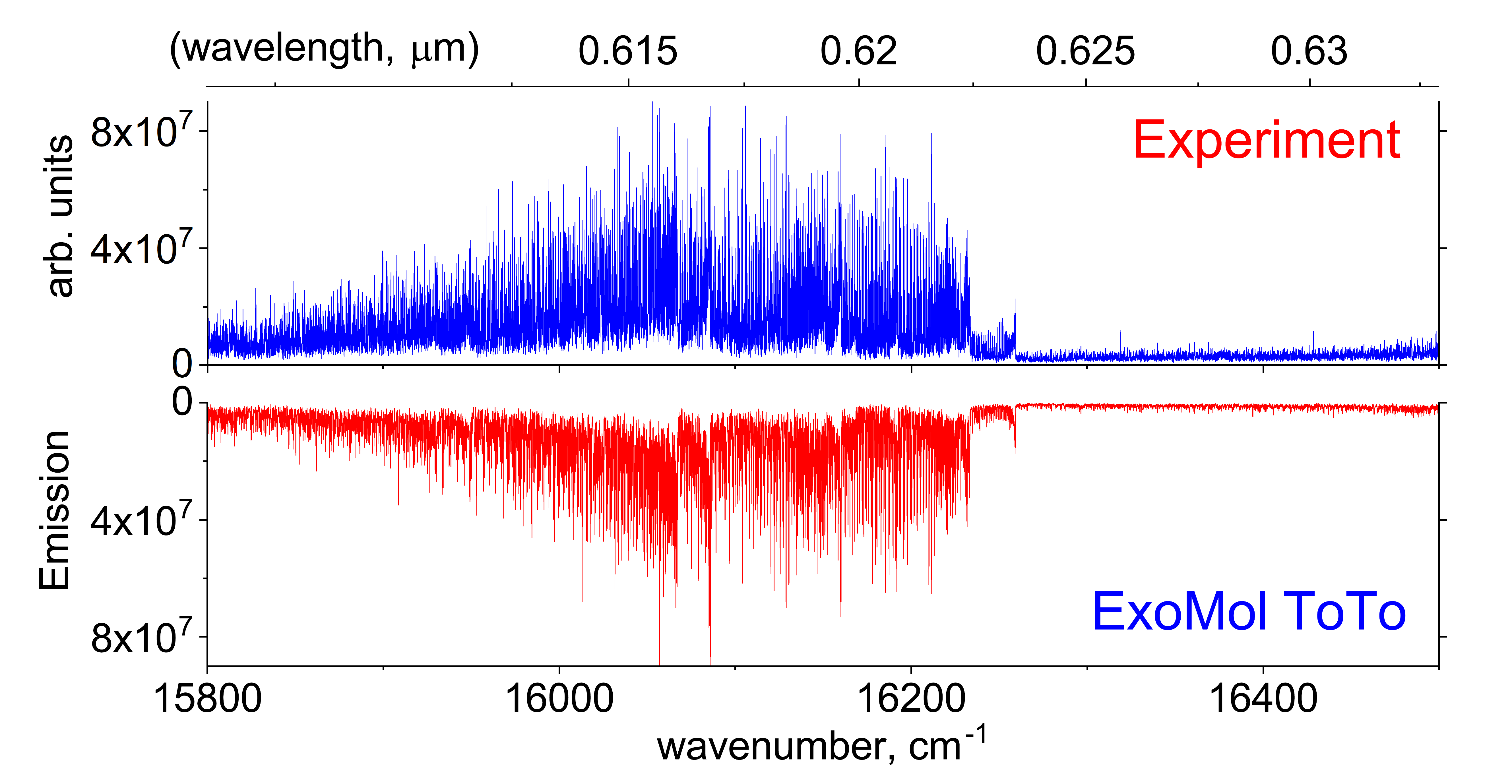}
\includegraphics[width=0.48\textwidth]{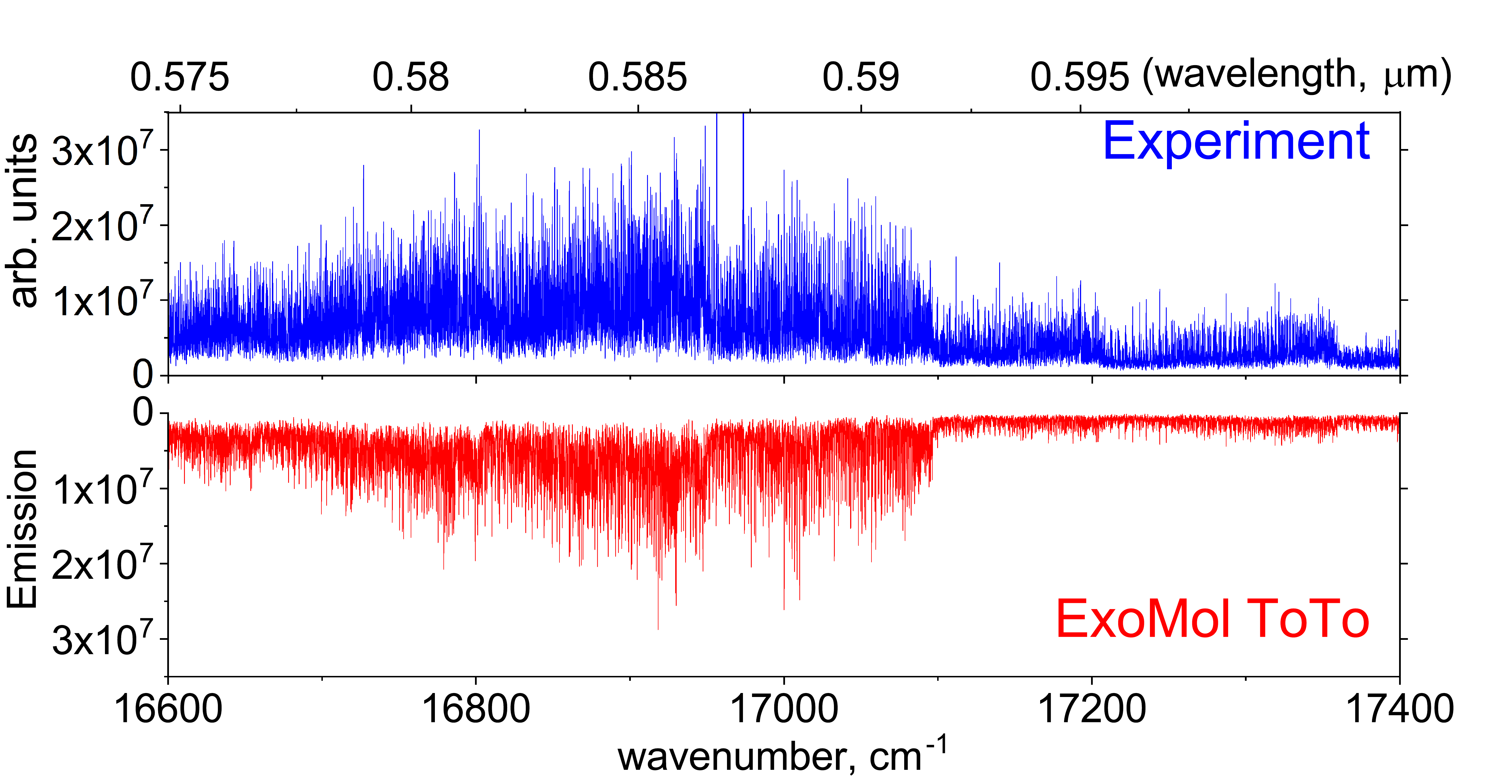}
\hspace{2em}

\includegraphics[width=0.48\textwidth]{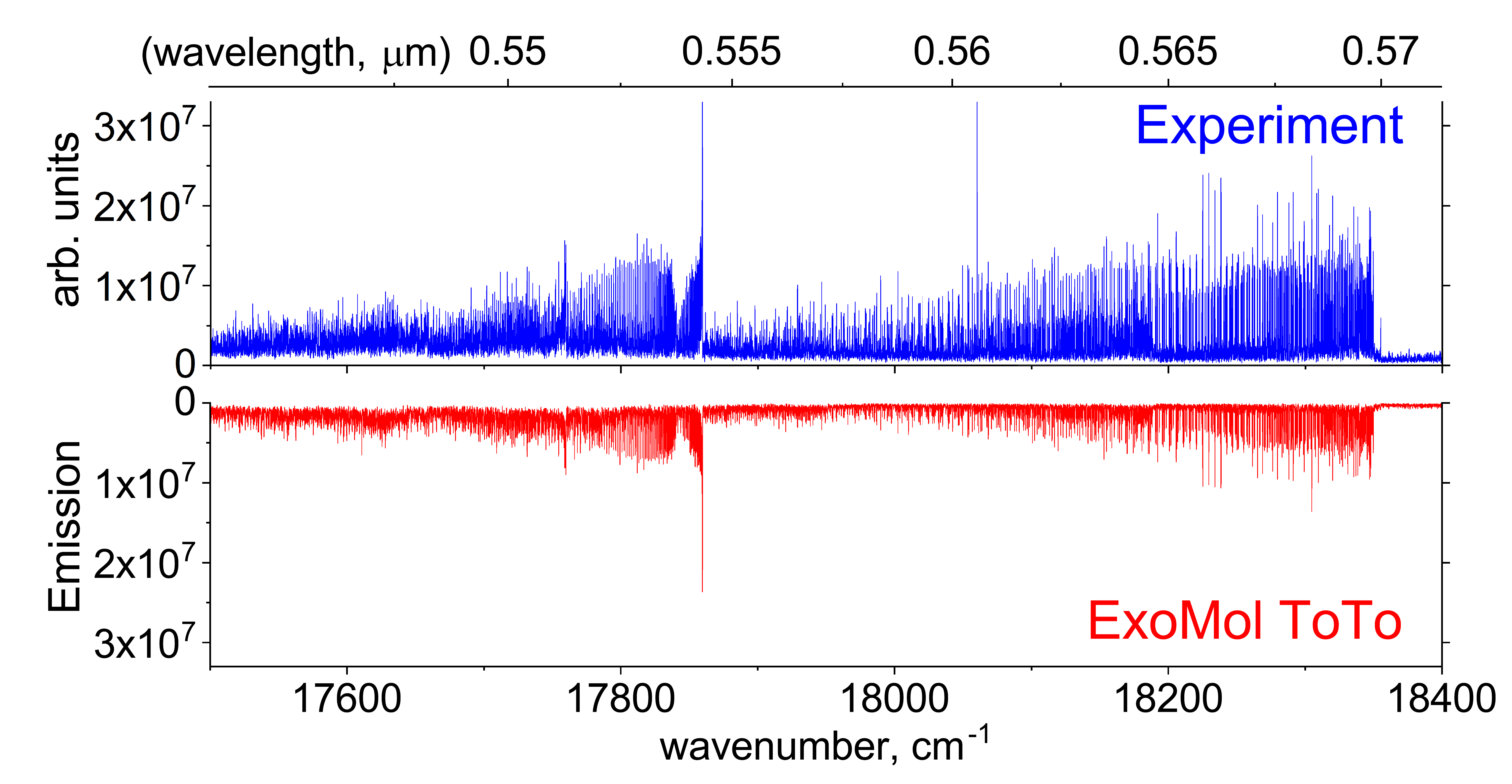}
\includegraphics[width=0.48\textwidth]{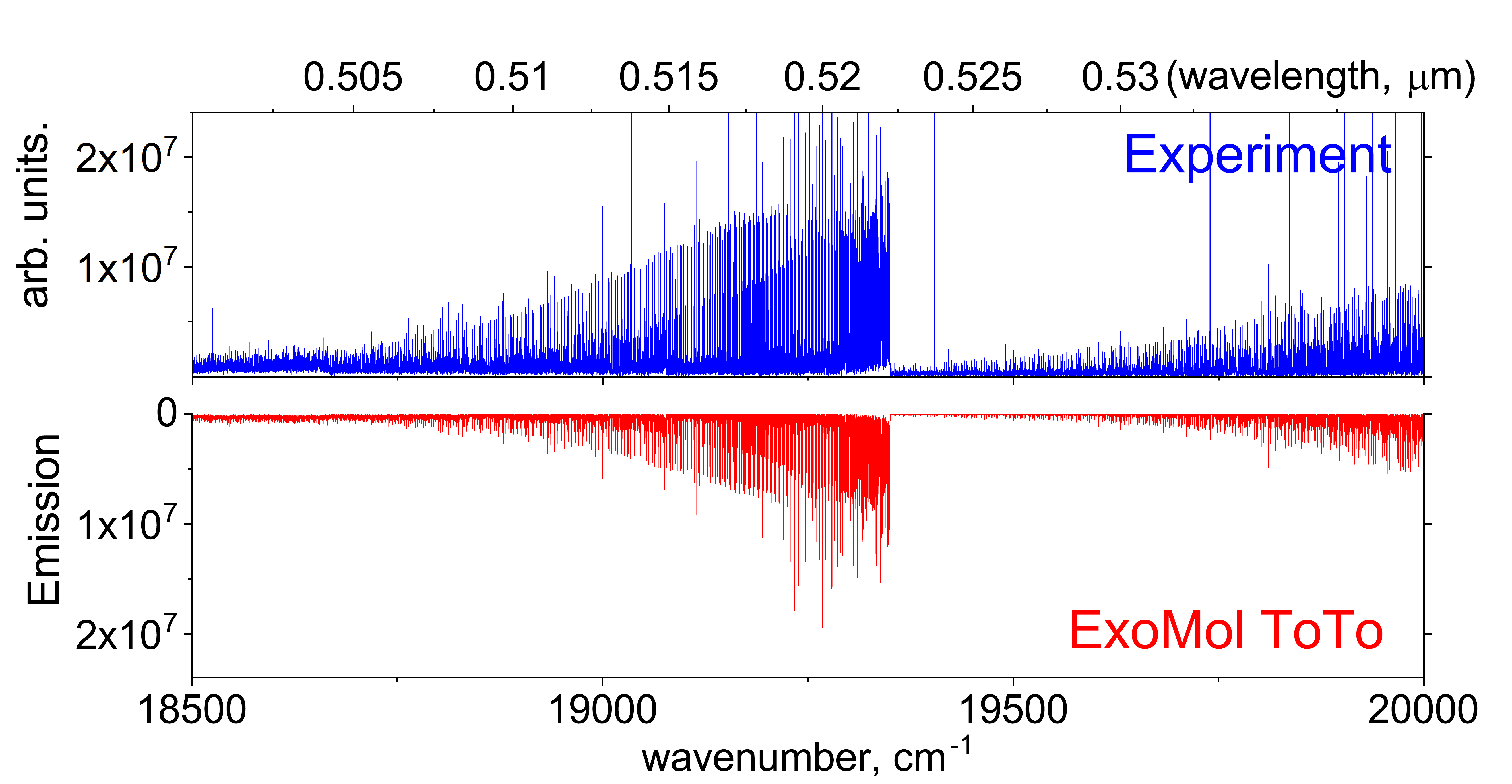}

\caption{\label{fig:Bernath} Comparison between the \LLname{} line list and experimental spectra from the Kitt Peaks archive at high resolution in bands with widths around 500 - 900 \cm{}. Details as for \Cref{fig:Bernathoverview}.
}
\end{figure*}

\begin{figure*}
\includegraphics[width=\textwidth]{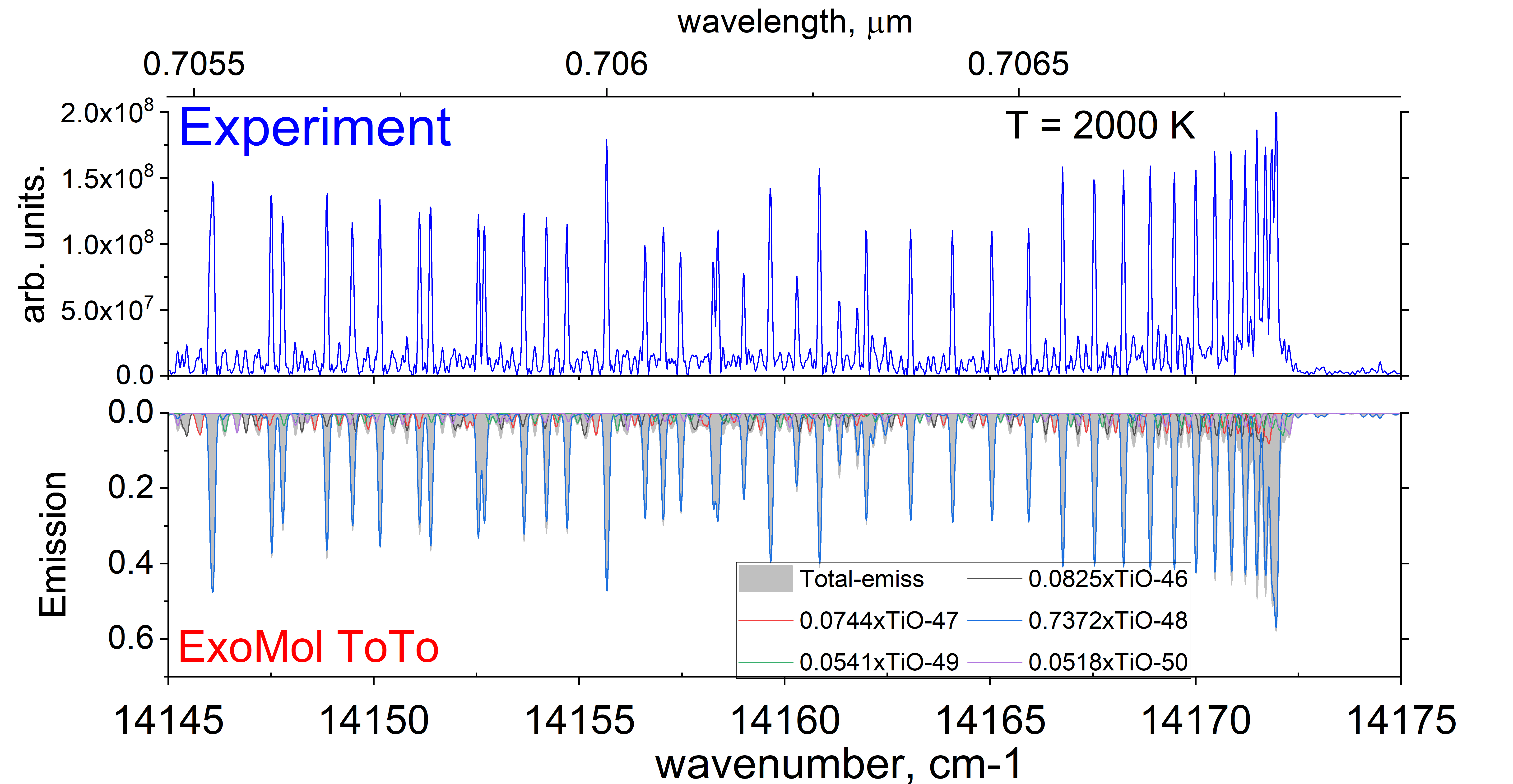}
\hspace{2em}

\caption{\label{fig:Bernathzoomed} Comparison between the \LLname{} line list and experimental spectra from the Kitt Peaks archive in smaller wavelength regions of less than 100 \cm{} at high resolution. Details as for \Cref{fig:Bernathoverview}, except for this figure also incorporates minor isotopologues in the \LLname{} spectra whose intensity is scaled by the standardised relative abundance of these isotopologues.
}
\end{figure*}

\subsubsection{Experimental Spectra}

\Cref{fig:Bernathoverview}, \Cref{fig:Bernath} and \Cref{fig:Bernathzoomed} compare experimental spectra in different wavelength bands obtained from the Kitt Peak archive (Bernath, \emph{Personal Communication} 2018) against cross-sections calculated using ExoCross using the \LLname{} line list.

\Cref{fig:Bernathoverview} shows the overview of all spectra. The bands are the same between the two spectra. The experimental spectra are stronger than the \LLname{} line list in the region above 16,500 \cm{}; while there are reasons to be uncertain of the relative intensities of the \LLname{} bands, the intensity calibration of the experimental spectrum is not sufficiently accurate for any changes to be made to \LLname{}. 

\Cref{fig:Bernath} compares the Kitt Peak and \LLname{} spectra in wavenumber bands of just under 1000 \cm{}. Overall, the spectra show very good agreement. The band heads are clearly very well reproduced in all cases. There are a few lines in the experimental spectra that are clearly experimental errors or impurities, e.g. the anomalous peak at 18,050 \cm{}. The experimental data has stronger lines than the \LLname{} model in the region 17,100 to 17,400 \cm{} with some structure; looking at the bands in this region, it is possible that the \LLname{} C-X band intensity should be slightly higher.

\Cref{fig:Bernathzoomed} compares the experimental spectra with that obtained using the \LLname{} line list in a specific spectral region at higher resolution. At this higher resolution, the contribution of the minor TiO isotopologues becomes clear, with these peaks forming the weaker underlying structure of the spectra while absorption from \TiO\ accounts for the strong lines and band structure.

\subsubsection{Comparison against Schwenke Line List}

\begin{figure}
\includegraphics[width=0.48\textwidth]{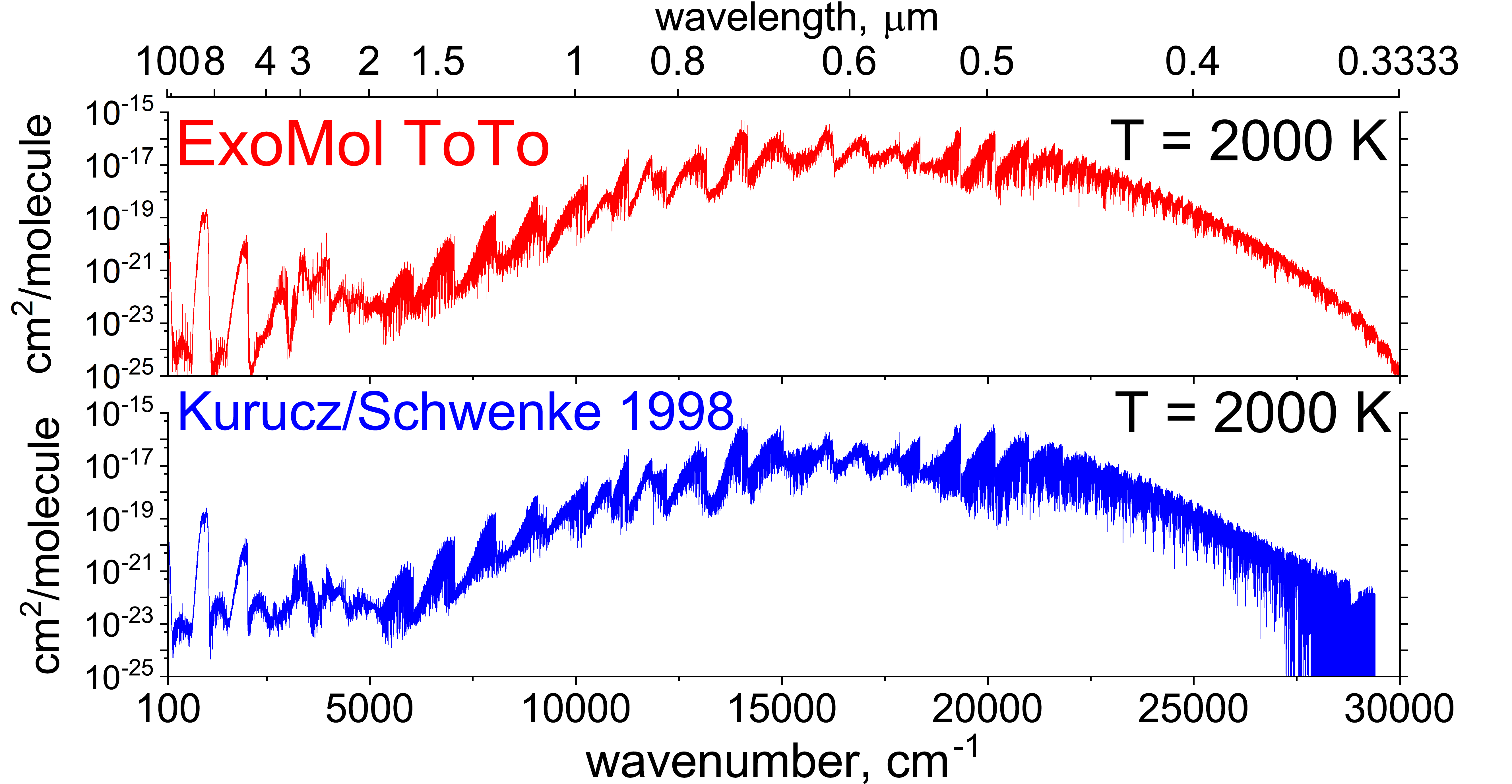}
\caption{\label{fig:Kurucz} Comparison between \LLname{} line list and the \citet{98Scxxxx.TiO} line list. Both cross sections were computed using Doppler broadening at $T=1000$~K. }
\end{figure}

\paragraph{Low resolution: } \Cref{fig:Kurucz} compares the ExoCross cross-sections of the \cite{98Scxxxx.TiO} and \LLname{} line lists from 0 to 30,000 \cm{} at 1000 K. There is good agreement across the whole spectra range, with the most significant differences occurring in the mid IR region from about 3000-5000 \cm{}, involving weaker transitions that are sensitive to parameters not yet well constrained experimentally or theoretically such as the parameters of the \D{} electronic state and spin-orbit coupling. The \LLname{} extends more smoothly to the high frequency region of the spectra.

\subsubsection{High resolution M-dwarf spectra}
\begin{figure}
\includegraphics[width=0.27\textwidth,angle=90]{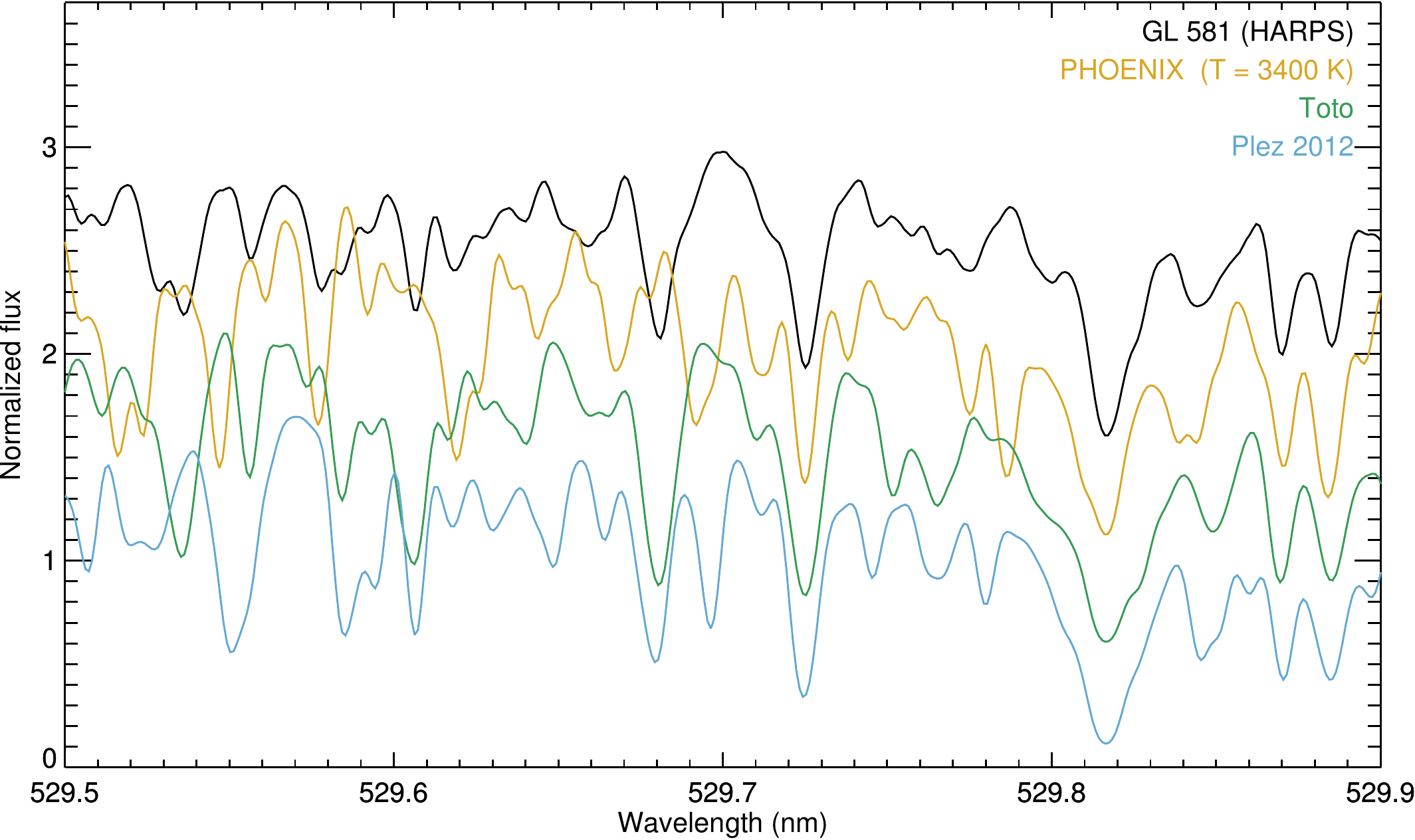}
\caption{\label{fig:killerpic} Stellar atmosphere models computed with the current Toto linelist, the list by Plez (2012) and PHOENIX, compared with HARPS observations of the M3V dwarf GL 581. The spectral region is chosen to match a key figure from \protect\cite{15HoDeSn.TiO}.}
\end{figure}

Past searches for the presence of TiO in the atmospheres of exoplanets have been impeded by inaccuracies in the line positions in available TiO linelists. This was first noted by \citet{15HoDeSn.TiO} who searched for the presence of TiO absorption in the transmission spectrum of the hot gas giant HD 209458 b using the cross-correlation technique \citep{Snellen2010}, and a TiO template based on the line-list used by \citet{Freedman2008}, itself a modification of the list by \citet{98Scxxxx.TiO}. Upon comparison with high-resolution observed spectra of cool main-sequence M-dwarf stars, these authors noted strong discrepancies between the observed spectra and the theoretical TiO template, leading them to the conclusion that the accuracies of the tabulated line positions were inadequate for the application of the cross-correlation technique at optical wavelengths.

\citet{17NuKaHa.TiO} noted that the accuracy of the line-list appeared to increase at wavelengths greater than $\sim630$ nm. They proceeded to observe the day-side of the extremely irradiated hot-Jupiter WASP-33 b at near-infra red wavelengths, and applied the cross-correlation technique in a similar manner, this time using the line-list provided by \citet{98Plxxxx.TiO}. This line-list exhibited similar inaccuracies at optical wavelengths, but was observed to match observed M-dwarf spectra in a number of bands, allowing \citet{17NuKaHa.TiO} to robustly detect the signature of TiO absorption in the day-side spectrum of an exoplanet for the first time.

Since 1998, the various TiO line lists have periodically been improved, mainly driven by the availability of new experimental constraints \citep[see e.g.][]{Freedman2008}. However, these improvements have not been comprehensively documented in the literature, and they have generally not been adopted in models of exoplanet atmospheres and stellar photospheres. The latest of these was published by Bertrand Plez in 2012, which we sourced from the VALD database \citep{VALD3}. In the following, this version of Plez's line list from 2012 will be referenced to as Plez-2012.

We compared synthetic spectra generated using the \LLname{} line list presented in this work and the line list Plez-2012, to high-resolution observations of M-dwarfs in a similar fashion as was done by \citet{15HoDeSn.TiO} and \citet{17NuKaHa.TiO}, in order to assess their (relative) accuracy and to determine whether it is suitable for application of the cross-correlation technique to high-resolution observations of exoplanets. The observed M-dwarf spectra are taken from the ESO data archive, and consist of an observation of GL 581 by the HARPS spectrograph at optical wavelengths (PID 072.C-0488, P.I. Mayor), and an observation of Wolf 359 by the UVES spectrograph (PID 082.D-0953, P.I. Liefke), covering the near infra-red.


\Cref{fig:killerpic} shows a comparison between synthetic spectra generated using the present \LLname{} line list data with Plez 2012, a PHOENIX stellar photosphere model that uses a previous version of the Plez line list \citep{Husser2013}, and the HARPS observation of GL 581. The wavelength range is chosen to match the range shown in \citep{15HoDeSn.TiO}. The list Plez-2012 and the present \LLname{} line list show comparable performance in many regions with most lines matching the observed stellar spectrum both in terms of relative depth and position. However, at multiple locations in the 529.5 to 529.6 nm range, the \LLname{} line list is superior to the Plez-2012 line list, with the former matching the stellar intensity pattern and line positions in this region more closely (e.g. the 529.56 and 529.53 lines).  The PHOENIX stellar photosphere model appears to use an older line list, as most lines show significant mismatches with the data, consistent with what was observed in \cite{15HoDeSn.TiO}.  Overall, it is clear that the \LLname{} data significantly outperforms any line list data discussed in \cite{15HoDeSn.TiO}.

\begin{figure}
\includegraphics[width=0.35\textwidth,angle=-90]{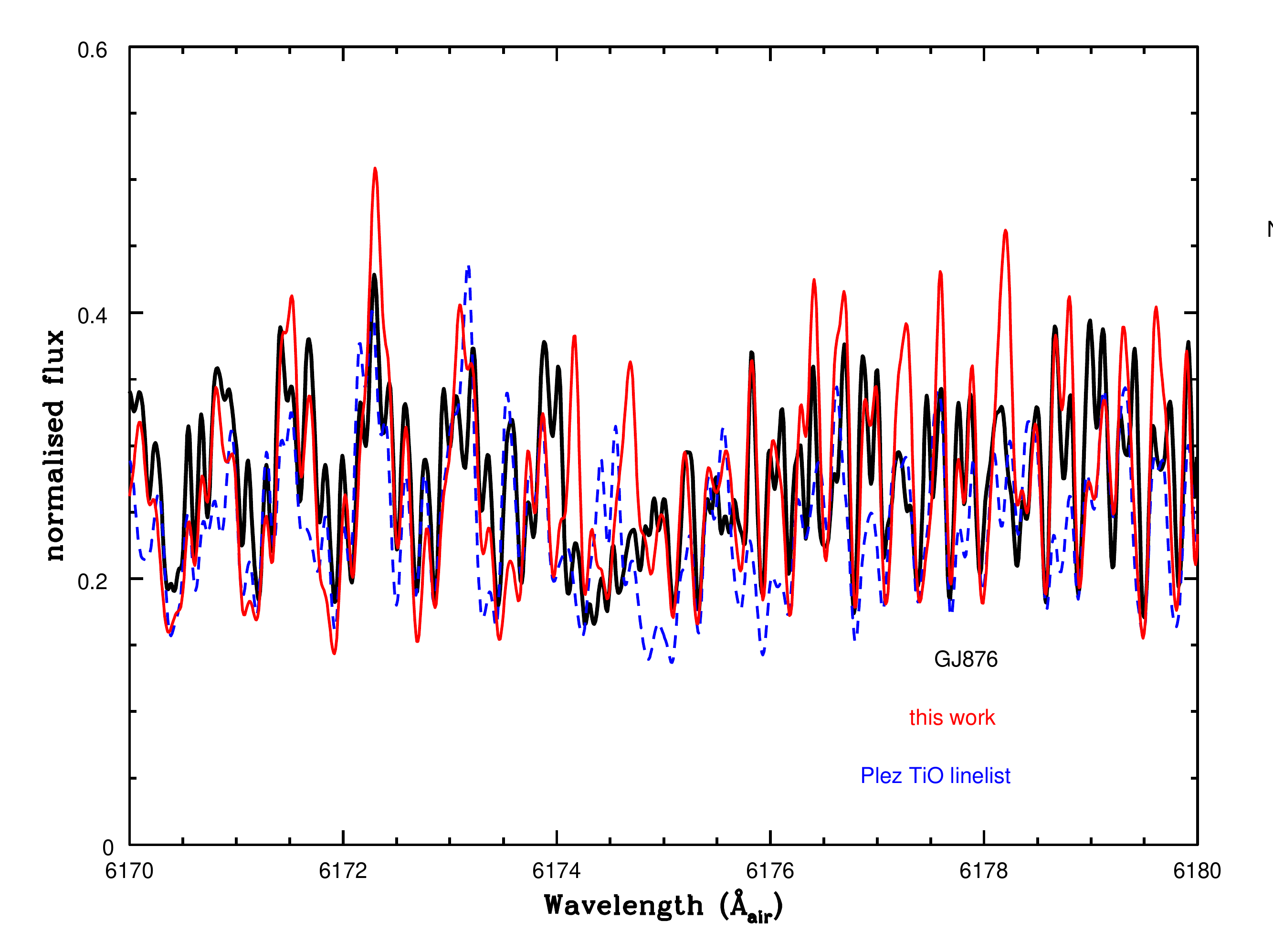}

\includegraphics[width=0.35\textwidth,angle=-90]{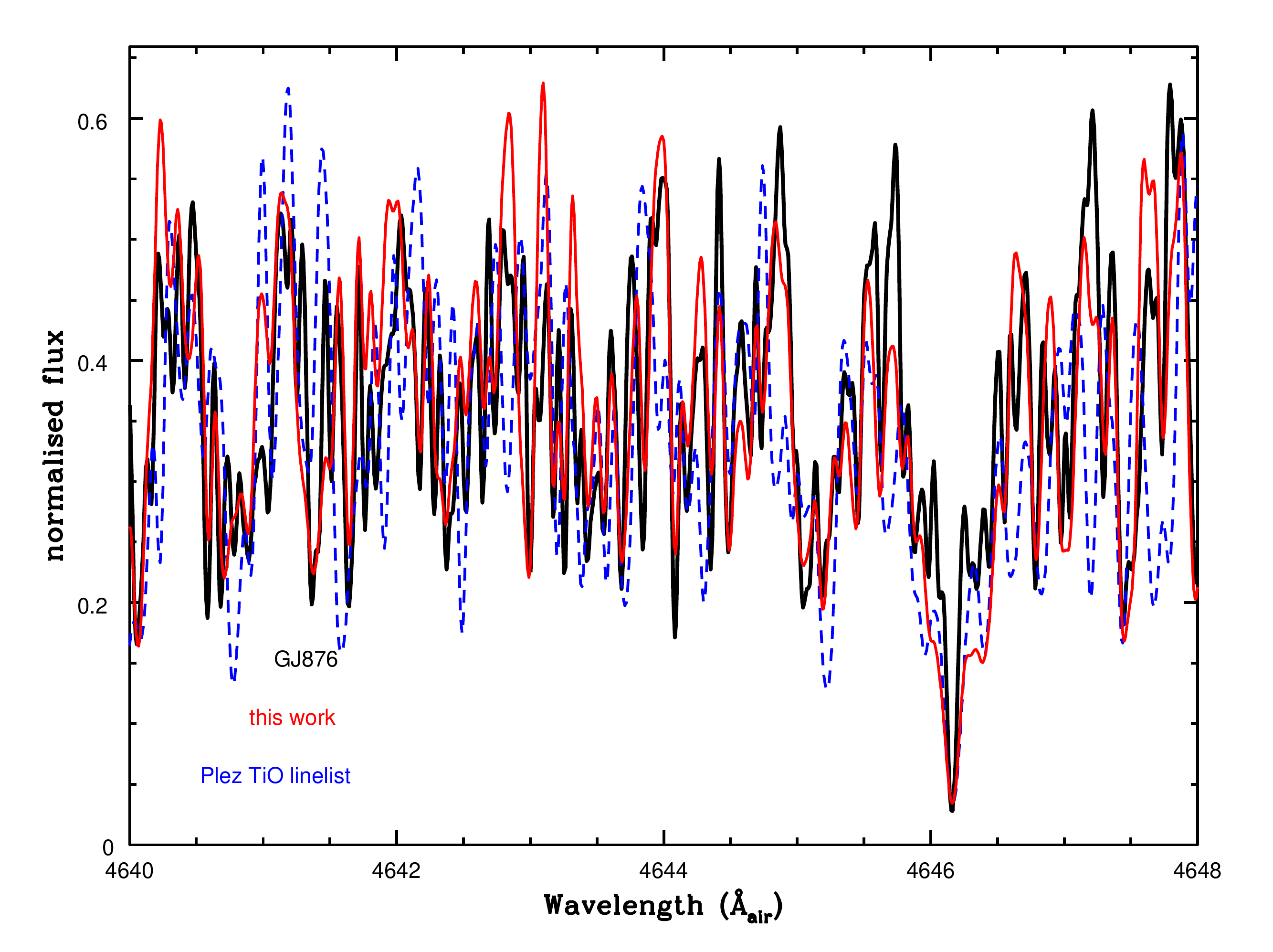}

\includegraphics[width=0.35\textwidth,angle=-90]{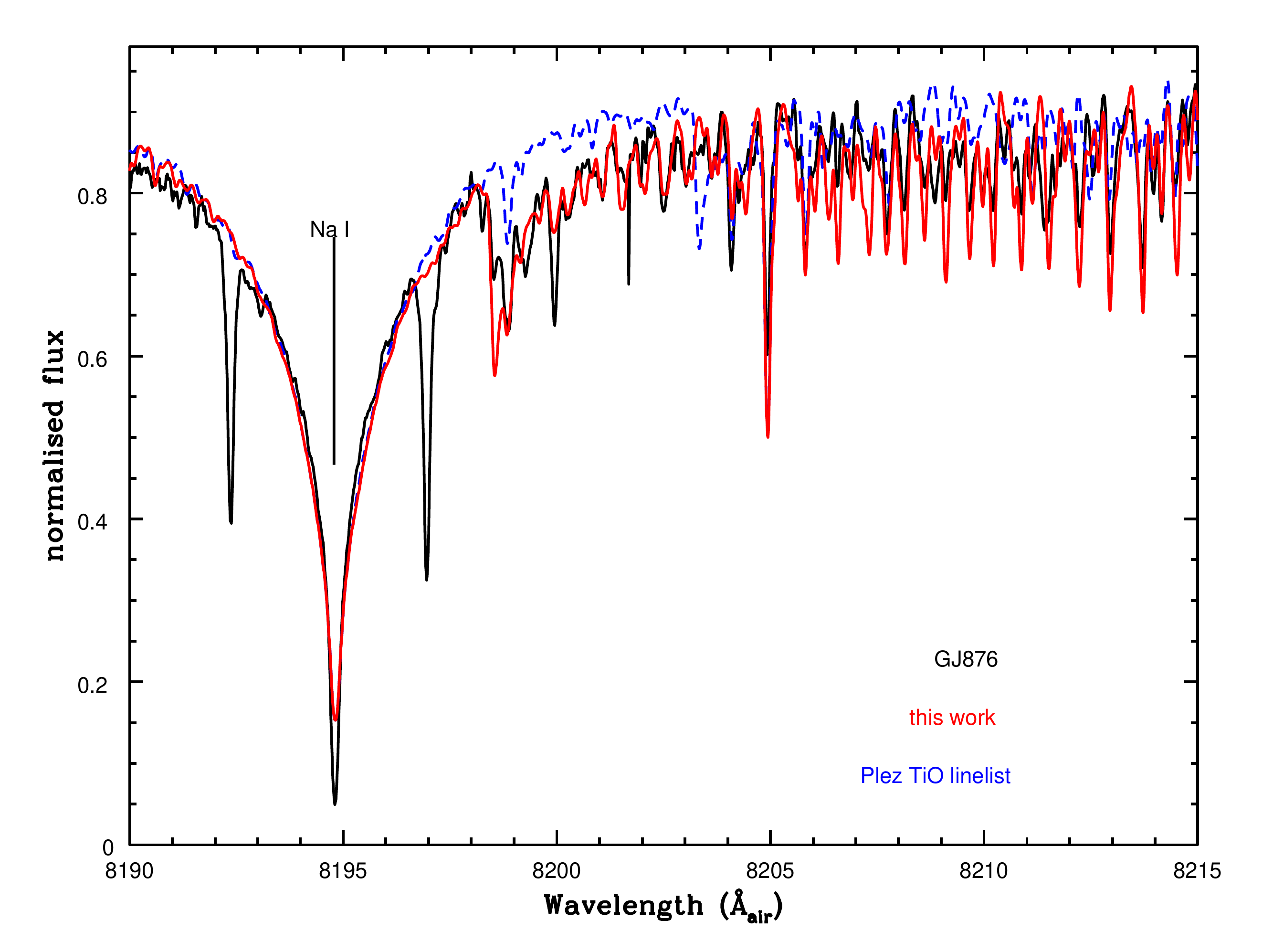}
\caption{\label{fig:GJ876} High resolution comparison of M dwarf spectra of GJ876 against full stellar photosphere models using  new \LLname{} linelist and older Plez linelist, in the region of the \B--\X{} transition (top), \C--\X{} transition (middle) and \Sb--\Sa{} transition (bottom).}
\end{figure}


For a more broad picture, \Cref{fig:GJ876} compares the high resolution spectra of the M-dwarf GJ~876 obtained with the HARPS spectrograph against synthetic spectra including all molecular and atoms species and using the Plez-2012 TiO line list and the \LLname{} line list from this work. It is clear that also in these regions, the \LLname line list shows significant improvements, particularly in the regions around 6174 \AA{} (\B-\X{} band) and 8202 \AA{} (\Sb-\Sa{} band).

\begin{figure}
\includegraphics[width=0.506\textwidth]{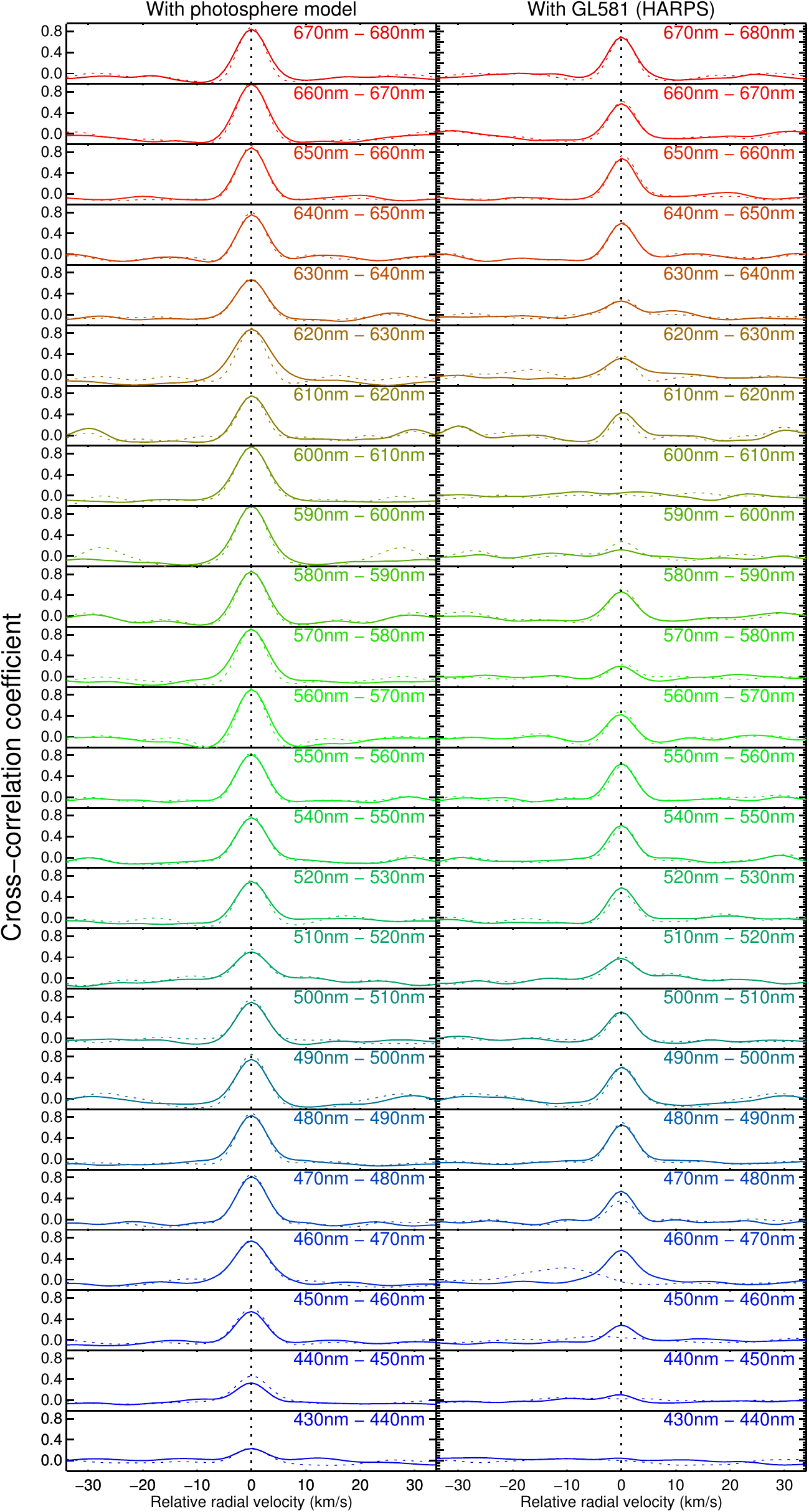}
\caption{\label{fig:cc1}
Band-wise cross-correlation in 10 nm bands between 430 and 680 nm  between the \LLname{} line list and the HARPS data for the M3V dwarf GL 581 (PID 072.C-0488, P.I. Mayor, right column), and with a stellar photosphere model that uses the Toto linelist to generate TiO opacity (left column). The peak value of the cross-correlation functions in the left column indicate the maximum correlation between an observed M-dwarf spectrum and the TiO spectrum that can be achieved with a perfect line list. Dashed lines indicate cross-correlation functions obtained with the equivalent approach using the Plez-2012 line list. }
\end{figure}

\begin{figure}
\includegraphics[width=0.5\textwidth]{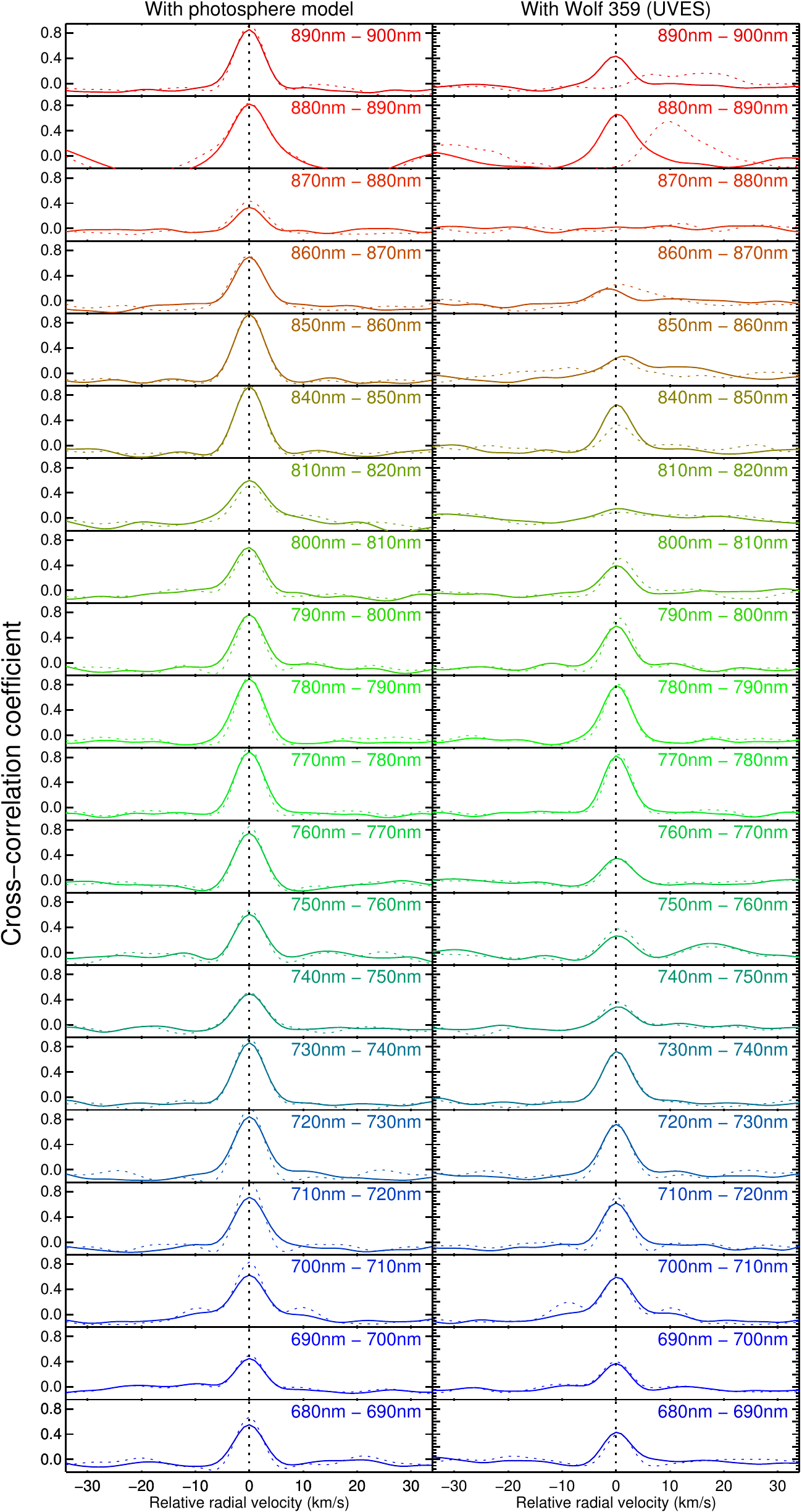}
\caption{\label{fig:cc2}
Band-wise cross-correlation in 10 nm bands between 680 and 900 nm between the \LLname{} line list and the UVES data for the Wolf 359, M6V (PID 082.D-0953, P.I. Liefke, right column), and with a stellar photosphere model that uses the Toto line list to generate TiO opacity (left column).}
\end{figure}


To quantify the performance of the \LLname{} and Plez-2012 line lists over a larger spectral range, we cross-correlate the synthetic M-dwarf spectra generated using these lists with the observed M-dwarf spectra, in a band-wise fashion similar to \citet{15HoDeSn.TiO} and \citet{17NuKaHa.TiO}. The cross-correlation templates are generated by simulating the emission spectrum of the M-dwarf without contribution from any species except \ce{^{48}Ti^{16}O}, and are cross-correlated with both a synthetic template that does include all elements (as to establish the maximal correlation in the case of a perfect line list), as well as the observed M-dwarf spectra. The data and the models are first continuum-subtracted using a high-pass Gaussian filter with a width of 0.2 nm, after which the cross-correlation coefficient is computed over a range of radial velocities, in bands with widths of 10 nm each. The resulting correlation functions are shown in Figures \Cref{fig:cc1} and \Cref{fig:cc2}. These show that the \LLname{} matches or outperforms the previous Plez-2012 line list in almost all bands. 

The left row of panels of \Cref{fig:cc1,fig:cc2} quantifies the contribution of \ce{^{48}Ti^{16}O} absorption to the full opacity of a model stellar spectra.  The plots show that TiO absorption dominates in most regions from 430 to 900 nm, apart from the regions near 870 -- 880 nm, 690 -- 700 nm and 430 -- 450 nm; in these regions, absorption from other atoms or molecules dominates resulting in lower peak cross-correlations.

The right columns of \Cref{fig:cc1,fig:cc2} show the cross-correlation of the \ce{^{48}Ti^{16}O}-only template with the observed M dwarf spectra. Cross-correlation functions that have substantially smaller peaks in the right than the left hand column indicate that the line list is imperfect at these wavelength regions. \LLname{} is generally of comparable quality as the Plez-2012 data across the full wavelength range, but with substantial improvements in 840 -- 850 nm and 450 -- 470 nm. Other areas of significant improvement are around 600 -- 670 nm (particularly the lower end of that range) and 845 -- 880 nm (particularly the middle part of this region). The 845 -- 880 nm (11,800 -- 11,360 \cm{}) range corresponds to the E-X 0-0 band with underlying A-X 0-2 bands, while the 570 -- 670 nm (17,540 -- 14,900 \cm{}) range corresponds to the A-X 2-0 and B-X 0-0 bands. Looking at the existing experimental data as compiled in \cite{jt672}, it is apparent that existing measurements in these spectral regions often originate from experiments performed by Phillips  in the 1970s (e.g. \cite{73Phxxxx.TiO}) which carry uncertainties of around 0.1 \cm{}. These uncertainties are too large for accurately computing the line lists in these regions. Therefore new laboratory measurements in these regions would be highly desirable.

Comparing with similar Figures in \cite{15HoDeSn.TiO} and \cite{17NuKaHa.TiO}, the recent Plez-2012 line list appears to be far superior to earlier line lists by Plez and Schwenke, which show very little correlation with observed M-dwarf spectra over the 500 to 630 nm range, and only small correlations from 630 to 670 nm. This improvement overall is almost certainly due to the replacement of theoretical transition frequencies with experimental frequencies.  The \LLname{} does a more fundamental replacement in correcting the theoretical energy levels with experimentally-derived energy levels from a \Marvel{} analysis.

\section{Conclusion and Future Directions}

Our complete \LLname{} line list for \ce{^{46}Ti^{16}O}, \ce{^{47}Ti^{16}O}, \ce{^{48}Ti^{16}O}, \ce{^{49}Ti^{16}O} and \ce{^{50}Ti^{16}O} can be accessed online at www.exomol.com in the ExoMol format described by \mbox{\cite{jt631}}. The main isotopologue \TiO{} line list contains \noELs{} energy levels and \notrans{} transitions.
It includes the transition energies and Einstein coefficients A(f $\leftarrow$ i).

Our investigations have highlighted that for high resolution applications such as molecular detection through cross-correlation of exoplanet spectra against templates, it is critical to "\Marvel ising" line lists by replacing theoretical levels by experimentally derived energy levels (e.g. from \Marvel{}).

The \LLname{} line list presented here is the best available line list for TiO and should be suitable for current and future exoplanet characterisation studies. However, there are significant areas for improvement in this line list in both the theoretical treatment of the electronic structure problem (which will influence the intensities) and experimental high resolution spectroscopy (which will improve the frequencies required for high resolution cross-correlation studies). For the former, it is yet unclear which area of improvement is presently most urgent - transitions from the ground electronic state have the biggest influence on the final spectra, but transitions between higher electronic states have much larger uncertainties with current methods (they are often not qualitatively trust-worthy). We performed a validation with the real-world TiO absorption spectrum by cross-correlating synthetic templates generated using the \LLname{} with observed high-resolution spectra M dwarf spectra. This analysis demonstrates that future improvements on the TiO line list should focus on the wavelength regions of 570 -- 640, 810 -- 820 and 850 -- 880 nm where only little correlation is currently observed. Improvements in either the theoretical treatment or the availability of new assigned experimental data can be readily incorporated into the line list. The Exomol team is open to future collaborations on this front.

\section*{Acknowledgements}
We thank Peter Bernath for supplying the Kitt Peak spectra quoted above and other helpful
discussions, Richard Freedman for useful comments on an initial line list, and Anna-Maree Syme for quantifying the number of spectral lines at different intensities and temperatures.
This work was supported by the European Research Council (ERC) under the
Advanced Investigator Project 267219, the UK Science and Technology Research Council (STFC) under grants No. ST/M001334/1 and ST/R000476/1, the COST action MOLIM
No. CM1405 and the European Union Horizon 2020 research and innovation programme under the Marie Sklodowska-Curie grant agreement No 701962.
T.M. and V.P.M. acknowledge support by the Spanish Ministry of Economy and Competitiveness (MINECO) under grant AYA-2017-88254-P. V.P.M. acknowledges the financial support from the Spanish Ministry of Economy and Competitiveness (MINECO) under the 2011 Severo Ochoa Program MINECO SEV-2011-0187. H.J.H. acknowledge the financial support of the European Research Council (ERC) under the European Union's Horizon 2020 research and innovation programme (project {\sc Four Aces}; grant agreement No 724427) and the National Centre for Competence in Research PlanetS, supported by the Swiss National Science Foundation. Based on observations collected at the European Organisation for Astronomical Research in the Southern Hemisphere under ESO programmes 072.C-0488 and 082.D-0953, and data products created thereof.




\bibliographystyle{mnras}




\bsp	
\label{lastpage}
\end{document}